\documentclass{article}
\usepackage[final]{neurips_2023}

\setcitestyle{authoryear,round,citesep={;},aysep={,},yysep={;}}

\usepackage[utf8]{inputenc} 
\usepackage[T1]{fontenc}    
\PassOptionsToPackage{hyphens}{url}
\usepackage{hyperref}
\usepackage{url}            
\usepackage{booktabs}       
\usepackage{amsfonts}       
\usepackage{nicefrac}       
\usepackage{microtype}      
\usepackage[table]{xcolor}  
\usepackage{placeins}       
\usepackage{ragged2e}
\usepackage{longtable}

\usepackage[most]{tcolorbox} 
\colorlet{lightgray}{gray!8}
\colorlet{darkgray}{gray!55}
\definecolor{myyellow}{HTML}{fff2cc}

\newcolumntype{C}[1]{>{\centering\arraybackslash}p{#1}}%
\newcolumntype{D}[1]{>{\centering\arraybackslash}m{#1}}%

\usepackage{CJKutf8}

\usepackage[labelfont=bf]{caption}
\usepackage{pdfpages} 

\usepackage{tablefootnote}

\usepackage{changepage}

\usepackage{tocloft}
\usepackage[capitalize,noabbrev, nameinlink]{cleveref}

\makeatletter
\renewcommand*{\@fnsymbol}[1]{\ensuremath{\ifcase#1\or \or * \or \dagger \or \ddagger\or
   \mathsection\or \mathparagraph\or \|\or **\or \dagger\dagger
   \or \ddagger\ddagger \else\@ctrerr\fi}}
\makeatother

\title{Governing Through the Cloud:\\The Intermediary Role of Compute Providers in AI Regulation}

\author{%
  \parbox{\linewidth}{\centering\bfseries\large%
  Lennart Heim,\!\thanks{Each author contributed ideas and/or writing to the paper. However, being an author does not imply agreement with every claim made in the paper.}\thanks{Denotes primary authors who contributed most significantly to the direction and content of the paper. Both primary authors and other authors are listed in approximately descending order of contribution.}\enskip \thanks{Correspondence to \href{mailto:lennart.heim@governance.ai}{\texttt{lennart.heim@governance.ai}}.}\enskip $^{1}$ Tim Fist,\!$^{\ast\,2}$ Janet Egan,\!$^{\ast\,3}$ Sihao Huang,\!$^{\ast\,4}$\\ Stephen Zekany,$^{\ast}$\,\thanks{Work completed whilst a Summer Research Fellow at the Centre for the Governance of AI, Oxford.} \;$^{1}$ Robert Trager,\!$^{\ast\,1, 6, 7}$ Michael A Osborne,\!$^{5, 7}$ Noa Zilberman$^{5}$
  \parbox{\linewidth}{\vspace{1em}\mdseries\centering\small
  $^{1}$Centre for the Governance of AI (GovAI),
  $^{2}$Center for a New American Security, Institute for Progress,
  $^{3}$Harvard Kennedy School,
  $^{4}$Department of Politics and International Relations, University of Oxford,
  $^{5}$Department of Engineering Sciences, University of Oxford,
  $^{6}$Blavatnik School of Government, University of Oxford,
  $^{7}$Oxford Martin AI Governance Initiative
  }
}
}

\begin{document}

\includepdf[pages={1}]{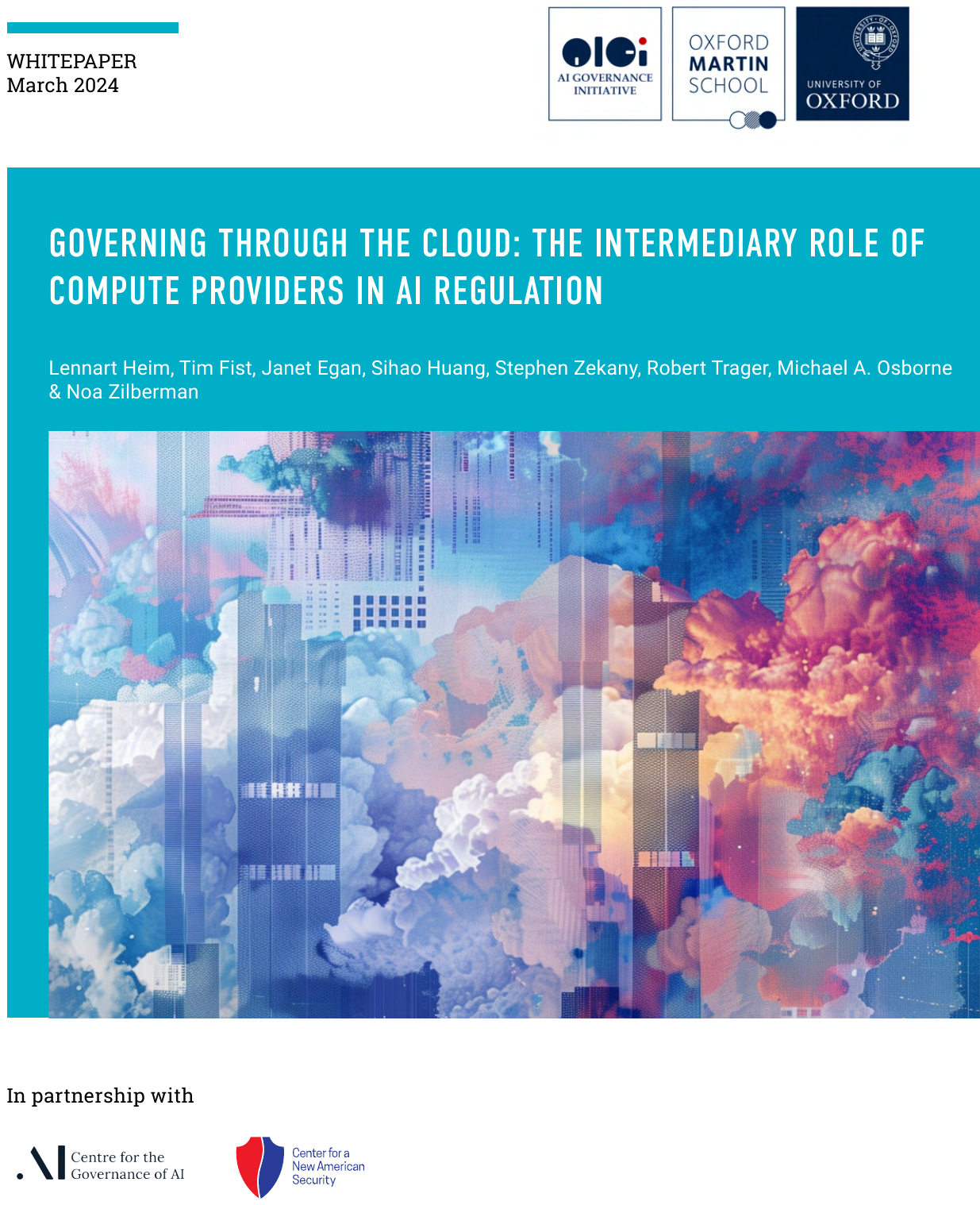}

\maketitle
\setcounter{footnote}{0}

    \begin{abstract}
    As jurisdictions around the world take their first steps toward regulating the most powerful AI systems, such as the EU AI Act and the US Executive Order 14110, there is a growing need for effective enforcement mechanisms that can verify compliance and respond to violations. We argue that compute providers should have legal obligations and ethical responsibilities associated with AI development and deployment, both to provide secure infrastructure and to serve as intermediaries for AI regulation. Compute providers can play an essential role in a regulatory ecosystem via four key capacities: as \emph{securers}, safeguarding AI systems and critical infrastructure; as \emph{record keepers}, enhancing visibility for policymakers; as \emph{verifiers} of customer activities, ensuring oversight; and as \emph{enforcers}, taking actions against rule violations. We analyze the technical feasibility of performing these functions in a targeted and privacy-conscious manner and present a range of technical instruments. In particular, we describe how non-confidential information, to which compute providers largely already have access, can provide two key governance-relevant properties of a computational workload: its type---e.g., large-scale training or inference---and the amount of compute it has consumed. Using AI Executive Order 14110 as a case study, we outline how the US is beginning to implement record keeping requirements for compute providers. We also explore how verification and enforcement roles could be added to establish a comprehensive AI compute oversight scheme. We argue that internationalization will be key to effective implementation, and highlight the critical challenge of balancing confidentiality and privacy with risk mitigation as the role of compute providers in AI regulation expands.
    \end{abstract}

\newpage

\section*{Executive Summary}\addcontentsline{toc}{section}{Executive Summary}

\textbf{Introduction} --- Jurisdictions around the world are taking
their first steps toward regulating AI, such as the EU AI Act and the US
Executive Order 14110. While these regulatory efforts mark
significant progress, they lack robust mechanisms to verify compliance
and respond to violations. We propose compute service providers (\emph{compute providers} below) as an important node for AI safety, both in
providing secure infrastructure and acting in an intermediary role for
AI regulation, leveraging their unique relationships with AI developers
and deployers. Our proposal is not intended to replace existing
regulations on AI developers but rather to complement them. (\cref{introduction})

\textbf{Compute Providers' Intermediary Role} --- Increasingly large
amounts of computing power are necessary for both the development and
deployment of the most sophisticated AI systems.
Consequently, advanced AI models today are trained, and deployed in
data centers, housing tens of thousands of ``AI accelerators''
(specialized computers for AI applications). Because of the large
upfront cost of building this infrastructure and economies of scale, AI developers
often access large-scale compute through models like
Infrastructure as a Service (IaaS), also often described as \emph{cloud
computing}. (\cref{compute-providers-intermediary-role})

\begin{figure}[!h]
    \centering
    \centerline{\includegraphics[width=1.05\linewidth]{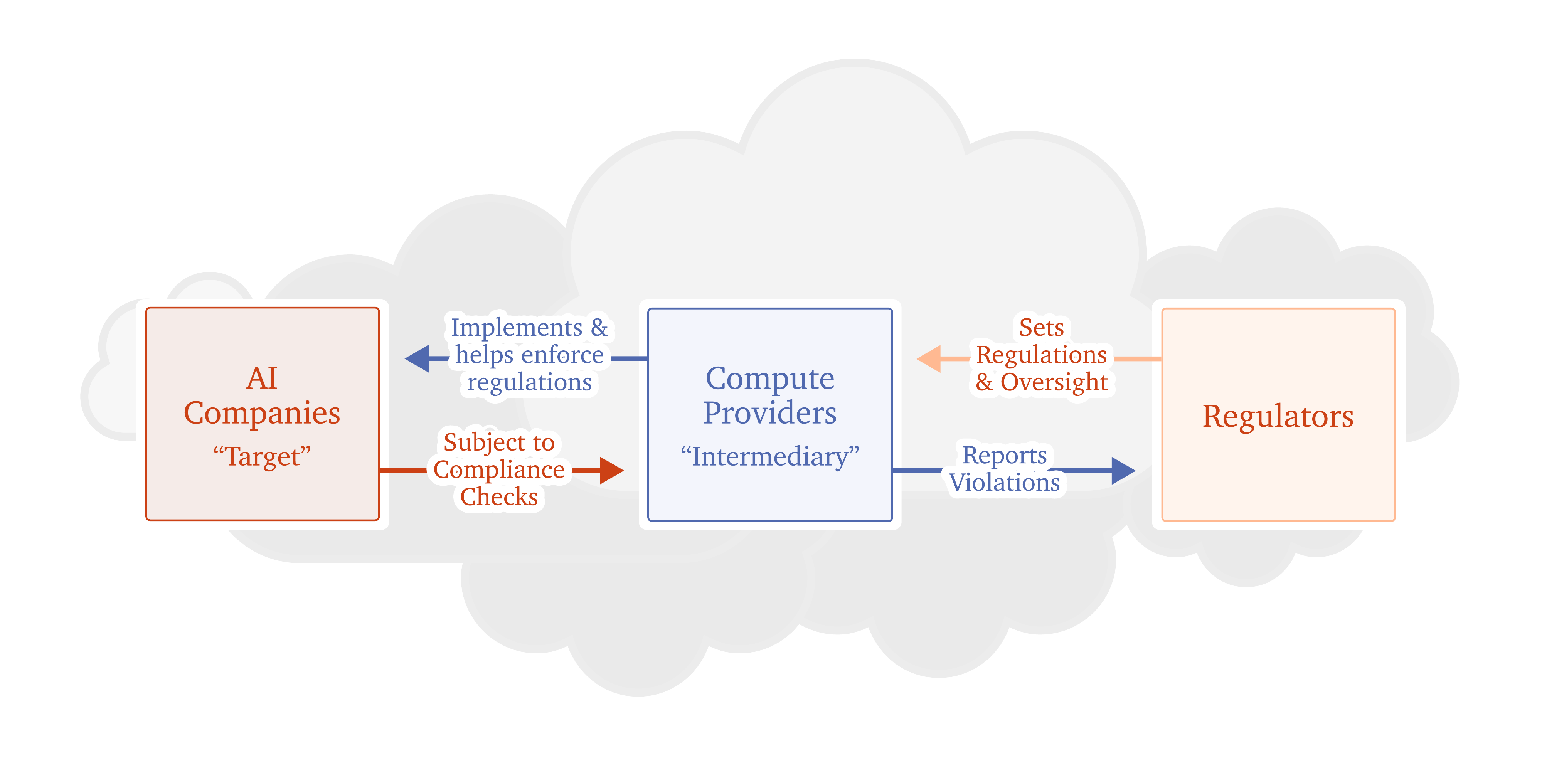}}
    \caption{The intermediary role of compute providers in relation to AI companies and regulators.}
    \label{fig:ES-Intermediary-role}
\end{figure}

Some leading AI firms currently manage their own data centers or
maintain exclusive partnerships with leading entities in this domain,
known as \emph{hyperscalers}. Notably, the most advanced AI research is
currently being conducted at or with these hyperscalers (e.g., Microsoft
Azure, Amazon Web Services (AWS), Apple, Bytedance, Meta, Oracle,
Tencent, and Google Cloud). While this situation presents complex
challenges for regulatory oversight, our discussion also encompasses
scenarios in which compute providers are internal to or closely linked
with an AI firm.

We focus on frontier AI systems, which
have the potential to give rise to dangerous capabilities and pose
serious risks. As these systems necessitate extensive amounts of compute
to train and deploy at large scales, targeting compute providers becomes
a promising method to oversee the development and deployment of such
systems. Furthermore, this target narrows the regulatory scope to the
smaller set of key customers who are building AI systems at the
frontier, thereby minimizing the burdens associated with regulatory
compliance and enforcement. (\cref{focus-on-frontier-ai})

\textbf{Governance Capacities} --- We propose that compute providers can
leverage their crucial role in the AI supply chain to secure
infrastructure and serve as the intermediate node in support of
regulatory objectives while maintaining customers' privacy and rights.
They can facilitate effective AI regulation via four key capacities: as
\emph{securers}, \emph{record keepers}, \emph{verifiers}, and, in some
cases, even \emph{enforcers}. Reporting represents a related yet
distinct dimension, wherein compute providers provide information to
authorities as mandated by law or regulations. (\cref{governance-capacities-of-compute-providers})

\clearpage

\begin{table}[ht]
\renewcommand{\arraystretch}{2}
\vspace{2em}
\centering
\begin{adjustwidth}{-0.05\linewidth}{-0.05\linewidth}
\resizebox{1.0\linewidth}{!}{%
    \begin{tabular}{>{\itshape}C{0.25\linewidth}>{\columncolor{lightgray}\itshape}C{0.25\linewidth}>{\itshape}C{0.25\linewidth}>{\columncolor{lightgray}\itshape}C{0.25\linewidth}}
    \toprule
    \multicolumn{4}{c}{\bfseries\Large Governance Capacities} \\[3pt]\hline
    \multicolumn{1}{c}{\bfseries\large Security} & \multicolumn{1}{c}{ \cellcolor{lightgray}\bfseries\large  Record Keeping} & \multicolumn{1}{c}{\bfseries\large  Verification} & \multicolumn{1}{c}{ \cellcolor{lightgray}\bfseries\large  Enforcement} \\\hline
    Helping provide physical and cybersecurity measures to secure the AI model, related intellectual property, and personal and confidential data.%
    & The selective collection, organization, and maintenance of high-level information of a compute provider’s infrastructure usage, such as a customer’s compute usage data.\footnotemark%
    & Actively verifying customer identities, specific activities, and high-level AI systems’ properties.%
    & Restriction or limitation of compute access to customers or workloads for non-compliant customers. \\\hline
    \multicolumn{4}{c}{\bfseries\large Enables} \\\hline
    Enables shared security standards to protect the public good, such as safeguarding critical infrastructure and helping prevent model theft. & Increases visibility into AI development, links customers and their usage to real-world actors, and enables post-incident attributions and forensics. & Ensures that the deployment and development of AI systems adhere to regulations or company policies and reported properties. & Directly impacts the capability of customers to develop or deploy advanced AI systems, ensuring adherence to rules.\\\hline
    \multicolumn{4}{c}{\bfseries\large Examples} \\\hline
    \begin{tabular}[t]{@{}>{\centering\arraybackslash}p{\linewidth}@{}}Help prevent IP (e.g., algorithms), model weights, and training data from being stolen by malicious actors.\\ Help prevent attacks on large-scale deployments of foundational models that could shut down dozens of critical services nationwide.\end{tabular} &
    \begin{tabular}[t]{@{}>{\centering\arraybackslash}p{\linewidth}@{}}Obtain insights into national compute use trends for policy formulation, such as compute distribution (e.g., US NAIRR).\\ Enable monitoring for suspected violations of the reporting requirements under Executive Order 14110.\\ Collect information on the environmental impact of AI compute use.\end{tabular} &
    \begin{tabular}[t]{@{}>{\centering\arraybackslash}p{\linewidth}@{}}Confirm compliance with mandatory reporting over training compute thresholds.\\ Verify compliance with data usage guidelines for frontier AI training.\\ Verify if the deployed frontier AI system has an adequate license or certification.\end{tabular} & 
    \begin{tabular}[t]{@{}>{\centering\arraybackslash}p{\linewidth}@{}}Restrict access to customers lacking licenses as an AI developer for their system.\\ Refuse to deploy an unlicensed or non-compliant AI model.\\ Disable AI systems that demonstrate activity that is undesirable, uncontrollable, or in violation of regulations (e.g., computer worm-like AI system).\end{tabular}\\\bottomrule
    \end{tabular}%
    }
\end{adjustwidth}
\caption{Summary of the key governance capacities that compute providers can enable.}
\label{tab:Governance-Capacities-ES}
\end{table}

\FloatBarrier
\footnotetext{Focusing on essential data that informs without compromising privacy and confidentiality.}

\textbf{Technical Feasibility} --- Our analysis indicates these
governance capabilities are likely to be technically feasible and
possible to implement in a confidentiality- and privacy-preserving way
using techniques available to compute providers today. Compute providers
often collect a wide range of data on their customers and workloads, for
the purposes of billing, marketing, service analysis, optimization, and
fulfilling legal obligations. Much of this data could also be used to
support identity verification, as well as verifying technical properties
of workloads. At a minimum, providers have access to billing information
and can access basic technical data on how their hardware is used. This
likely makes it possible for compute providers to develop techniques to
detect and classify certain relevant workloads (e.g., whether a workload
involves training a frontier model) and to quantify the amount of
compute consumed by a workload. Verification of more detailed properties
of a workload, such as the type of training data used, or whether a
particular model evaluation was run, could be useful for governance
purposes but is not currently possible without direct access to customer
code and data. With further research and development efforts, compute
providers may be able to offer ``confidential computing'' services to
allow customers to prove these more detailed properties without
otherwise revealing sensitive data. (\cref{technical-feasibility-of-compute-providers-governance-role})

\textbf{Constructing an Oversight Scheme} --- Via Executive Order 14110 the US
government is already beginning to implement record keeping roles for
compute providers by requiring them to implement a Customer
Identification Program (essentially a Know-Your-Customer (KYC) program) for foreign
customers, and to report foreign customer training of highly capable
models to government. Expanding the role of compute providers to also
record and validate domestic customers using compute at frontier AI
thresholds could enable the US government to identify and address AI
safety risks arising domestically. Complementing these measures with
verification and enforcement roles for compute providers could further enable
the construction of a comprehensive compute oversight scheme, and ensure
that AI firms and developers are complying with AI regulations. (\cref{constructing-an-oversight-scheme})

\begin{figure}[ht]
    \centering
    \centerline{\includegraphics[width=1\linewidth]{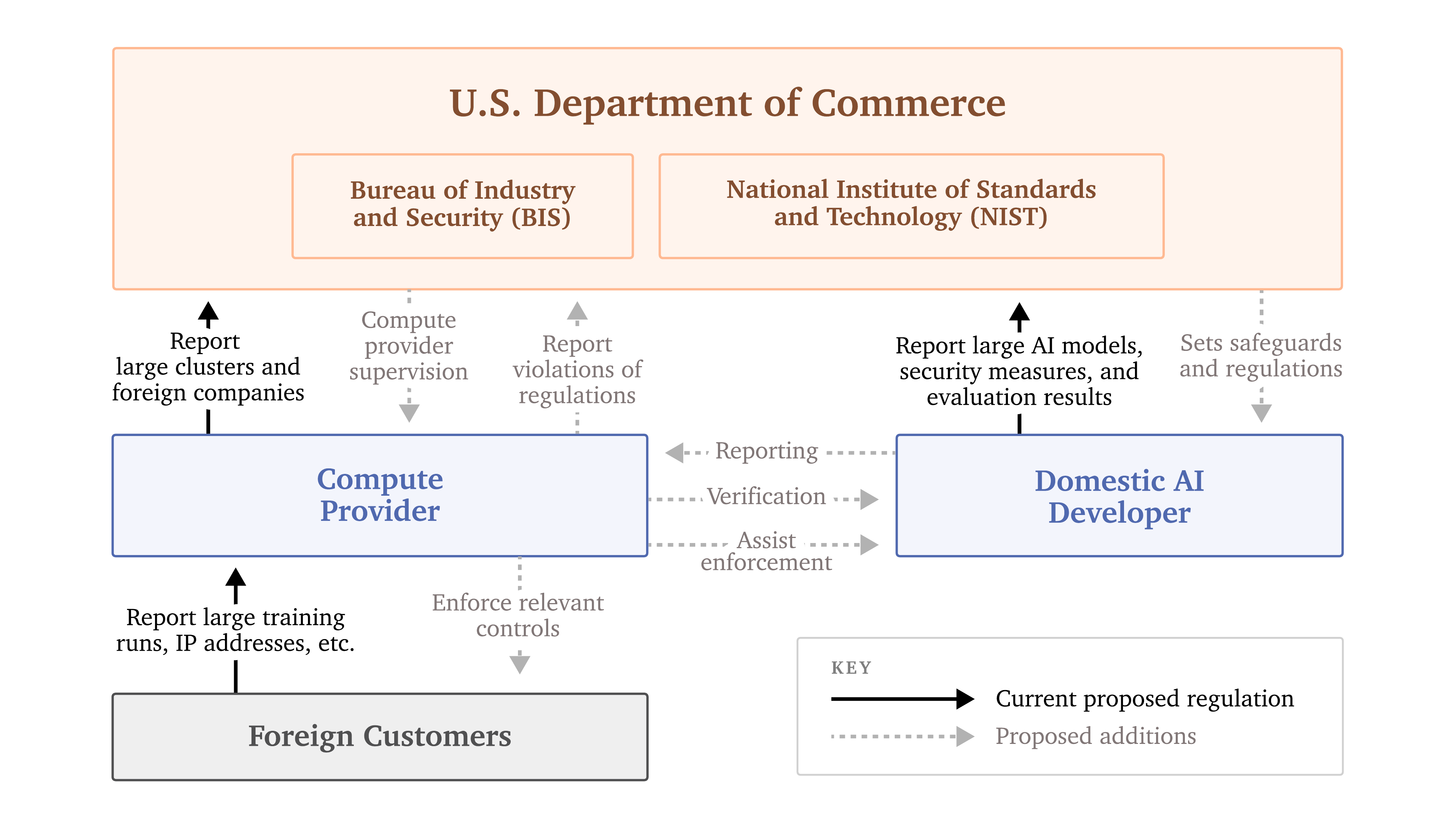}}
    \caption{Additional measures, implemented by the Department of
Commerce, would strengthen the intermediary role of compute providers
and enable a compute oversight scheme.}
    \label{fig:ES-Additional-Measures-Commerce}
\end{figure}

\textbf{Technical and Governance Challenges} --- To realize a robust
governance model, several technical and governance challenges remain. These include identifying additional measurable properties of
AI development that correspond to potential threats, making workload classification methods robust to potential evasion, and
formulating privacy-preserving verification protocols. (\cref{technical-challenges})

The success of our proposed oversight scheme hinges on its multilateral
adoption to prevent the migration of AI activities to jurisdictions with
less stringent oversight. For an international framework to be durable
and effective, it must address concerns from non-US governments.
Cooperation will need to account for complex privacy and oversight
issues associated with globally spread data centers. Compute provider
oversight may affect competition in the AI ecosystem and raise concerns
about issues of national competitiveness, and, consequently, this may
influence the ability of US providers to offer products globally,
including to foreign public-sector customers. Industry-led
privacy-preserving standards could help ensure trust, but further
research is needed to incentivize broad
international buy-in to a global framework. (\cref{limitations-and-future-research-directions} and \cref{governance-challenges})

\textbf{Conclusion} --- Compute providers are well-placed to support
existing and future AI governance frameworks in a privacy-preserving
manner. Many of the interventions we propose are feasible with the
current capabilities of compute providers. However, realizing the
full potential necessitates addressing technical and governance
challenges, requiring concerted efforts in research and international
cooperation. As governments and regulatory bodies move to address AI
risks, compute providers stand as the intermediate node in ensuring the
effective implementation of regulation. (\cref{conclusion})

\clearpage

\tableofcontents

\clearpage

\section{Introduction}\label{introduction}

As governments, international organizations, and regional bodies
formulate approaches for governing advanced AI, we ask: how can
authorities gain visibility into development and deployment practices
and enforce rules? The visibility is currently obscured because frontier
AI development largely takes place within the private sector and often
relies on self-reporting by AI companies, which may not always be
reliable due to growing incentives to obfuscate the results
\citep{anderljung2023, mulani2023, whittlestoneWhyHowGovernments2021}.
Ongoing governance processes worldwide require answers to this question.

Jurisdictions including China, the European Union (EU), the United
Kingdom (UK), and the United States (US) are attempting to impose new
reporting requirements on AI firms and compute providers
\citep{futureoflifeinstituteEUArtificialIntelligence,
jiangChinaCreateImplement2023,
secretaryofstateforscienceinnovationandtechnologyProinnovationApproachAI2023,
SenatorWienerIntroduces2024, sheehanTracingRootsChina2024,
thewhitehouseExecutiveOrderSafe2023}. They are also beginning to
regulate the development and deployment of models with potentially
harmful capabilities. Yet, enforcing these rules proves challenging;
without appropriate mechanisms, it is difficult to detect
violations \citep{hacker2023, whittlestone2023}. Governance
approaches, such as the recent US Executive Order
\citep{thewhitehouseExecutiveOrderSafe2023} and the EU's proposed AI
Act \citep{council_of_the_european_union_proposal_2024}, have not
developed practical means such as robust spot-checking (randomized
inspections to ensure compliance) and evidence-gathering mechanisms for
achieving these goals.

This paper demonstrates how compute providers---firms who make
computing resources (``compute'' below)\footnote{For a discussion of
  compute as a governance node, see
  \cite{sastryComputingPowerGovernance2024}.} available 
for AI development and deployment---can effectively serve as an
intermediary for frontier AI governance between governments and the
firms developing and deploying AI. In this role, compute providers can
act as a first line of detection of violations of a governance regime
and even defend against violations.

\subsection{Compute Providers' Intermediary
Role}\label{compute-providers-intermediary-role}

Large amounts of computing power are necessary for both the development
and deployment of frontier AI systems. Consequently, advanced AI models
are trained, and deployed, in \emph{data centers}\footnote{Our focus is
  on the entities that own and operate data centers, prioritizing the
  ``legal entity'' level of abstraction over the physical locations or ``data
  centers'' themselves.}, housing tens of thousands of AI accelerators.
Because of the large upfront cost of building this infrastructure
, AI developers often access large-scale compute through
models like IaaS\footnote{This discussion extends
  to services that offer hardware access with sufficient flexibility for
  customer-defined usage, which may occasionally encompass certain
  Platform-as-a-Service (PaaS) offerings. However, our emphasis is on
  scenarios in which usage surpasses certain AI compute thresholds that
  are of relevance for frontier AI. In contrast, services providing
  access to pre-configured AI models (Software-as-a-Service, or SaaS)
  fall outside our defined scope of compute providers. (The suggested
  governance capacities could still help if the regulation of these
  services is desired.)}, also often described as \emph{cloud
computing}.\footnote{The term ``cloud'' is more associated with a specific
  business model rather than the underlying activity of providing
  compute resources. For this context, we prefer the term ``compute
  providers'' to accurately reflect the focus on the provision of
  computing power. This choice also allows us to include entities that
  predominantly provide their computational resources internally (e.g., Meta~\citep{janardhanReimaginingOurInfrastructure2023}) within the scope of our discussion.} Throughout this
paper, we refer to entities that provide access to this computational
power as \emph{compute providers}.

Some AI firms currently manage their own data centers or maintain
exclusive partnerships with leading compute providers, known as
\emph{hyperscalers}.\footnote{The most notable hyperscalers include
  Microsoft Azure, Amazon Web Services (AWS), Apple, Bytedance, Meta,
  Oracle, Alibaba, Tencent, and Google Cloud
  \citep{vailsheryWorldwideInfrastructureService2024}.} Notably, the
most advanced AI research is currently being conducted at or with these
hyperscalers.\footnote{Many prominent AI companies either operate as
  hyperscalers themselves or maintain strategic partnerships with them.
  For example, OpenAI's collaboration with Microsoft
  \citep{microsoftcorporateblogsMicrosoftOpenAIExtend2023}, and
  Anthropic's associations with AWS and Google Cloud
  \citep{amazon.cominc.AmazonAnthropicAnnounce2023,
  anthropicAnthropicPartnersGoogle2023} , exemplify such
  relationships.} While this situation introduces complex challenges for
regulatory oversight, our discussion also encompasses scenarios in which
compute providers are internal to, or closely linked with an AI
firm.\footnote{These relationships and the concentrated market for
  large-scale compute have given rise to antitrust and competition
  concerns, drawing scrutiny from regulators, including an ongoing
  inquiry by the US Federal Trade Commission
  \citep{federaltradecommissionFTCLaunchesInquiry2024} and an
  investigation by the UK Competition and Markets Authority
  \citep{milmoCMAInvestigateUK2023}. While we acknowledge the need to
  carefully analyze how enhanced regulatory measures may impact
  competition issues in the sector, we expect the impact to be limited
  and not broadly significant. This is due to the focused scope of the
  measures, targeting only those compute providers capable of supporting
  large-scale AI infrastructure and customers who are running
  large-scale workloads costing tens to hundreds of millions of dollars
  or more. A detailed discussion of these antitrust concerns requires
  broader analysis and is beyond the scope of this paper.} For example,
an AI company should not be able to circumvent the proposed guidelines
by categorizing its usage as internal provisions or failing to identify
itself as a customer. This would ensure comprehensive coverage of all
relevant forms of compute provision for frontier AI.

\begin{figure}[ht]
    \centering
    \centerline{\includegraphics[width=1\linewidth]{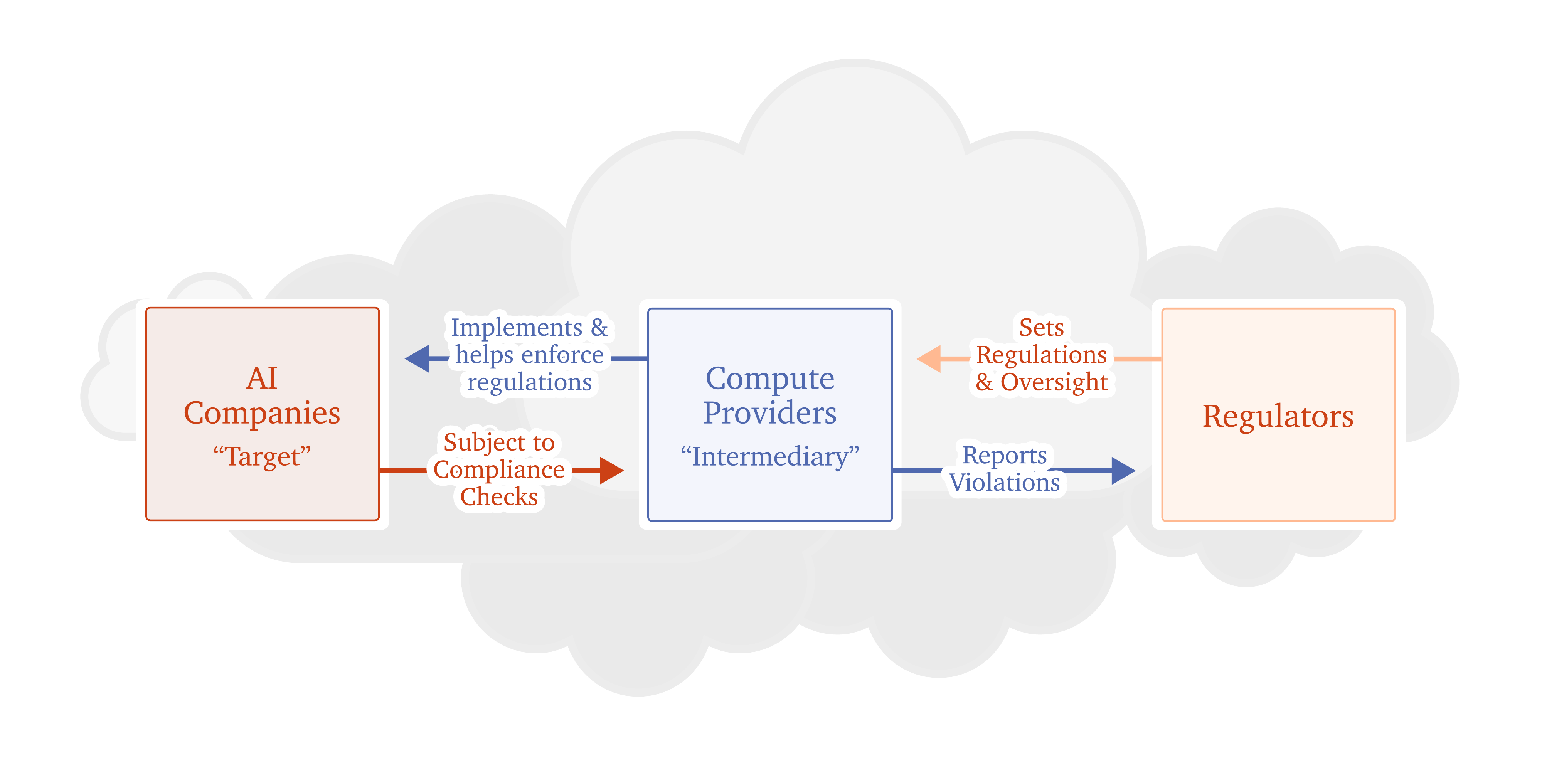}}
    \caption*{\textbf{Figure 1:} The intermediary role of compute providers in relation to AI companies and regulators.}
    \label{fig:Intermediary-role}
\end{figure}

\subsubsection{Regulatory Intermediaries}\label{regulatory-intermediaries}

The use of private sector entities as regulatory intermediaries is not a
new concept \citep{abbottRegulatoryIntermediariesAge2017,
abbottTheorizingRegulatoryIntermediaries2017,
hayPrivateEnforcementPublic1998}. Regulations for the aviation
industry require airline operators to maintain security plans, keep
records on passengers, verify passenger identities, and deny boarding in
response to specified violations
\citep{AcceptanceScreeningIndividuals2024,
AviationTransportSecurity2005}. In the context of communication
service providers, some governments have enacted data retention policies
that require the collection and retention of information on customers'
activities for a specified amount of time
\citep{australianattorney-generalsdepartmentDataRetentionGuideline2015,
countrylegalframeworksresourceFrance2023}. Internationally coordinated
anti-money laundering and counter-terrorism financing standards require
financial institutions to keep records, verify identities, and act to
prevent illicit flows of money
\citep{financialactiontaskforceInternationalStandardsCombating2023}.

Compute providers are already subject to regulations that dictate
operational standards and data management practices. For example, laws
targeting Child Sexual Abuse Material (CSAM) impose obligations on
compute providers to detect and report such content
\citep{amazon.cominc.2023, biden1738110thCongress2008,
europeanunion2011}. Data protection regulations, such as the General
Data Protection Regulation (GDPR) in the EU, set stringent requirements
for data handling, privacy, and user consent, which significantly affect
how compute providers manage and secure user data \citep{aws,
europeanunion2016, googlecloud2021}. Additionally, legislation aimed
at combating terrorism and extremism may require compute providers to
monitor and restrict the dissemination of harmful content
\citep{europeanunion2021}.

\subsection{Focus on Frontier AI}\label{focus-on-frontier-ai}

The customer base of AI compute providers is diverse, ranging from
individuals and small enterprises to large corporations and government
agencies. However, the primary focus of this paper is on a subset of AI
systems known as frontier AI. Frontier AI systems are defined as
``\emph{highly capable general-purpose AI models that can perform a wide
variety of tasks and match or exceed the capabilities present in today's
most advanced models}''
\citep{departmentforscienceinnovationandtechnologyCapabilitiesRisksFrontier2023}.\footnote{Also
  see Section 2.1 of \citet{anderljungFrontierAIRegulation2023}.}
These systems represent the frontier of current AI capabilities and
currently require substantial compute resources for their development
and operation \citep{anderljungFrontierAIRegulation2023,
sastryComputingPowerGovernance2024, sevillaComputeTrendsThree2022}.

Given the significant compute and research demands, only a small number
of entities possess the necessary resources to develop frontier AI
systems. This paper addresses primarily large corporations for
regulatory consideration. The intent of the proposed regulatory
frameworks is not to encompass the entirety of compute
providers\textquotesingle{} customer base indiscriminately. Rather, the
focus is limited to those actors who are in a position to develop or
deploy frontier AI systems---based on their compute usage---warranting
closer scrutiny and potential regulation. This targeted approach ensures
that regulatory measures are both effective and proportional, avoiding
unnecessary encumbrances on smaller entities or individual users who do
not fall within the relevant frontier AI activities.\footnote{In
  \cref{governance-challenges}, we elaborate on privacy and confidentiality
  considerations, emphasizing that our proposed regulatory measures are
  specifically aimed at key actors in frontier AI development, rather
  than being broadly applied to the entire customer base of compute
  providers, such as individuals.
  \citet{heimAccessingControlledAI2023} and
  \citet{lennartOversightFrontierAI2023} also discuss the idea of
  ``above-threshold compute usage.''} 
\citet{sastryComputingPowerGovernance2024} discuss the importance of
compute for frontier models in more detail and examine the conditions
under which compute, and by extension, compute providers, serve as
effective policy levers, as well as scenarios and conditions where their
impact is limited.

\subsection{Overview of our
Contributions}\label{overview-of-our-contributions}

We suggest that compute providers can play a key role in governance
regimes similar to the examples listed above via four key capacities: as
\emph{securers,} \emph{record keepers}, \emph{verifiers}, and
\emph{enforcers}. Security can restrict access to AI-related IP, such as
the model weights, from bad actors. Record keeping allows selected
information gathering for insights into AI activities and allows for
post-incident attribution. Verification includes checking customer
activities and AI systems to verify compliance with regulations and
standards. Enforcement entails the restriction or limitation of compute
access for non-compliant customers.

We analyze the technical means of performing each of these functions.
Concerning \emph{security}, we outline the role many compute providers
already serve in establishing baseline levels of physical,
infrastructure, and network security, and briefly discuss how this role
should be expanded to address more sophisticated threats. Regarding
\emph{record keeping}, we describe the information that is likely
already available to compute providers and the potential insights it can
offer. Regarding \emph{enforcement}, we review the capabilities of data
center providers to facilitate the enforcement of standards, either of
an external governance regime or of their own terms of service.

We focus much of our technical analysis on the role compute providers
could play in \emph{verification} in a targeted and privacy-conscious
manner. We describe information that is already available to compute
providers that can be used for four sub-categories of verification:
workload classification, compute accounting, verification of properties
of code and data (``detailed workload verification''), and identity
verification, such as Know-Your-Customer (KYC) regimes
\citep{lennartOversightFrontierAI2023, smith2023}. Several of these
techniques could be used in real-time to establish whether a customer is
engaging in an activity that might be subject to regulatory oversight.
If compute providers were to implement these techniques, they could
likely distinguish whether their customers are engaging in activities
such as training a large model, or engaging in inference at scale, as
well as quantify the amount of compute consumed by large workloads

After describing the technical possibilities, we focus on the US,
explaining how some of the techniques we describe can facilitate
compliance with the mandates of the Biden Administration's 2023
Executive Order 14110 on Safe, Secure, and Trustworthy AI (hereafter
``the AI Executive Order'') and support its overall objectives
\citep{thewhitehouseExecutiveOrderSafe2023}. We identify a set of next
steps for compute providers to verify that model developers have met
their Section 4 requirements under the AI Executive Order. We further
argue that the broad goals of the AI Executive Order could benefit from
the full range of techniques described here, particularly when
internationalized. In the final section, we highlight the technical and
governance challenges and opportunities for technical governance and
policy research.

\subsection{Limitations and Future Research
Directions}\label{limitations-and-future-research-directions}

In this paper, we propose a conceptual model wherein compute providers
improve the efficiency and effectiveness of the developing regulatory AI
ecosystem. Our discussion is not a comprehensive policy blueprint ready
for implementation. While certain aspects of our proposal may be 
directly actionable, other aspects require more evaluation.
Our primary objective is to argue for particular roles for compute
providers and to stimulate a debate: Should compute providers embrace
the roles we have outlined? Which of the proposed activities are most viable, and which
require further research?

While we emphasize the need to internationalize our proposed concept, we
acknowledge that this aspect is not explored in detail, especially the
legal aspects. The complexity of such an analysis extends beyond the
scope of this paper, calling for detailed legal and policy analysis. Our
aim is to motivate more research in this domain, recognizing the need
for a more comprehensive investigation into the intricacies involved in
applying these regulatory concepts across different jurisdictions. This
is particularly relevant given the intricate challenges that have
emerged in the past, such as those surrounding the EU-US Privacy Shield,
and others \citep{buttarelli2018}.

Some of the suggested regulatory measures may have implications on
privacy and confidentiality for customers of compute providers and we
recognize that increasing legal reporting requirements must be carefully weighed
against the potential for misuse or government overreach. However, our
focus---frontier AI---largely pertains to compute-intensive workloads
undertaken by a limited number of corporate entities. This simplifies
the regulatory landscape with respect to information describing these
workloads and the companies running them, as stringent privacy
regulations like the GDPR safeguard the personal data of individuals
(``natural persons''), while typically imposing fewer constraints on data
related to companies (``juridical person'').

The urgency of this discourse is magnified by the KYC requirements outlined in the US's AI Executive Order, with
implementation currently being progressed through the Department of
Commerce's proposed rule \citep{federalregister2024}.\footnote{Especially,
  as some of the currently proposed measures may be overly broad,
  potentially violating confidentiality principles.} Reaching a
comprehensive and thoughtful regulatory framework requires going beyond
unilateral measures, which could lead to unintended adverse
consequences. International coordination is crucial to address the
questions arising from the intersection of AI regulation with
international trade law, confidentiality, and privacy. This paper
contributes to the dialogue and advocates for an AI regulatory framework
that is collaborative, nuanced, effective, and globally
inclusive and responsive.

\section{Governance Capacities of Compute Providers
}\label{governance-capacities-of-compute-providers}

In the current ecosystem, frontier AI models are trained and deployed
using large compute clusters consisting of thousands to tens of
thousands of AI accelerators \citep{pilz2023,
sevillaComputeTrendsThree2022}. The concentrated supply chain, in
addition to the detectability, excludability, and quantifiability of
physical computing hardware, makes compute a particularly effective node
of governance compared to other inputs to AI development
\citep{belfield2022,pilz2023}.

Therefore, the physical infrastructure required for AI development and
deployment can be used as an instrument to enhance existing AI
regulations---making them more efficient and effective while enabling
new policies. Compute providers can perform these functions via four key
capacities: as \emph{securers}, to protect IP; \emph{record keepers},
enabling data collection for analysis and future reference; as
\emph{verifiers} of customer activities, enabling appropriate oversight;
and as \emph{enforcers}, taking actions against norm and rule
violations. We discuss these capacities in this section in more detail.

This approach offers a nuanced alternative to broad measures such as
chip export controls \citep{allen2022}, positioning compute providers
as a more precise and adaptable governance mechanism within the AI
supply chain. Unlike controls on physical computing hardware, the
dynamic model of compute provision affords a higher degree of
flexibility and specificity (Table 1 in \citet{heimAccessingControlledAI2023}).

\subsection{Compute Providers in the AI Compute Supply
Chain}\label{compute-providers-in-the-ai-compute-supply-chain}

\begin{figure}[ht]
    \centering
    \centerline{\includegraphics[width=1.3\linewidth]{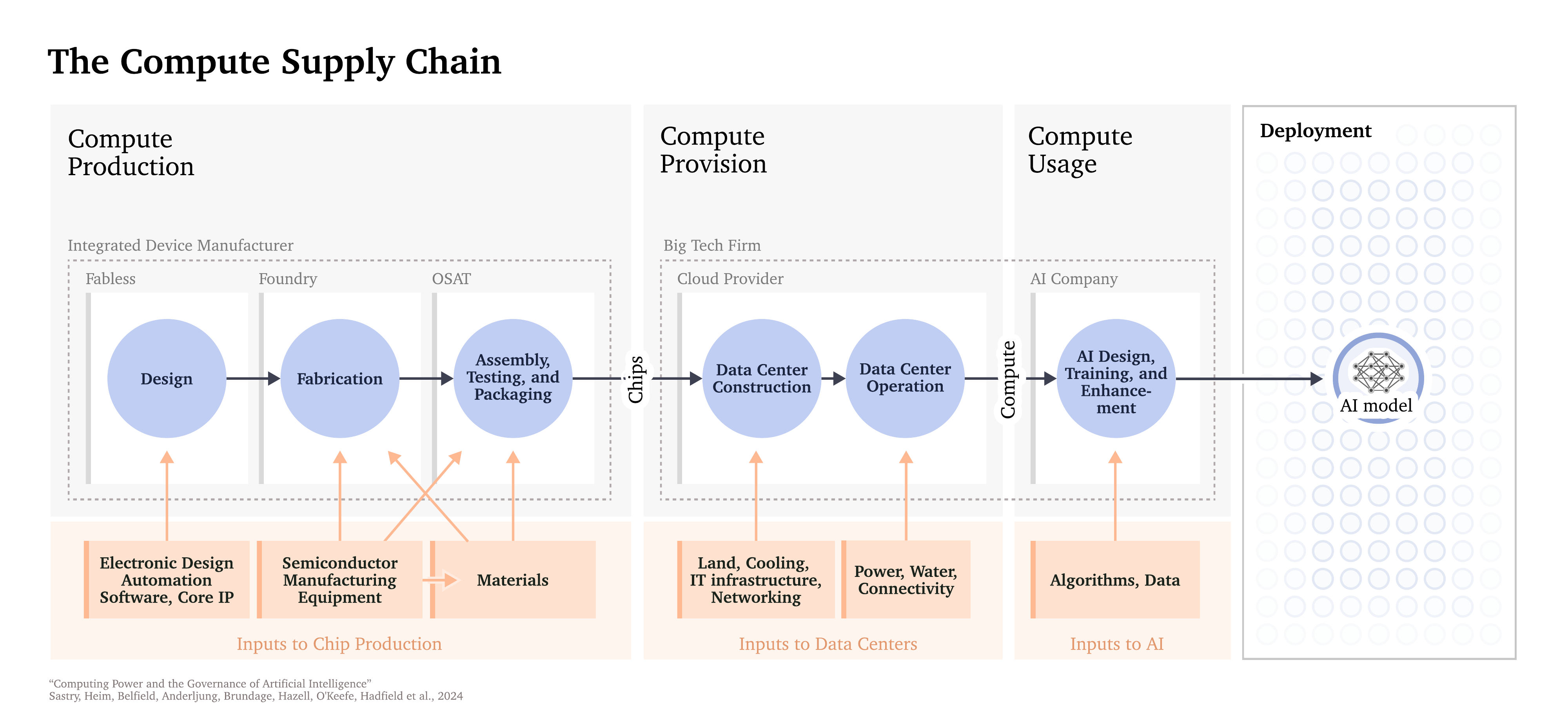}}
    \caption{The compute supply chain including compute providers
in the middle. Like the production of state-of-the-art AI chips, compute
providers' market shares are concentrated. (Figure from \citet{sastryComputingPowerGovernance2024}.)}
    \label{fig:Compute-Supply-Chain}
\end{figure}

In the AI supply chain, compute providers act as an intermediary,
offering computational resources to customers. These providers house
large numbers of chips in their data center facilities, operating them
cost-effectively in large quantities with necessary elements such as
power, land, cooling, and connectivity, and optimizing them for
developing and deploying AI models (\cref{fig:Compute-Supply-Chain}). With current
technologies, training large AI systems requires physically co-located
chips. This has caused much of contemporary AI deployment and
development to occur in large facilities, wherein compute is made
available to customers digitally and remotely, often through models like
Infrastructure as a Service (IaaS) or cloud computing. The compute
provider industry has seen significant consolidation in recent years due
to the economic advantage of scale \citep{InfographicAmazonMaintains2024}.

By capitalizing on their position as infrastructure providers, compute
providers can play a key role in regulating AI companies. Leveraging
compute in this manner can help reduce regulatory burdens because each
compute provider (i) typically services multiple AI firms,\footnote{The
  size of the customer base of compute providers varies significantly.
  Typically, compute providers offer their services to a wide range of
  entities, with most usage falling outside the scope of our primary
  concern in this paper (see \cref{introduction}).
  Nonetheless, our analysis also includes large compute owners who
  exclusively use their resources internally. For example, a major
  technology company cannot circumvent the guidelines proposed in this
  discussion by merely categorizing its usage as ``internal provisions''
  or not identifying itself as a ``customer.'' This approach ensures
  comprehensive coverage of all relevant forms of compute provision for
  frontier AI.} thereby streamlining the regulatory process, and (ii)
screens potential targets for regulation by the scale of compute usage
(which can be supplemented by further criteria), whether it is used for
training or deployment. This follows the model of how financial
institutions operate under KYC schemes
\citep{financialcrimesenforcementnetwork,
financialcrimesenforcementnetworka, lennartOversightFrontierAI2023},
although compute providers with the capacity for frontier AI development
and deployment are much fewer in number than banks.

\subsection{Governance Capacities}\label{governance-capacities}

\begin{figure}[ht]
    \centering
    \centerline{\includegraphics[width=1.3\linewidth]{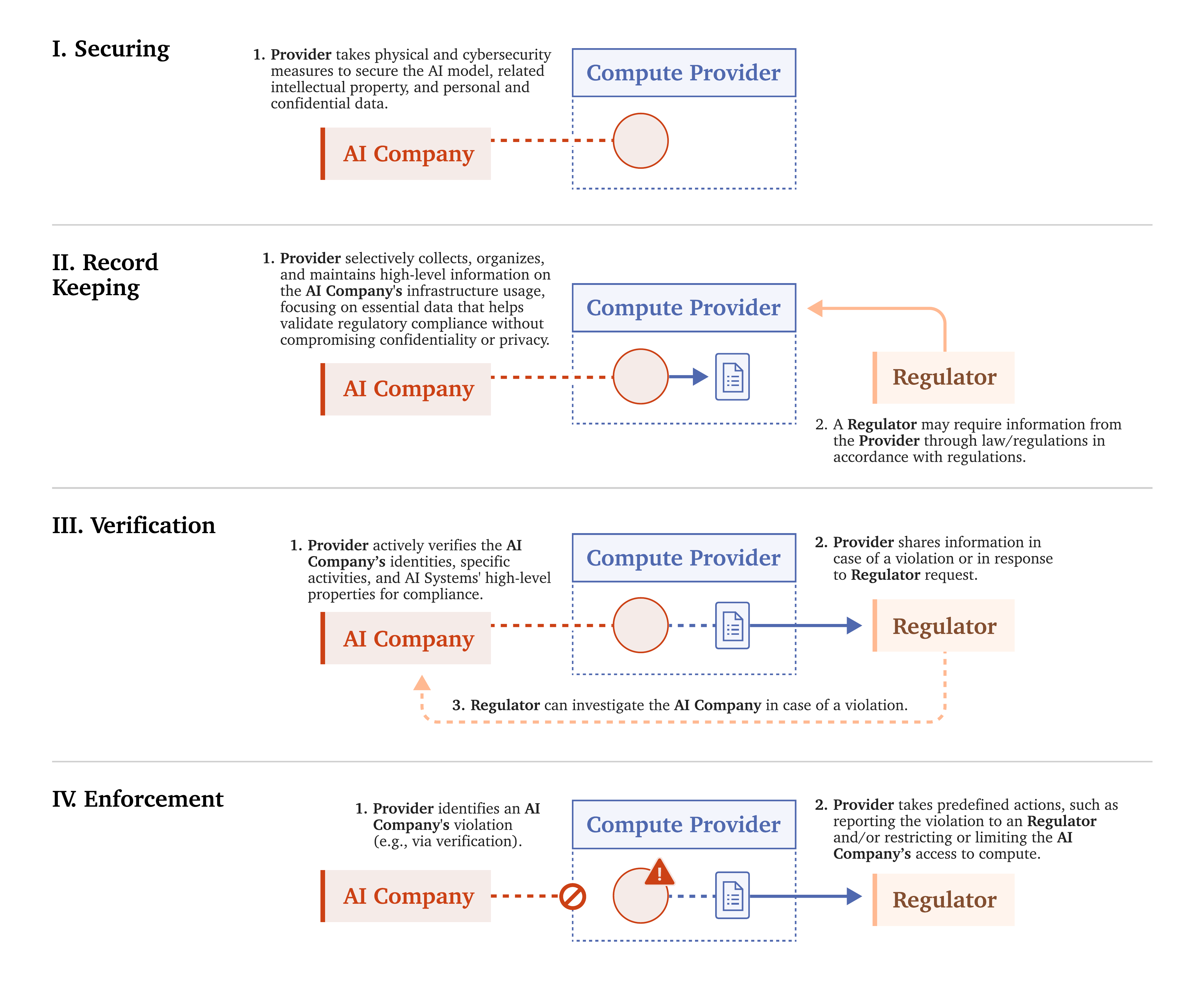}}
    \caption{Overview of the different governance capacities
and how they relate to three actors: customers (the AI developers and
deployers), compute providers, and regulators.}
    \label{fig:Visualization-governance-capacities}
\end{figure}

Compute providers can facilitate effective AI regulation via four key
capacities. They can be \emph{securers}, \emph{record keepers},
\emph{verifiers}, and, in some cases, even \emph{enforcers} (\cref{tab:Governance-Capacities,fig:Visualization-governance-capacities}).

\begin{table}[ht]
\renewcommand{\arraystretch}{2}
\centering
\begin{adjustwidth}{-0.05\linewidth}{-0.05\linewidth}
\resizebox{1.0\linewidth}{!}{%
    \begin{tabular}{>{\itshape}C{0.25\linewidth}>{\columncolor{lightgray}\itshape}C{0.25\linewidth}>{\itshape}C{0.25\linewidth}>{\columncolor{lightgray}\itshape}C{0.25\linewidth}}
    \toprule
    \multicolumn{4}{c}{\bfseries\Large Governance Capacities} \\[3pt]\hline
    \multicolumn{1}{c}{\bfseries\large Security} & \multicolumn{1}{c}{ \cellcolor{lightgray}\bfseries\large  Record Keeping} & \multicolumn{1}{c}{\bfseries\large  Verification} & \multicolumn{1}{c}{ \cellcolor{lightgray}\bfseries\large  Enforcement} \\\hline
    Helping provide physical and cybersecurity measures to secure the AI model, related intellectual property, and personal and confidential data.%
    & The selective collection, organization, and maintenance of high-level information of a compute provider’s infrastructure usage, such as a customer’s compute usage data.\footnotemark%
    & Actively verifying customer identities, specific activities, and high-level AI systems’ properties.%
    & Restriction or limitation of compute access to customers or workloads for non-compliant customers. \\\hline
    \multicolumn{4}{c}{\bfseries\large Enables} \\\hline
    Enables shared security standards to protect the public good, such as safeguarding critical infrastructure and helping prevent model theft. & Increases visibility into AI development, links customers and their usage to real-world actors, and enables post-incident attributions and forensics. & Ensures that the deployment and development of AI systems adhere to regulations or company policies and reported properties. & Directly impacts the capability of customers to develop or deploy advanced AI systems, ensuring adherence to rules.\\\hline
    \multicolumn{4}{c}{\bfseries\large Examples} \\\hline
    \begin{tabular}[t]{@{}>{\centering\arraybackslash}p{\linewidth}@{}}Help prevent IP (e.g., algorithms), model weights, and training data from being stolen by malicious actors.\\ Help prevent attacks on large-scale deployments of foundational models that could shut down dozens of critical services nationwide.\end{tabular} &
    \begin{tabular}[t]{@{}>{\centering\arraybackslash}p{\linewidth}@{}}Obtain insights into national compute use trends for policy formulation, such as compute distribution (e.g., US NAIRR).\\ Enable monitoring for suspected violations of the reporting requirements under Executive Order 14110.\\ Collect information on the environmental impact of AI compute use.\end{tabular} &
    \begin{tabular}[t]{@{}>{\centering\arraybackslash}p{\linewidth}@{}}Confirm compliance with mandatory reporting over training compute thresholds.\\ Verify compliance with data usage guidelines for frontier AI training.\\ Verify if the deployed frontier AI system has an adequate license or certification.\end{tabular} & 
    \begin{tabular}[t]{@{}>{\centering\arraybackslash}p{\linewidth}@{}}Restrict access to customers lacking licenses as an AI developer for their system.\\ Refuse to deploy an unlicensed or non-compliant AI model.\\ Disable AI systems that demonstrate activity that is undesirable, uncontrollable, or in violation of regulations (e.g., computer worm-like AI system).\end{tabular}\\\bottomrule
    \end{tabular}%
    }
\end{adjustwidth}
\caption*{\textbf{Table 1:} Summary of the key governance capacities that compute providers can enable.}
\label{tab:Governance-Capacities}
\end{table}

\FloatBarrier
\footnotetext{Focusing on essential data that informs without compromising privacy and confidentiality.}

These governance capacities are distinct from but related to obligations
to \emph{report} information to governments. In many cases, it will be
appropriate for compute providers to collect and retain information
internally and only provide information to governments in response to
existing legal authorities (for example, identified violations of
sanctions, or in response to legal warrants). In other cases, for
example, where a customer is undertaking a large training run,
regulators may see fit to mandate proactive reporting. Record keeping
can ensure that compute providers are aware of, and able to comply with,
broader regulations to increase visibility and oversight.

\subsubsection{I. Security}\label{i.-security}

Compute providers, as custodians of sensitive data and AI-related IP,
have a distinct capacity for governing and implementing information
security measures that protect AI systems.

\begin{figure}[ht]
    \centering
    \centerline{\includegraphics[trim={11cm 43cm 14cm 0},clip,width=.75\linewidth]{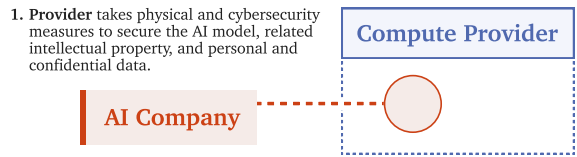}}
    \caption{The security measures implemented by compute providers to help protect AI company's models,
intellectual property, and confidential data.}
    \label{fig:I-Securing}
\end{figure}

Frontier AI models represent considerable financial, computational,
research investments, and powerful tools that could be misused for
financial or political gain; this makes them highly attractive to cyber
attackers and other adversarial actors
\citep{epoch2023trendsinthedollartrainingcostofmachinelearningsystems,
sevillaComputeTrendsThree2022}. Security measures should extend beyond
safeguarding model weights to include the protection of the model's
architecture, its algorithmic innovations, training data, and other
related intellectual property (IP) \citep{nevo2023}. Therefore, it is
important to take robust cybersecurity measures proportional to these
risks. These measures could be legally mandated, as with the physical
and cybersecurity precautions currently required from data centers that
handle HIPAA- and ITAR-compliant health
\citep{officeforcivilrights2016} and defense data\footnote{ITAR
  compliance requires arms-related data to be secured from foreign
  persons. There is no formal certification process for cloud providers,
  although many choose to be audited by a third-party organization
  certified under the Federal Risk Authorization Management Program
  (FedRAMP) \citep{codeoffederalregulations22CFRPart}.},
as preventing the theft of potentially dual-use AI intellectual property
and disruption to critical AI infrastructure are matters of public
good.\footnote{Such regulations could be enforced by chartering or
  registration (for example, firms cannot accept federally insured
  deposits unless chartered as a bank, credit union, or thrift)
  \citep{congressionalresearchservice2023}, or tied to a license to
  acquire specialized AI compute hardware, as is already in place for
  the export of certain equipment from US companies \citep{amd}.}

Such a measure should take into account existing industry practice:
Large-scale compute providers already allocate significant resources to
physical and cybersecurity, given strong business incentives to keep
clients' sensitive data secure (including, in some cases, classified
government data \citep{lohr2023}. When multiple AI firms utilize a
single compute provider, they inherently benefit from the
provider's comprehensive physical and cybersecurity
measures, which benefit from the economies of scale and significant
investment. Nevertheless, compute providers remain vulnerable to
malicious cyber actors and can be the subjects of successful cyber
attacks \citep{gatlan2023, vanian2024}. The potential for frontier AI
to be stolen, misused, or to become critically important infrastructure,
may warrant the development of stronger security requirements for the
compute providers that train and deploy them. This necessity was
reflected in the voluntary White House Commitments
\citep{thewhitehouse2023} and the Hiroshima International Guiding
Principles for AI, which advocate for strengthened safeguards in this
domain \citep{europeancomission2023,europeancomission2023a}.

It is crucial to acknowledge that the responsibility for ensuring
information security does not rest solely with compute providers. The
efficacy of their security measures requires robust security practices
and collaborative efforts by AI companies themselves. Without these
companies' proactive engagement in safeguarding their operations, the
protective mechanisms implemented by compute providers could prove
ineffective. Therefore, these efforts are not substitutive but rather
complementary, with both compute providers and AI companies sharing
responsibility.

\subsubsection{II. Record Keeping}\label{ii.-record-keeping}

Record keeping describes the process of collecting, organizing, and
maintaining information on a compute provider's
customers and their infrastructure usage. Compute providers are
inherently record keepers by virtue of their role and technical
necessity. They store and process valuable technical data during large
AI deployments and training runs for billing purposes, resource
management, and service-level agreement tracking (see \cref{technical-feasibility-of-compute-providers-governance-role} for
more detail). Provided that robust privacy protections are in place,
this information could be useful to regulators in overseeing the
development of advanced AI systems. We recommend that regulators and
providers focus on essential data that informs AI regulation without
compromising privacy and confidentiality (which we discuss in more
detail in \cref{governance-challenges}).

\begin{figure}[ht]
    \centering
    \centerline{\includegraphics[trim={11cm 28.7cm 0 13cm},clip,width=1\linewidth]{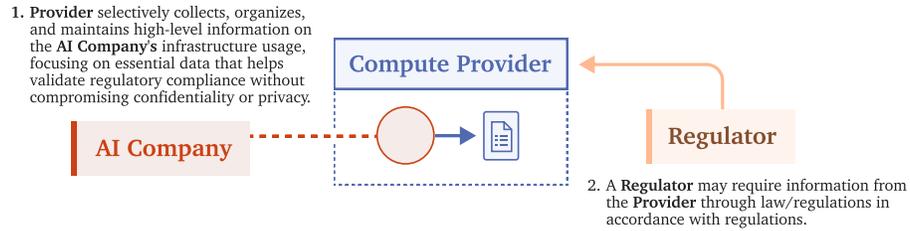}}
    \caption{The compute provider collects and manages essential
usage data on the AI company and its infrastructure usage, focusing on key data that helps validate regulatory compliance without compromising privacy. This facilitates greater transparency into AI advancements,
could link compute use to real-world actors, and enables effective
post-incident response and forensics.}
    \label{fig:II-Record-Keeping}
\end{figure}

Record keeping serves three main purposes: it allows more visibility
into the developments of AI generally, helps link customers and their
usage to real-world actors, and enables post-incident attributions and
forensics. First, it provides visibility into AI developments, which is
important for monitoring the trajectory of AI systems and their compute
requirements across various sectors
\citep{sastryComputingPowerGovernance2024}. As we have seen
demonstrated by recent national compute initiatives
\citep{centerforopenscience2019,departmentforscienceinnovationandtechnology2023,theeuropeanhighperformancecomputingjointundertaking2023,u.s.nationalsciencefoundation2024}, compute provision can be a
mechanism to guide AI development and formulate policies for a
beneficial distribution of compute resources \citep{besiroglu2024,sastryComputingPowerGovernance2024}. By gathering aggregated and
anonymized compute usage data, governments are better positioned to
formulate policies that mitigate unequal access, reduce market
monopolization, address potential negative impacts, and boost beneficial
innovations aligned with national objectives
\citep{sastryComputingPowerGovernance2024}.

Second, record keeping enables compute providers and regulators to
identify their customers and the corresponding legal entities (KYC).
With the development of frontier AI regulations, it will likely become
necessary to verify the identity of compute customers utilizing large
amounts of compute and ensure that they have the necessary
certifications and safeguards in place, if they plan to build and deploy
advanced AI systems. Furthermore, countries may wish to enforce export
controls on compute, restricting the sales of data center capacity to
actors with improper sources of funding (e.g., terrorist organizations)
or originating from sanctioned geographical areas
\citep{bureauofindustryandsecurity2023,heimAccessingControlledAI2023}.

Third, and related to customer verification, record keeping enables
post-incident attributions and forensics
\citep{obrienDeploymentCorrectionsIncident2023}.\footnote{An example
  of this could be in the event of an AI system malfunctioning and
  causing financial loss or physical harm, record keeping allows for the
  traceability of the AI's development and deployment
  process, helping to identify the origin of the fault
  (\emph{forensics}) and parties responsible (\emph{attribution}).}
Information about developers and their activities is critical for
assigning liability or enhancing existing systems in the aftermath of
incidents. This is akin to mandatory record keeping in the financial and
telecommunications industries for law enforcement purposes
\citep{legislation.gov.ukDataRetentionInvestigatory,u.s.securitiesandexchangecommissionRetentionRecordsRelevant2003}. We
propose careful legislation that requires compute providers to maintain
relevant records while taking into account privacy concerns and the
regulatory burden. This is especially important in cases of potential
unlawful use where the developers may not keep adequate records in the
absence of legislation. For example, high-level compute utilization data
could be made available to regulators upon request, while detailed
retained information may be kept confidential unless required for
enforcement or legal proceedings.

In contrast to verification, which is introduced below, record keeping
does not require active processing by the compute provider. Instead, it
describes the collection of specific data, which is already happening to
some extent, to make more information available to governments. For
example, it could inform national policy decisions and strategies
\citep{whittlestoneWhyHowGovernments2021}, and help prevent and
respond to serious incidents (where required under warrants and/or by
regulation). The use of aggregated and anonymized compute usage
statistics could offer valuable insights into trends in AI development
and corresponding compute demands (e.g., to learn more about the impacts
of the \emph{compute divide} \citep{ahmedDedemocratizationAIDeep2020,besiroglu2024}). This approach allows regulators to understand and
respond effectively to the evolving needs of the AI economy without
encroaching upon the privacy and confidentiality of companies. This
could operate on a need-to-know basis, ensuring that only pertinent
information is gathered and utilized in a way that minimizes regulatory
burden.

Furthermore, transparency requirements for environmental accountability,
particularly in the context of compute providers' substantial energy
consumption, have been suggested
\citep{oecdMeasuringEnvironmentalImpacts2022}. Mandating reports on
the environmental impact of compute providers could equip regulatory
bodies with the insights necessary to fulfill environmental objectives.
The EU already has implemented reporting requirements.\footnote{``\emph{Under
  the directive, owners and operators of data centers with 500 kilowatts
  or more of installed IT capacity will need to report their 2023 energy
  performance by May 15, 2024. That includes statistics about installed
  power, incoming and outgoing data traffic, total data stored and
  processed, energy consumption, power usage, temperature set points,
  waste heat utilization, and use of renewable energy.}''
  \citep{korolov2023}} Concurrently, numerous compute providers are
already advancing toward greater environmental transparency and have
pledged to undertake climate initiatives. \citep{amazon.cominc.Cloud,azureAzureSustainability, googlecloudClimateSustainability}.

\subsubsection{III. Verification}\label{iii.-verification}

Compute providers can also actively verify customer compliance with
regulatory requirements, providing AI firm oversight. Similar to banks
and other financial intermediaries, compute providers can actively
verify the identity of customers and key customer activities, checking
that the properties of AI systems being deployed or developed match
customer reporting. This might include verifying the type of
computational workload run by the customer (e.g., training an AI model,
or deploying a model at scale) as well as claims about the total amount
of compute used, or the type of data used in the training process.\footnote{Hardware-enabled mechanisms for verification and enforcement
  of governance regimes are discussed in
  \citet{aarneSecureGovernableChips2024} and
  \citet{kulpHardwareEnabledGovernanceMechanisms2024}.}

\begin{figure}[ht]
    \centering
    \centerline{\includegraphics[trim={12.5cm 13.4cm 4cm 27.6cm},clip,width=1\linewidth]{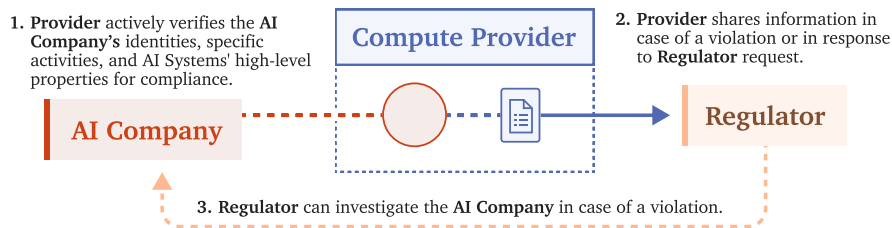}}
    \caption{The compute provider actively verifies the AI
company's identities, specific activities, and/or
properties of AI systems for regulatory compliance.}
    \label{fig:III-verification}
\end{figure}

As we will discuss in \cref{technical-feasibility-of-compute-providers-governance-role}, some of these capabilities, such as
identifying whether a customer is training or deploying an AI model, can
likely be implemented using existing data collected by compute
providers. Providers could use this information to catch regulatory
violations, including breaches of model training requirements, wherein
developers avoid reporting requirements and the unauthorized deployment
of AI models at scale.

Verification rules should balance between regulatory requirements,
customer privacy laws, and safety considerations, similar to record
keeping practices. Spot checks (including those triggered by reports or
events that raise suspicion) may be adequate for less risky activities,
while continuous monitoring to ensure that customer-submitted records
are accurate may be mandated for riskier ones. Providers can then check
if mandated or adequate risk practices are applied. AI safety
regulations could be applied only to those using large amounts of
compute and waived for smaller customers. Additional factors may warrant
further examination, like a history of non-compliance or inclusion on
restrictive lists like the US Department of Commerce's
Entity List, which serves to protect national security interests. The
standardization and aggregation of reporting processes across providers
is crucial for ensuring effective cross-provider verification. This
uniformity prevents AI developers from evading regulations by splitting
their training or deployment activities among multiple providers.

Verification becomes vital in cases of behavior that may indicate
non-compliance, such as a failure to report details of a training run
mandated by regulations like the AI Executive Order. For example, if
there is credible evidence of a sophisticated AI model having been
trained, regulators could request certain verifications from compute
providers to confirm that model developers or providers are in
compliance. This approach is similar to selective tax audits, providing
the regulation with more enforcement capability \citep{advani2023}.
Even though immediate action might not be taken, having access to such
information allows for inspection if necessary, especially in response
to suspicious activity. Compute providers could respond to such
regulatory requests by verifying customer activities and providing an
additional check that model developers are following rules. This process
may involve asking customers to provide justifications or evidence for
their activities in the same way that financial institutions can inquire
about suspicious transactions from their clients.

Unlike record keeping, verification demands more proactive engagement
from compute providers. Verification could span a wide range of
different specific activities and may include confirming the identities
of customers who use substantial compute resources or checking that the
workload being executed matches their declared type. In complex cases
(such as frontier model training) or direct violations of reporting
requirements, the compute provider may flag the case and refer it
directly to regulators. It is important to note that several open
questions regarding privacy and technical feasibility must be
investigated and potentially addressed prior to implementation. These
considerations are discussed in the \cref{key-challenges}.

For illustrative purposes, imagine a scenario in which regulations
require AI developers to report their compute usage, or notify a
government prior to training a model above a certain compute threshold,
as has been outlined in the AI Executive Order. In such cases, an AI
developer could be required to provide information about compute usage
to the provider. This allows customers to demonstrate that their
computing power was employed for specific purposes, such as training
several smaller models rather than a single larger model, which might be
subject to different compliance standards. The provider then checks that
the actual usage is consistent with the developer's
representations. If inconsistencies are identified, the compute provider
can flag these events for further investigation and compliance checks.
This arrangement, which uses compute providers as an intermediary, helps
minimize regulatory burdens on the AI industry. Similarly, a compute
provider could require a model developer to provide proof that they have
appropriately notified the government of a threshold-exceeding training
run prior to the compute provider allowing access to related compute.

\subsubsection{IV. Enforcement}\label{iv.-enforcement}

Compute providers can also aid regulatory enforcement. By virtue of
controlling the AI data centers themselves, providers have the ability
to directly deny access to rule-breaking customers, and, therefore,
prevent the customer from developing or deploying certain kinds of AI
systems with that provider. The compute provider might limit compute
resources devoted to workloads that raise red flags pending further
investigation. Similarly, record keeping and verification processes
could trigger regulatory enforcement measures by other actors, such as
the Department of Justice in the US.

\begin{figure}[ht]
    \centering
    \centerline{\includegraphics[trim={12.5cm 0.4cm 0 44cm},clip,width=1\linewidth]{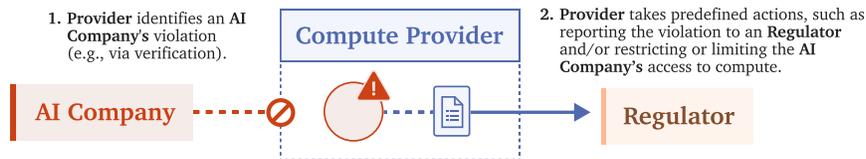}}
    \caption{The compute provider detects violations by the AI
company, e.g., via the verification process, and can take appropriate
enforcement actions, such as restricting or limiting access to compute.}
    \label{fig:IV-Enforcement}
\end{figure}

In addition to assisting governmental enforcement, compute providers can
enforce their terms of service (ToS). By expanding their ToS to include
stipulations on AI development and usage to promote safety, providers
can autonomously enforce compliance. This self-regulation approach goes
beyond merely reacting to government regulations; it could be an active
promotion of responsible AI practices---reflecting the
providers\textquotesingle{} commitment to ethical standards in AI
development.

Direct enforcement by compute providers can come in multiple forms. On
the most basic level, regulation could be crafted to require providers
to refrain from providing compute to developers to train a model above a
certain threshold until the developer proves they have taken appropriate
regulatory steps, such as notifying the government of a large training
run or obtaining relevant approvals. Providers could also help
regulators enforce rulings (e.g., when a license is denied, or when a
government is trying to prevent a malicious actor from accessing
compute) or automatically deny the execution of AI software depending on
whether the model is approved. More fine-grained restrictions, such as
slowing access or limiting certain workloads until the proper regulatory
approvals are issued (similar to financial service providers temporarily
flagging a transaction for further review), could also be enforced.
Compute providers could monitor the activity of customers using large
amounts of compute that have stated they are not engaging in model
training for signs they are circumventing the types of requirements laid
out above.

Enforcement works most effectively in conjunction with verification.
Detected violations can be promptly acted upon through the compute
providers\textquotesingle{} capabilities. However, it is crucial to keep
these two capacities distinct. Enforcement actions can be based on
information beyond what is verified by compute providers, for example,
in response to intelligence provided by law enforcement agencies.
Likewise, violations identified through the verification process may be
addressed through means other than restricting compute access. For
example, if a customer is found to be out of compliance with existing
regulations, or the compute provider's ToS, the provider can restrict
their access to the computational resources completely and report them
to the relevant authorities. The provider can also use less onerous
enforcement mechanisms, e.g., turning off the workload of concern,
reducing the amount of available compute resources, or fining the actor.

\section{Technical Feasibility of Compute Providers' Governance
Role}\label{technical-feasibility-of-compute-providers-governance-role}

This section examines the technical feasibility of activities within
each of the governance capacities introduced in the previous section.

We assess feasibility in the context of the technical capacities
available to compute providers today, and the capacities that could
foreseeably be developed. An overview of the technology stack available
to compute providers can be found in
\cref{a.-overview-of-compute-provider-technologies}
and is recommended for readers unfamiliar with data center technology.

\begin{tcolorbox}[breakable,boxrule=1pt,enhanced jigsaw, sharp corners,pad at break*=1mm,colbacktitle=lightgray,colback=lightgray,colframe=black,coltitle=black,toptitle=1mm,bottomtitle=1mm,width=\linewidth,fonttitle=\bfseries\large,parbox=false,title=Key terms used in this section
]

\textbf{Workload:} A computational task, defined by a specific instance
of software (i.e., written or compiled code) that will be run on a
hardware configuration. For example, training an AI model is a type of
workload. The input data to the workload may or may not be known in
advance.

\textbf{Operation (OP):} A single calculation run on computer hardware,
typically a multiplication or addition of two numbers. AI workloads
usually involve large numbers of matrix-multiply operations, which are
implemented as repetitive multiplications and additions (known as
multiply-accumulate).

\textbf{AI accelerator:} A hardware device (typically a specialized
chip) designed for fast and efficient execution of AI workloads.
Typically AI accelerators excel at simple, parallel calculations, which
can also be useful for graphical and scientific computing workloads.
This is an umbrella term for both current hardware paradigms (such as
GPUs and TPUs), as well as novel future designs.

\textbf{Node:} A single computer within a data center. Each node is a
distinct unit with its own processing power, memory, and storage,
capable of running a workload by itself, or in collaboration with other
nodes. Each node may contain multiple AI accelerators.

\textbf{(Computing) cluster:} A group of linked nodes that can work
together to process workloads. Typically used for workloads requiring
significant computational resources, such as large-scale AI training.

\textbf{Data center:} A facility that hosts computing clusters (or other
computing infrastructure) and provides the supporting infrastructure
needed to operate them efficiently.    

\end{tcolorbox}

In summary, compute providers are responsible for maintaining secure
premises, and physical and network infrastructure. They also operate
controls over that infrastructure (at both the hardware and software
level) that allow them to grant and revoke hardware access to particular
customers, store and maintain customer data, track hardware performance,
and debug technical issues. These same tools could allow compute
providers to engage in security, record keeping, verification, and
enforcement for frontier AI regulation, while preserving existing
industry norms surrounding confidential and private data. \cref{tab:Overview-Relevant-Technical-Capacities}
maps these governance capacities onto the relevant technical capacities
available to compute providers, and indicates the current feasibility of
using them.


\begin{table}[ht]
\vspace{3mm}
\centering
\begin{adjustwidth}{-0.05\linewidth}{-0.05\linewidth}
\small
\renewcommand{\arraystretch}{1.9}
\rowcolors{1}{white}{lightgray}
\resizebox{1.0\linewidth}{!}{%
\begin{tabular}{>{\itshape}C{0.11\linewidth}p{0.41\linewidth}p{0.39\linewidth}}\toprule
\normalfont\begin{tabular}[c]{@{}C{\linewidth}@{}}\bfseries \normalsize Governance capacity\end{tabular} & %
\begin{tabular}[c]{@{}C{\linewidth}@{}}\bfseries\normalsize Relevant technical capacities of compute provider\end{tabular} & %
\begin{tabular}[c]{@{}C{\linewidth}@{}}\bfseries\normalsize Current technical feasibility\end{tabular} \\[9pt]\hline
\vbox to -3.4\baselineskip{Security} & \begin{tabular}[t]{@{}p{\linewidth}@{}}Physical security (e.g., locks, guards, surveillance).\\ Infrastructure-level security (e.g., access controls, secure firmware, resource isolation).\\ Network security (e.g., firewalls, network authentication).\\ Providing additional cybersecurity services (e.g., user access management, encrypted storage).\end{tabular} & Feasible for low-end threats (in a supporting role). Many compute providers provide decent default levels of physical, infrastructure, and network security. With adequate customer investments in cybersecurity, these measures are likely sufficient against opportunistic attackers, but not against well-resourced and persistent expert attackers. \\
\vbox to -.5\baselineskip{\begin{tabular}[t]{@{}C{\linewidth}@{}}Record keeping\end{tabular}} & \begin{tabular}[t]{@{}p{\linewidth}@{}}Data record collection and maintenance (e.g., for required verification activities).\\ Securing data records (e.g., encryption at rest and in transit, managing access).\end{tabular} & Highly feasible. Compute providers can (and do) collect a wide variety of records on customers and their activities. These records are typically kept secure, and retained for as long as necessary to comply with relevant regulations or support internal business use cases. \\
\vbox to -6\baselineskip{Verification} & \begin{tabular}[t]{@{}p{\linewidth}@{}}Identity verification: ensuring a customer and/or user is who they say they are.\\ Workload classification: determining whether a customer workload falls within a category relevant to regulatory requirements (e.g., training a very large AI model).\\ Compute accounting: estimating the number of operations consumed by a workload (e.g., to validate compute threshold-based reporting requirements\tablefootnote{For example, see the training compute threshold-based reporting requirement in the AI Executive Order \citep{thewhitehouseExecutiveOrderSafe2023}.}).\\ Detailed workload verification: verifying aspects of the specific code or data used in a workload (e.g., to validate whether a customer is complying with reporting requirements for certain high-risk training data\tablefootnote{For example, the AI Executive Order \citep{thewhitehouseExecutiveOrderSafe2023} creates specific reporting requirements on developers for models primarily trained with biological sequence data. It may become useful for these kinds of reporting requirements to also be validated by compute providers, or for compute providers to offer verification tools to developers (e.g., “confidential computing” services).}).\end{tabular} & \begin{tabular}[t]{@{}p{\linewidth}@{}}Identity verification: likely feasible with a sufficiently rigorous process, focused on customers accessing large-scale compute resources.\\ Workload classification: likely feasible for detecting large-scale pre-training and inference workloads.\\ Compute accounting: feasible, with multiple approaches possible for large-scale pre-training/inference workloads.\\ Detailed workload verification: currently not possible without directly observing customer code or data. “Confidential computing” techniques could change this.\end{tabular} \\
\vbox to -1.7\baselineskip{Enforcement} & \begin{tabular}[t]{@{}p{\linewidth}@{}}Account-level enforcement: revoking service access to particular customers/accounts/users.\\ Model-level enforcement: revoking service access where services are used to deploy particular models, or where models are displaying dangerous behavior (e.g., a computer worm-like AI system).\end{tabular} & \begin{tabular}[t]{@{}p{\linewidth}@{}}Account-level enforcement: highly feasible, compute providers have both physical and software management-based control over their hardware; this is widely used.\\ Model-level enforcement: currently not possible without directly observing customer code or data. This may become possible with some technical effort.\end{tabular}\\\bottomrule
\end{tabular}}
\end{adjustwidth}
\caption{Overview of relevant technical capacities available to
compute providers and an assessment of their technical feasibility.}
\label{tab:Overview-Relevant-Technical-Capacities}
\end{table}

\FloatBarrier


\subsection{Security}\label{security}

Compute providers usually provide security at certain technology layers,
whereas other layers are left up to customers
\citep{googlecloudCloudDataProcessing,awsSecurityAWSInfrastructure,coreweaveSecurityCompliance}.
The compute provider is typically responsible for:

\begin{itemize}
\item
  \textbf{Physical security}, which includes protecting data center
  premises with locks, cameras, guards, and surveillance.

\item
  \textbf{Infrastructure security}, which includes ensuring that
  hardware is up-to-date with the latest firmware security patches,
  securely disposing of old hardware, restricting physical/virtual
  access to infrastructure to approved personnel for management and
  maintenance purposes, and ensuring appropriate isolation of critical
  system resources across different customers and workloads

\item
  \textbf{Network security}, which includes operating firewalls and
  providing other forms of network-level security and isolation.

\end{itemize}

These physical, infrastructure, and network security measures are
sufficient to provide customers with a baseline level of information
security, one that many customers could not achieve on their own.

Customers are then generally held responsible for the parts of the
technology stack they have control over, which encompasses many aspects
of cybersecurity, including protecting data generated or collected by
their workloads, ensuring their employees are well-trained in security
best practices, and implementing access control policies based on
different permission levels. Many compute providers, especially larger
providers, offer cybersecurity software-as-a-service products for their
customers to help them implement these measures; many of these products
are free and/or standard with infrastructure offerings. These additional
services typically provide the customer the ability to:

\begin{itemize}
\item
  Securely manage user access to resources in their account

\item
  Work with encrypted data storage in transit and at rest

\item
  Manage and secure inbound/outbound traffic from nodes

\item
  Define different levels of security within different regions of their
  infrastructure

\end{itemize}

In the context of protecting frontier AI workloads, if a customer is
working with a security-conscious compute provider, and the customer has
systematically implemented industry best practices for cybersecurity,
they are likely well-protected against most opportunistic attackers.
However, these measures are almost certainly inadequate to defend
against well-resourced, expert attackers, such as nation-state-backed
hacking groups (also known as ``advanced, persistent threats (APTs)'').
Such threats are of significant concern for frontier AI, given the
potential economic returns of model theft, and risks of misuse
\citep{nevo2023}.

As discussed in \cref{i.-security}, it is therefore important to strengthen
security standards for frontier AI workloads. In addition to requiring
strong cybersecurity standards for frontier AI developers, regulators
could define enhanced security standards for compute providers who offer
infrastructure capable of training frontier models to close security
gaps.\footnote{For an example of what such standards might look like,
  see security requirements in the Federal Risk Authorization Management
  Program (FedRAMP). FedRAMP assigns different levels of requirements
  depending on the sensitivity of the use case
  \citep{Baselines}.}

\subsection{Record Keeping}\label{record-keeping}

Record keeping is highly feasible using tools and metrics currently
available to compute providers, who already collect a wide range of data
on customers and service usage for:

\begin{itemize}
\item
  Accurately billing customers

\item
  Marketing new services to customers

\item
  Maintaining and optimizing service provision

\item
  Detecting and responding to fraud, abuse, security risks, and
  technical issues

\item
  Complying with legal obligations, such as financial record keeping

\end{itemize}

Compute providers also share these records with third parties for
activities such as:

\begin{itemize}
\item
  Exchanging information with other companies for fraud prevention,
  detection, and credit risk reduction

\item
  Providing third-party vendors with information for promotional and
  marketing purposes

\item
  Complying with legal obligations, such as an enforceable government
  request

\end{itemize}

Compute providers typically have well-defined privacy policies around
these records, including specific retention and security policies based
on the sensitivity and business- or legal use cases for different kinds
of records. Some of this data will likely be useful for verification
activities relevant to frontier AI governance. The specific data
attributes generally collected by compute providers can be found in
\cref{tab:Overview-Categories-Data-Attributes} below, mapped onto specific use cases for governance
purposes. This information is based on conversations, public data
collection, and privacy policies available from a representative sample
of large and small compute providers \citep{awsPrivacyNotice2024,
coreweavePrivacyPolicy2022, fluidstackFluidStackPrivacyNotice2022,
googlecloudGoogleCloudPrivacy2024, lambdaLambdaPrivacyPolicy2022,
microsoftMicrosoftPrivacyStatement2024}.

\subsection{Verifying}\label{verifying}

There are a range of verification activities that compute providers
could perform to support frontier AI governance. Primarily, it will be
useful for compute providers to verify the identity of any customer
seeking to access a hardware configuration capable of efficiently
training a frontier model (``identity verification''), as discussed in
\citet{lennartOversightFrontierAI2023} and by Microsoft
\citep{smith2023}. It may also be useful for compute providers to
serve as an independent form of validation for different properties of
frontier AI workloads. We find that data attributes already widely
available to compute providers can likely enable them to adequately
verify two key properties of a workload that are currently highly
relevant for frontier AI governance:

\begin{itemize}
\item
  The stage of the AI lifecycle the workload fits into, e.g.,
  large-scale model training or inference (``workload classification'')

\item
  The quantity of compute consumed by the workload (``compute accounting'')

\end{itemize}

In the future, it may also be useful for compute providers to verify aspects
of the particular code or data used in a workload, such as the specific
model that was deployed, or the type of data used to train a model. We
describe such activities as ``detailed workload verification.''
Currently, this is largely not possible without directly observing
confidential customer code or data. However, with some technical
development work, it may become possible to implement wider use of
``trusted execution environments'' to allow customers to prove certain
properties of their workloads to their compute provider (or directly to
a regulator) without revealing other sensitive data
\citep{aarneSecureGovernableChips2024,
nvidiaConfidentialComputeNVIDIA2023}. We now describe each of these
potential verification activities in more detail.

\subsubsection{Identity Verification}\label{identity-verification}

Identity verification (also commonly known as ``Know-Your-Customer
(KYC)'') is a useful measure to ensure customers are meeting applicable
rules, and enforce relevant penalties. This may include ensuring
particular developers are reporting relevant large-scale training runs
or enforcing export controls that prevent certain customers (e.g., those
with links to foreign military/intelligence organizations) from
accessing particular services. In the beginning of 2024, the US Department of Commerce
proposed new regulations that place explicit identity verification
requirements on US compute providers offering services to foreign
customers \citep{federalregister2024}.
\citet{lennartOversightFrontierAI2023} discuss in further detail the
mechanisms of a KYC scheme for customers accessing large-scale compute,
drawing on lessons from the financial sector.


\begin{table}[ht]
\centering
\small
\begin{adjustwidth}{-0.05\linewidth}{-0.05\linewidth}
\renewcommand{\arraystretch}{1.35}
\rowcolors{1}{white}{lightgray}
\resizebox{\linewidth}{!}{%
\begin{tabular}{p{0.26\textwidth}C{0.14\textwidth}C{0.175\textwidth}p{0.35\textwidth}}\toprule
\begin{tabular}[c]{@{}C{\linewidth}@{}}\bfseries \footnotesize Attribute category\end{tabular} & %
\begin{tabular}[c]{@{}C{\linewidth}@{}}\bfseries\footnotesize Uses (in terms of specific verification activities)\end{tabular} & %
\begin{tabular}[c]{@{}C{\linewidth}@{}}\bfseries \footnotesize Involves collection of data not already widely collected?\tablefootnote{Based on our current understanding and insights gathered from interviews, and given publicly available information.}\end{tabular} & \begin{tabular}[c]{@{}C{\linewidth}@{}}\bfseries \footnotesize Current state of collection, validation, and possible circumvention\tablefootnote{In this column, we collect information on whether each data attribute is already collected, and if not, how it could be collected. We also list whether it might be possible for customers to falsify each data attribute in order to avoid information being verified.}\end{tabular}\\\hline
\begin{tabular}[c]{@{}p{\linewidth}@{}}\emph{Customer information}\\ e.g., name, billing address, credit card data, IP addresses, date and time of access, device identifiers, language\end{tabular} & \begin{tabular}[c]{@{}C{\linewidth}@{}} Identity\\ verification\end{tabular} & \begin{tabular}[c]{@{}C{\linewidth}@{}}No, already collected.\end{tabular} & \begin{tabular}[c]{@{}p{\linewidth}@{}}Compute providers already collect a wide range of customer information.\\ Customers can potentially spoof much of this data to try to avoid identification.\end{tabular} \\
\addlinespace[10pt]
\begin{tabular}[c]{@{}p{\linewidth}@{}}\emph{Billing-related technical information}\\ e.g., hardware configuration requested by a customer, number of hours that hardware resources are used\end{tabular} & \begin{tabular}[c]{@{}C{\linewidth}@{}}Workload classification\\ Compute accounting\end{tabular} & \begin{tabular}[c]{@{}C{\linewidth}@{}}No, already collected.\end{tabular} & \begin{tabular}[c]{@{}p{\linewidth}@{}}Already collected by compute providers for billing purposes. Highly difficult or impossible for customers to alter to avoid monitoring.\end{tabular} \\
\addlinespace[10pt]
\begin{tabular}[c]{@{}p{\linewidth}@{}}\emph{Cluster-level technical information}\\ e.g., power consumption, network bandwidth utilization between nodes\end{tabular} & \begin{tabular}[c]{@{}C{\linewidth}@{}}Workload classification\\ Compute accounting\end{tabular} &\begin{tabular}[c]{@{}C{\linewidth}@{}} No, already collected. \end{tabular} & \begin{tabular}[c]{@{}p{\linewidth}@{}} Already collected by compute providers for service health monitoring and maintenance. Customers could modify their workloads to avoid certain forms of cluster-level verification, likely with performance penalties.\end{tabular}\\
\addlinespace[10pt]
\begin{tabular}[c]{@{}p{\linewidth}@{}}\emph{Node-level technical information}\\ e.g., AI accelerator core utilization, AI accelerator memory bandwidth utilization\end{tabular} & \begin{tabular}[c]{@{}C{\linewidth}@{}}Workload classification\\ Compute accounting\end{tabular} & \begin{tabular}[c]{@{}C{\linewidth}@{}}Potentially, already collected.\tablefootnote{While examining customer data is out-of-scope, collection and analysis of certain types of metadata could risk exposing details of the customer’s data.}\end{tabular} & \begin{tabular}[c]{@{}p{\linewidth}@{}}Possible to collect using existing tooling, and collected by some compute providers.\tablefootnote{See \citet{wengMLaaSWildWorkload2022} and \citet{GoogleClusterdata2024}.} Customers could modify their workloads to avoid certain forms of node-level verification, likely with performance penalties.\end{tabular} \\
\addlinespace[10pt]
\begin{tabular}[c]{@{}p{\linewidth}@{}}\emph{Workload-level technical information}\\ e.g., code, data, hyperparameters\end{tabular} & \begin{tabular}[c]{@{}C{\linewidth}@{}}Workload classification\\ Compute accounting\\ Detailed workload verification\end{tabular} & \begin{tabular}[c]{@{}C{\linewidth}@{}}Yes, currently not collected.\end{tabular} & \begin{tabular}[c]{@{}p{\linewidth}@{}}Compute providers cannot typically retain or inspect this information (by design). Confidential computing tools could potentially be developed to verify information in a privacy-preserving manner.\end{tabular}\\\bottomrule
\end{tabular}}
\end{adjustwidth}
\caption{An overview of the categories of data attributes
available to compute providers and how they can be used for different
verification activities.}
\label{tab:Overview-Categories-Data-Attributes}
\end{table}

Identity verification of this kind appears feasible, but we recommend
that policymakers consider several strong caveats. As described above,
compute providers typically collect a range of information relevant to
verifying customer identities. This includes
\citep{awsPrivacyNotice2024, coreweavePrivacyPolicy2022,
fluidstackFluidStackPrivacyNotice2022,
googlecloudGoogleCloudPrivacy2024, lambdaLambdaPrivacyPolicy2022,
microsoftMicrosoftPrivacyStatement2024}:

\begin{itemize}
\item
  Personal information, such as legal names, user names, email
  addresses, phone numbers, and government-issued identification
  documents

\item
  Information about customer organizations and the people in those
  organizations

\item
  Financial information, such as credit card and bank account
  information, and tax identifiers

\item
  Service-related information, such as the locations from which users
  are accessing the service, time zones, the type of device a user is
  using to access the service, the language used on that device, web
  cookies describing sites previously visited, and IP addresses used
  when accessing the service.

\end{itemize}

This equips compute providers with large amounts of useful information
for verifying the identity of customers seeking to access infrastructure
sufficient to efficiently run frontier AI workloads. Because such
customers will by definition be few in number, best practices for
identity verification could be drawn from more involved identity
verification activities such as those conducted in other industries. One
example is the ``enhanced due diligence'' process used in financial
sectors for higher-risk customers or transactions, which can involve
commissioning intelligence reports on customers or their ``beneficial
owners'' (the entity that ultimately owns or controls the customer
organization)
\citep{financialactiontaskforceFATF40Recommendations2003}. These kinds
of measures may be necessary to successfully perform identity
verification in situations where a well-resourced illicit actor is
actively trying to obfuscate their identity.

In performing identity verification, it is important for regulators and
compute providers to bear in mind potential trade-offs with user privacy
and data privacy regulation in different jurisdictions. It would likely
be useful for identity verification requirements to be standardized
across different jurisdictions. We discuss these challenges in \cref{international-coordination,governance-challenges}.

\subsubsection{Workload Classification}\label{workload-classification}

``Workload classification'' describes a scenario where an infrastructure
provider is attempting to classify a workload into a particular
category. We will consider, from a frontier AI governance perspective,
what these categories might be, and how they might be differentiated.

First, it is useful to know whether a workload relates broadly to AI.
Frontier AI workloads will generally all use AI accelerators, but not
all workloads that use AI accelerators will necessarily be AI workloads.
For example, graphics and scientific computing workloads sometimes use
AI accelerators. However, these workload categories can likely be
differentiated using observable properties of the workload. Examples of
such properties are outlined in \cref{b.-observable-data-attributes} in the Appendix. Within the
broad category of AI workloads, there are several sub-categories of
workload, corresponding to stages in the AI model's life cycle, that are
useful to differentiate from a governance perspective:

\begin{itemize}
\item
  \textbf{Design}, in which researchers and engineers experiment with
  different model designs, algorithms, and datasets.

\item
  \textbf{Training}, in which a model learns from a large dataset.
  Typically known as ``pre-training'' to distinguish from enhancement.

\item
  \textbf{Enhancement,} in which a trained model is further refined
  using a smaller data set (e.g., fine-tuning), sometimes using
  techniques such as reinforcement learning.

\item
  \textbf{Deployment} in which a trained model is used in an operational
  setting, e.g., to make new predictions (``inference'').

\end{itemize}

\begin{figure}[ht]
    \centering
    \centerline{\includegraphics[width=1.3\linewidth]{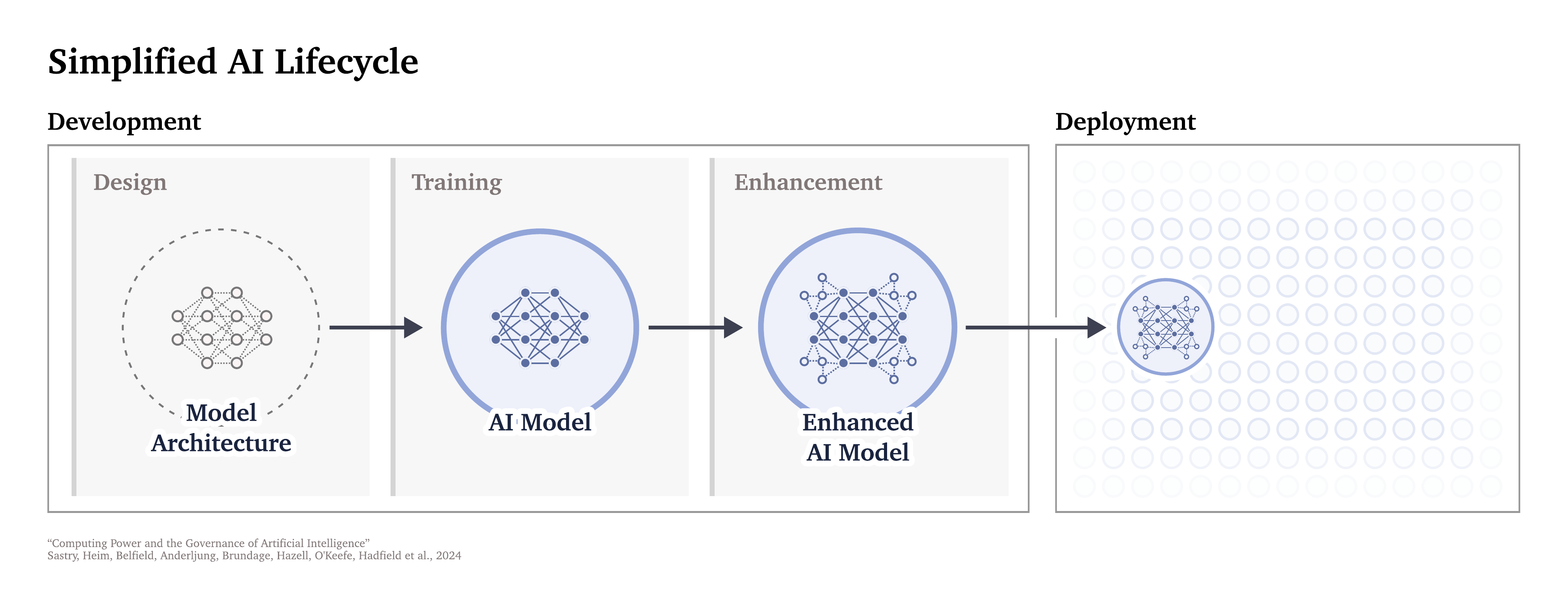}}
    \caption{Simplified AI Lifecycle including training,
enhancement (e.g., fine-tuning), and deployment (i.e., inference). (Figure from \citet{sastryComputingPowerGovernance2024}.)}
    \label{fig:Simplified-AI-Lifecycle}
\end{figure}

Each of these categories can be distinguished based on scale, among
other attributes. Given the large scale of the frontier AI workloads we
are interested in detecting and classifying, we can place a strong
initial filter on which specific workloads warrant attention, ignoring
the vast majority of workloads run on a compute provider's
infrastructure. As discussed by \citet{sastryComputingPowerGovernance2024}, the training
stage, where compute demands are especially high, is an especially
practical stage to monitor.\footnote{See Section 3.C of \citet{sastryComputingPowerGovernance2024}.}

The simplest method of workload classification might be based purely on
the hardware configuration (in terms of types and numbers of devices)
available to a customer. For example, in 2024, if a customer has
requested a hardware configuration involving tens of thousands of AI
accelerators, connected together using a high-bandwidth network fabric,
it becomes much more likely they intend to engage in large-scale
training, relative to other customers. Combining this information with
the amount of time the hardware is being used can tell us whether it was
possible for a customer to run a particular workload (e.g., pre-training
above a particular scale), which could provide grounds for a more
detailed investigation. These coarse methods are possible using data
already available to compute providers for billing purposes. See \cref{measuring-theoretical-compute-budget} below for more information on these
kinds of approaches.

More precise methods for frontier workload classification might involve
manually defining some technical characteristics of relevant workloads,
or training a machine learning classifier using cluster- and node-level
technical information to predict whether a workload falls into a
relevant category. Compute providers could also collect declarations
from customers about the workloads they are running, and use that
information as a reference point for classification. These approaches
are based on the assumption that different kinds of workloads will have
characteristic and learnable features. As an example, we expect
frontier-scale training in 2024 will likely have several distinguishing
features relative to other possible workloads. These could include the
number of accelerators used (tens of thousands), the peak operation
throughput utilization of accelerators over time (fairly constant), the
patterns of communication between and within nodes (following specific
patterns corresponding to different forms of parallelization), and
limited outbound/inbound communication to external networks, such as the
Internet, during the training run. According to six interviews with
commercial compute providers, including both large and smaller
providers, these kinds of cluster- and node-level characteristics are
often already collected and used to understand customer workloads to
optimize their services.\footnote{Interviews conducted between October
  2023 and February 2024} Research released by Google, Microsoft, and
Alibaba demonstrates some of the ways this technical information is
collected and analyzed for business purposes
\citep{jeonAnalysisLargeScaleMultiTenant2019,
tirmaziBorgNextGeneration2020, wengMLaaSWildWorkload2022}. Some of
this data has been released as public data sets that can be used to
develop workload classification techniques
\citep{AlibabaClusterTrace2024, GoogleClusterdata2024}.

Workload classification techniques using these kinds of data have been
studied in different contexts. \citep{tangMITSupercloudWorkload2022}
introduced the ``MIT Supercloud Dataset,'' containing node-level
technical information for over 3,000 AI accelerator-based workloads.
Workload classifiers trained on this data have reached 95\% accuracy at
distinguishing AI workloads across ten different model architectures
\citep{weissEvaluationLowOverhead2022}. Other research on workload
classification for different kinds of high-performance computing
workloads has reached similar levels of accuracy
\citep{banjongkanMultilabelClassificationHigh2018,
teraiWorkloadClassificationPerformance2017}, including classifiers
trained only to use data on power draw \citep{coposCatchMeIf2020,
kohlerRecognizingHPCWorkloads2021}.

However, there are several ways these findings may not be representative
for frontier AI workload classification in a production environment.
First, the authors mostly generated labeled data by running workloads
themselves, which likely involved a level of standardization in software
and datasets that would be unrealistic for real-world conditions.
Second, this research typically involved a small number of different
hardware configurations and scales, whereas these parameters will likely
vary further in production contexts. Lastly, even a 5\% error rate may
be prohibitively high in production, given the potential consequences of
reporting a false positive to a regulator.

Compute providers offering large-scale AI clusters are likely to have
the expertise to address these technical challenges. In doing so, we
recommend that compute providers---in collaborative efforts where
possible---consider a range of technical approaches for classification,
ranging from simple manually defined thresholds through to
machine-learning based classifiers. These methods could also be combined
with other useful data and processes, such as by soliciting customer
declarations on the intended use/purpose of a hardware configuration or
workload, and by conducting follow-up investigations in cases where
classification confidence is low. \hyperlink{box1}{Box 1} demonstrates what a workload
classification process combining these elements might look like.

\begin{tcolorbox}[breakable,boxrule=1pt,enhanced jigsaw, sharp corners,pad at break*=1mm,colbacktitle=lightgray,colback=lightgray,colframe=black,coltitle=black,toptitle=1mm,bottomtitle=1mm,width=\linewidth,fonttitle=\bfseries\large,parbox=false,title=Box 1: An example process for frontier AI workload classification.
,phantom={\phantomsection\hypertarget{box1}}]
\begin{enumerate}
    \vspace{-.5em}
    \item The compute provider lists the specific set of hardware configurations and scales they offer that are sufficient for efficient training of frontier AI workloads (i.e., within defined cost/time boundaries). Given that compute providers tend to specialize in particular hardware configurations (e.g., AI accelerator types and node configurations), this number could be quite small.
    \item The compute provider collects labeled cluster- and node-level data on workloads simulated or run on each of these configurations. Compute providers offering similar hardware configurations may benefit from coordinating to produce larger datasets.
    \item The compute provider creates technical thresholds to define relevant workload categories based on their identifying characteristics, and tests them on the collected data. This could include training ML-based workload classifiers. It may be the case that a single classification approach works well for a range of different hardware configurations and/or scales, or that more specific classifiers are required for different configurations.\footnotemark
    \item In operation, for any customer seeking access to or already using a relevant hardware configuration, the compute provider could then:
    \begin{enumerate}
        \item[a.] Ensure they have performed adequate identity verification.
        \item[b.] Collect declarations from the customer on the intended use of the hardware configuration, and/or declarations on the nature of any sufficiently large workload run on that configuration.
        \item[c.] Validate these declarations by running automated classification on workloads that use that configuration, and conduct follow-up analyses where useful, especially in cases where classification confidence is low.
    \end{enumerate}
\end{enumerate}
\end{tcolorbox}
\footnotetext{For an example of an approach for simulating workloads on large clusters, see \citet{sliwkoAGOCSAccurateGoogle2016}.}

The difficulty of classifying workloads increases if a compute customer
is actively trying to disguise the nature of their workload. This kind
of obfuscation may become likely in cases where a customer has a strong
financial, criminal, or political incentive to avoid regulatory
oversight. Such incentives are likely to grow when frontier AI models
become both more attractive for criminal activities and more
economically lucrative. Analogous practices can be observed in the
finance sector, where illicit actors have engaged in ``structuring''
(breaking up a single transaction into several smaller transactions) to
avoid automated transaction reporting from their bank to the regulator
\citep{linnRedefiningBankSecrecy2010}. We discuss and list these
challenges in \cref{technical-challenges}.

\subsubsection{Compute
Accounting}\label{compute-quantity-accounting}

We introduce ``compute accounting'': measurements and techniques to produce an estimation of the amount of compute consumed
by a customer running one or more workloads on a specific compute
cluster. These techniques are comparatively similar to the previous
section on workload classification (\cref{workload-classification}). However, rather than establishing the \emph{class} of a workload, compute accounting
aims to determine its \emph{magnitude}. Moreover, compute accounting is
useful even when the workload is not classified at all, as an estimate
of the total amount of compute used by a given customer is an upper
bound for a single (unknown) workload. From a practical governance
perspective, compute accounting could be used as an input to workload
classification, and/or as a standalone metric to determine whether a
particular workload has exceeded a compute-based reporting threshold.

The amount of compute used by a given workload is a useful metric from a
governance perspective. In the context of AI training, novel
capabilities (and related risks) are likely to first emerge in models
that require large amounts of compute
\citep{pilzIncreasedComputeEfficiency2024,
sevillaComputeTrendsThree2022}. In the context of AI inference,
compute is correlated with the scale and processing speed of the
deployment: how many copies of the model are being run, and how fast the
model is operating (the throughput, e.g., tokens per second for LLMs).
Insofar as the model is capable enough to potentially cause harm, these
factors could then be correlated with risk and the need for enhanced
oversight (Appendix I of \citet{obrienDeploymentCorrectionsIncident2023}).

\begin{figure}[ht]
    \centering
    \centerline{\includegraphics[width=1.3\linewidth]{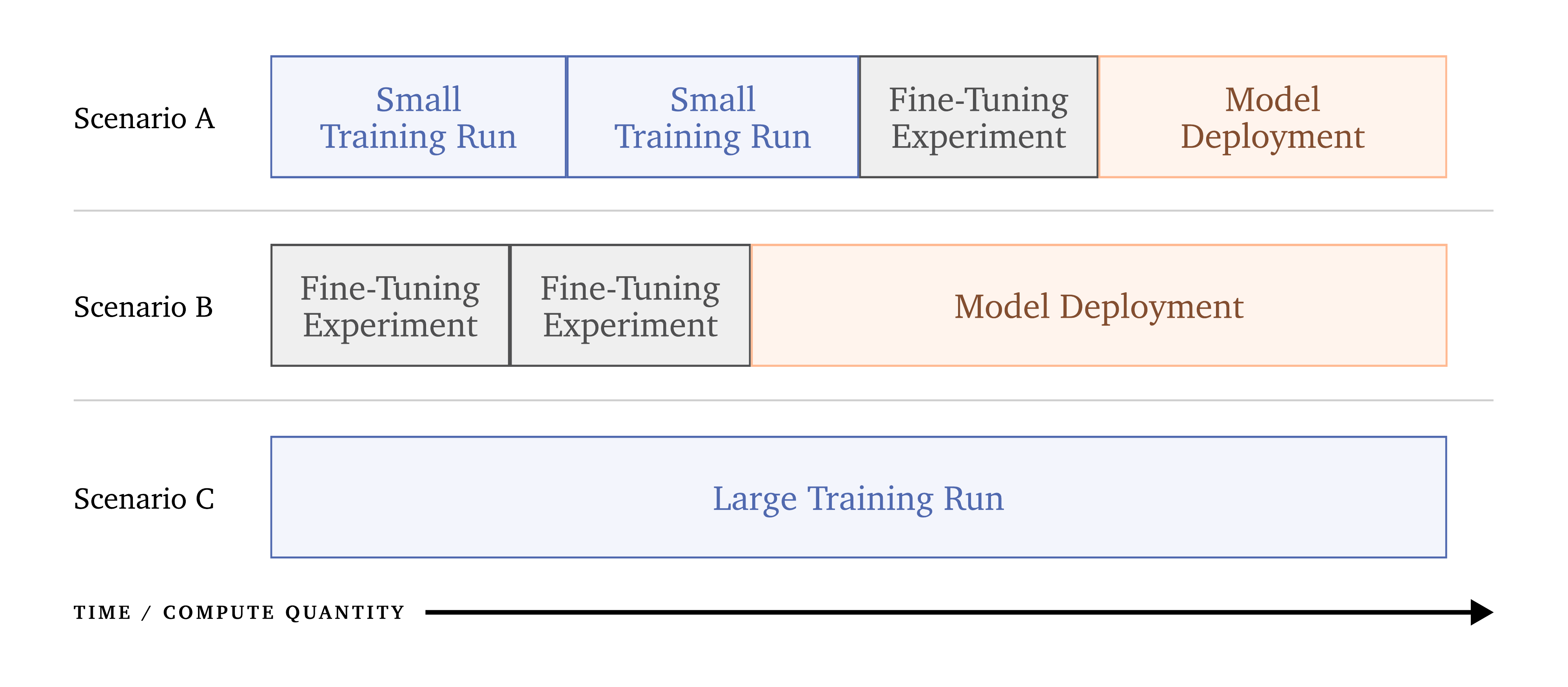}}
    \caption{Three example scenarios of a set of AI accelerator
nodes running different workloads over time. Compute accounting establishes the amount of compute used over
time, while workload classification can differentiate between these
three scenarios by mapping compute usage to specific workloads.}
    \label{fig:Three-Usage-Cases-AI-Accelerator-nodes}
\end{figure}

More formally, the total computing power of a rented cluster and how long a customer has
access to it results in a quantity of available compute---a ``compute
budget.'' The customer is then choosing how to allocate that budget
across different workloads. For example, a given set of AI accelerators
could be used for a single training run, multiple small training runs,
or model deployment (\cref{fig:Three-Usage-Cases-AI-Accelerator-nodes}). In cases where compute is being consumed by a customer (as opposed to hardware sitting idle), the amount of compute consumed can be attributed to at least one workload. The addition of
workload classification allows fractions of that usage to be ascribed to workloads of particular types.

We can estimate the compute budget via two different approaches:

\begin{enumerate}
\def\labelenumi{\arabic{enumi}.}
\item
  \textbf{Theoretical compute budget estimation:} calculated using the
  assumed throughput (measured in OP/s) of the hardware potentially
  involved in the workload, and multiplying it by the time the hardware
  is being used.

\item
  \textbf{Empirical compute budget estimation:} calculated using actual
  measurements from the hardware that can serve as more direct proxies for compute consumption. For example, aggregating AI
  accelerator core utilization and time-in-use data across all AI
  accelerators involved in a workload, and multiplying by the peak
  capacity of each core.

\end{enumerate}

\begin{figure}[ht]
    \centering
    \centerline{\includegraphics[width=1.3\linewidth]{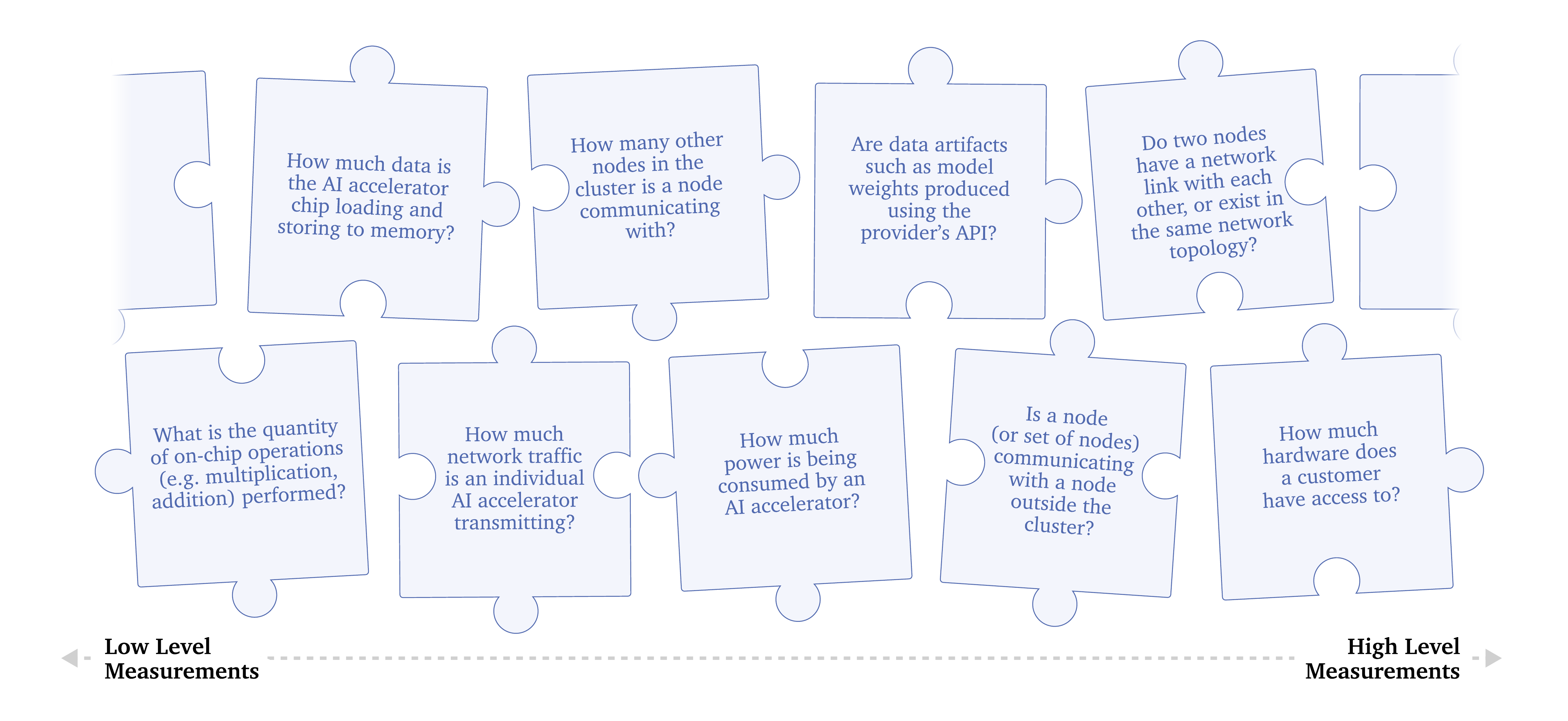}}
    \caption{A spectrum of possible compute usage metrics for AI workload
analysis, from low-level measurements, such as on-chip calculations, to
more high-level measurements, such as the hardware available to a
customer. Each of these metrics can be synergistically combined to enhance the accuracy and sensitivity of workload classification and compute accounting.}
    \label{fig:Spectrum-Usage-Metrics}
\end{figure}

Theoretical compute is a derivative of empirical compute, useful for
establishing an estimate in circumstances where empirical measurements
are not available. As exact circumstances and configurations differ between compute
providers, not all attributes of both theoretical and empirical compute
are likely to be observable. However, in practice, both kinds could be
used to inform an overall estimate of compute usage for a particular
instance of running a workload (\cref{fig:Spectrum-Usage-Metrics}).

Regardless of whether the approach is theoretical or
empirical, it will be important for compute providers to estimate both throughput (OP/s) and total quantity of
operations (OP). This provides crucial information to detect cases where a customer is evading a
reportable compute threshold by utilizing multiple compute providers or
accounts to sequentially perform partial training of a model. In this
case, the throughput available to the customer is necessary to identify
that a \emph{rate} of compute usage corresponding to a reportable
threshold has been reached, even if the reportable threshold itself has
not. We discuss this in \ref{technical-challenges}.

\textbf{Measuring Theoretical Compute
Budget}\label{measuring-theoretical-compute-budget}
--- Theoretical approaches measure the \emph{potential} for a certain amount
of compute to be used for one or more workloads within a given time
frame. This is easier to measure than empirical compute, and in the
simplest form is equivalent to hardware resources a customer has been
allocated to access within the cluster. For any given compute provider,
the number of customers with access to sufficient theoretical compute to
train a frontier model will be small.\footnote{For example, to meet the
  AI Executive Order's reporting requirement of 10\textsuperscript{26}
  operations for a training run, a customer will need access to around
  60,000 cutting-edge AI accelerators (Nvidia H100) for 90 days,
  assuming a utilization of 34\%.} This makes theoretical compute a
useful measure for determining which specific customers are relevant for
a frontier AI regulatory regime. This can be calculated using data
already available to compute providers for billing purposes (\cref{tab:Overview-Categories-Data-Attributes}). Relevant data for measurement of compute are:

\begin{itemize}
\item
  \textbf{Node assignment}: Compute providers can bill customers for
  \emph{on-demand} nodes (a full or partial node) at a granularity
  ranging from seconds to hours \citep{awsAmazonEC2OnDemand}, or
  \emph{reserved} nodes ranging from days to months
  \citep{awsAmazonEC2Reserved}. Theoretically, the used compute budget
  can be calculated using this information by summing the theoretical
  peak performance of the AI accelerators in each node, multiplying it
  by the time the node is available to the customer, and the assumed
  average utilization of the AI accelerators.\footnote{``Utilization''
    in this context refers to the usage as a proportion of the node's
    theoretical peak computational performance.}

\item
  \textbf{Data ingress/egress}: Data into and out of the cluster is
  metered and sometimes billed \citep{googlecloud, microsoftazure,
  pal2021}. The communication of nodes within the cluster to endpoints
  outside the cluster, as well as the amount of data transferred and
  time when communication occurs, can inform whether nodes outside the
  cluster participated in a training run or deployment.

\end{itemize}

The exact procedures to allocate, measure, and invoice customer usage
for billing purposes are not publicly available for any major provider.
However, every provider must have internal control systems and
diagnostics to record this information accurately, as well as status
reporting and other telemetry to maintain the health of their clusters
(such as the state of individual machines and network switches). While
billing information provides a widely-measured baseline for customer
compute usage, intra-cluster network information such as the network
topology can provide greater detail. Specifically, knowledge of whether
two nodes are capable of communicating within a cluster informs whether they may
participate in running the same parallel workload.

Theoretical compute, while relatively simple to calculate in most cases,
is useful primarily to establish an upper limit on the total compute
budget of a customer. To map compute usage onto a particular workload
requires additional information at the cluster or node level, as
previously covered in \cref{workload-classification}.

\textbf{Measuring empirical compute
budget}\label{measuring-empirical-compute-budget}
--- Measurements of empirical compute usage involve observations of a
cluster's hardware-level characteristics, often measurements of the node
itself (perhaps from a hypervisor or other privileged software) or
inter-node communication fabric. Empirical compute, in contrast to
theoretical, can provide a highly precise and accurate accounting of the
amount of compute, though some limitations exist. We consider that two
categories of measuring empirical compute exist: \emph{operations} and
\emph{data transfer}. \emph{Operations} refers to the mathematical
calculations performed as part of a workload (most frequently
multiplication and addition for contemporary AI workloads). \emph{Data
transfer} refers to the movement of the data necessary to perform those
calculations: network links from node-to-node, or chip-to-chip within a
node or loaded from an AI accelerator's memory.

While these properties are essentially metadata, compute providers would
need to detail collection and observation of this within their terms of
service with very clear guidelines about how the data will be collected,
stored, and used. While strict internal policies are
necessary to ensure the integrity of such metadata, this kind of usage
policy would likely not require any significant deviation from existing
policies for sensitive customer data handling, or deviation from the kinds of data that are already often collected
\citep{awsPrivacyNotice2024, coreweavePrivacyPolicy2022,
fluidstackFluidStackPrivacyNotice2022,
googlecloudGoogleCloudPrivacy2024, lambdaLambdaPrivacyPolicy2022,
microsoftMicrosoftPrivacyStatement2024}.

Within a given node, opportunities to measure empirical compute include:

\begin{itemize}
\item
  \textbf{Operations performed on AI accelerators}: Individual chips
  contain \emph{performance counters} to measure information such as the
  number of instructions executed \citep{enwiki:1169157291}. A vendor
  tool may be required to access this information
  \citep{nvidiaNVIDIANsightPerf}.

\item
  \textbf{Data flow to/from AI accelerator's memory:} The rate at which
  data is written to or read from the AI accelerator's memory can be
  observed over time, allowing measurement of throughput and quantity,
  and can inform an estimation of the total number of operations
  performed \citep{nationalenergyresearchscientificcomputing,
  williams2008}.

\item
  \textbf{Data traffic between accelerators and other nodes}:
  Node-to-node and chip-to-chip communication is an indicator of
  participating in the same workload, even if the workload itself cannot
  be classified \citep{merritt2023, nvidia, shoeybi2020}.

\end{itemize}

Even without privileged software access to the node, other measurements
of cluster operations are useful to inform an estimate of empirical
compute:

\begin{itemize}
\item
  \textbf{Power consumption}: In cases where precise chip utilization is
  not observable, measurements of power consumption (of a node or
  individual AI accelerators within a node) can help inform an estimate.
  The amount of power consumed by each node is considerably higher when
  a node (or even an individual AI accelerator \citep{zotero-3386}) executes a workload compared to idle. However, power consumption does
  not simply scale linearly with performance
  \citep{patelPOLCAPowerOversubscription2023}, though specific
  calibration for a device may enable improved estimation. Power consumption will typically be a way of measuring both operations and data transfer, as both these activities consume energy within a node.

\item
  \textbf{Data traffic between nodes}: Granular information such as the
  number of data sent to and from the node, the source and destination of
  this data, and the timing with which they are sent can inform how
  multiple nodes are cooperatively executing the same workload.

\end{itemize}

While many of these measurements alone provide limited insight, combining
measurements can provide more insight into
the compute usage of a particular workload (\cref{fig:Spectrum-Usage-Metrics}). Empirical measurements form an upper bound for compute usage, as not every
operation can be known to have contributed to a workload.

\subsubsection{Detailed Workload
Verification}\label{detailed-workload-verification}

To verify compliance with regulations on the development or deployment
of frontier AI systems, it may be useful for a compute provider to
validate more fine-grained features of a workload, such as whether a
particular training dataset was used, the model architecture, or whether
a particular model evaluation was run. We call such activities
``detailed workload verification.'' This form of verification differs
from workload classification in that it will almost always require
knowing certain properties of the code and/or data used by the customer.

One undesirable form of workload verification would simply require
compute providers to have direct access to customer code and data.
However, this level of access is not acceptable, as compute providers
will not access customer data without permission unless required to
maintain the health of their cluster or legally compelled
(see \citet{awsAWSCustomerAgreement2024}, and other terms of services from
compute providers). Internal risk management processes, such as auditing
access to customer data by employees, typically govern details such as
when access occurred, by whom, and whether it was authorized.

\begin{figure}[ht]
    \centering
    \centerline{\includegraphics[width=1.3\linewidth]{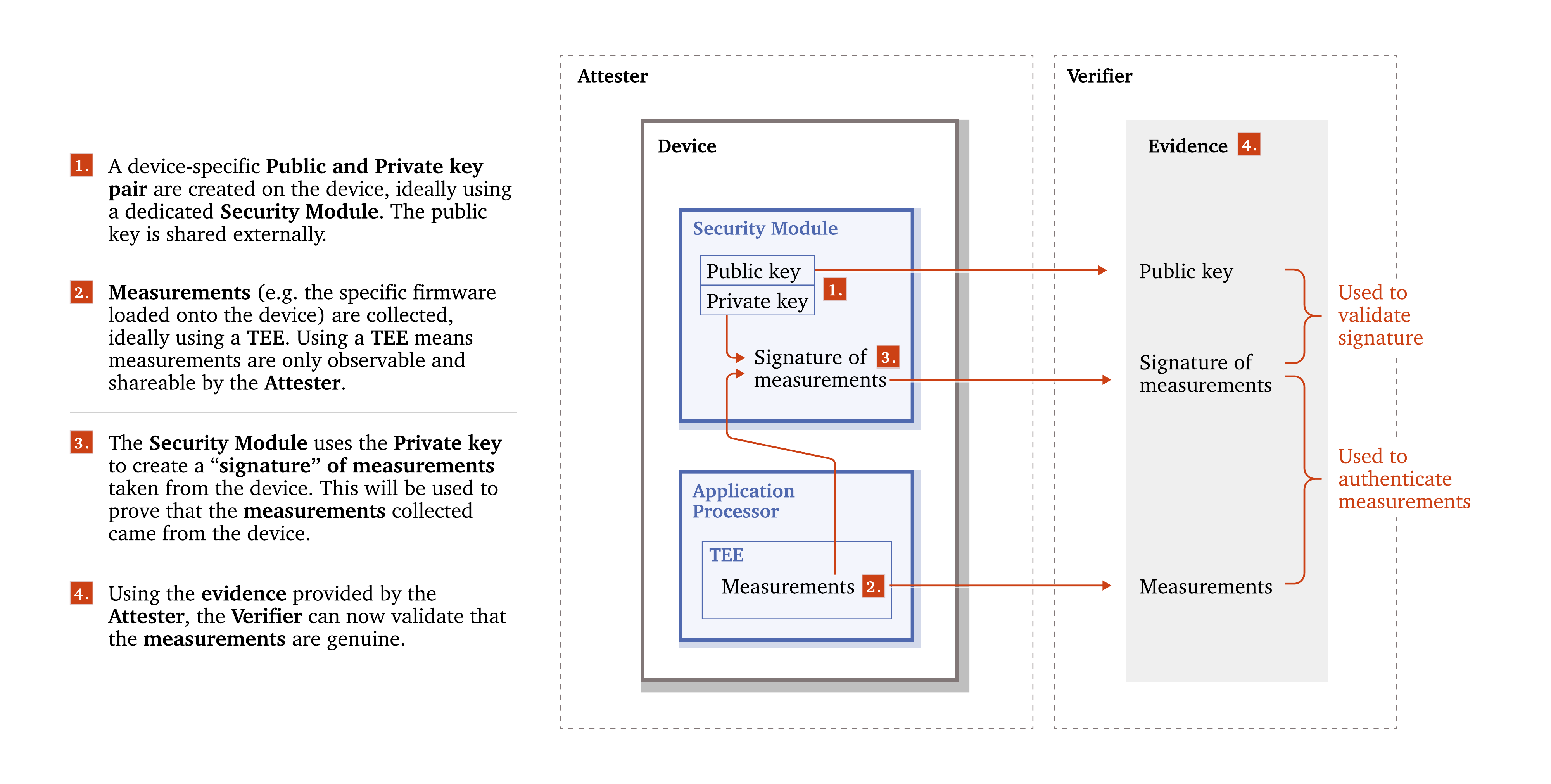}}
    \caption{Using confidential computing techniques allows an
``attester'' (customer) to share high-level information about a workload
with a ``verifier'' (e.g., a compute provider or a regulator) such that
the verifier can trust the information, without the attester sharing any
additional code or data. (Adapted from
\citet{aarneSecureGovernableChips2024}.)}
    \label{fig:Attester-Verifier}
\end{figure}

However, using privacy-preserving technologies built into data center
hardware, it may become possible for a compute provider, in
collaboration with a customer, to verify particular properties of a
workload without observing any other information---only the required
verification result needs to be shared
\citep{aarneSecureGovernableChips2024}. As one example, many modern
CPUs and AI accelerators, such as NVIDIA's H100, and data center CPUs
from AMD and Intel, come equipped with a ``trusted execution
environment'' (TEE), allowing the AI accelerator/CPU's customer to
assert the confidentiality and integrity of code/data, while exposing
only the code/data they choose to, and having full control over who they
expose it to (\cref{fig:Attester-Verifier}). Techniques that leverage a TEE in this way
are often known as ``confidential computing.'' Compute providers are
increasingly making these features available to customers
\citep{awsAWSNitroEnclaves,
microsoftlearnConfidentialComputingAzure2024,
microsoftlearnWhatConfidentialComputing2023}.

Using confidential computing techniques, customers may be able to
provably verify particular governance-relevant properties of their
workloads to their compute provider or directly to a regulator. For
example, a customer may wish to demonstrate that they ran a particular
model evaluation, obtained a particular result on a model evaluation, or
did (not) use a particular dataset during training. However, these
techniques have yet to be fully validated and implemented in production contexts. Several organizations are
actively researching and developing software for using confidential
computing to allow privacy-preserving auditing of models
\citep{MithrilsecurityBlindai2024, HowAuditAI2023,
OpenMinedPySyft2024}. There has also been some work on expanding these
techniques to allow privacy-preserving auditing of training workloads
(e.g., the dataset used, or quantity of compute consumed), though this
area is less well-explored
\citep{choiToolsVerifyingNeural2023,MithrilsecurityAicert2024}.\footnote{\citet{choiToolsVerifyingNeural2023} propose a method for verifying training data, but it requires
  sharing of sensitive data with the verifier. Making the scheme fully
  privacy-preserving is discussed but left for future work by the
  authors.} If regulatory
requirements on compute providers end up requiring them to validate more
fine-grained properties of workloads, these kinds of methods could be
used to achieve this in a way that preserves customer confidentiality
and privacy. In the meanwhile, we encourage compute providers and
developers to explore and develop these techniques to ensure they can be
implemented without meaningful performance penalties, and while
preserving other aspects of customer experience and confidentiality.


\section{Constructing an Oversight
Scheme}\label{constructing-an-oversight-scheme}

Incorporating roles for compute providers in security, record keeping,
verification, and enforcement, accompanied by appropriate reporting,
could form the basis of an effective compute oversight scheme. Such a
scheme could enable greater visibility of AI development and help ensure
the adoption of appropriate safeguards. There is an opportunity to build
on and complement existing policies around compute governance occurring
around the globe.

This section looks at the US as a case study. We examine the Biden
Administration's 2023 Executive Order 14110 on Safe, Secure, and
Trustworthy AI (``the AI Executive Order'')
\citep{thewhitehouseExecutiveOrderSafe2023}: outlining the ways in
which it progresses record keeping requirements for foreign customers
and signals a need for greater verification and enforcement
capabilities. We then explore what additional steps might be required to
enlist compute providers in administering a more comprehensive compute
oversight scheme, and where further research is needed. We highlight the
importance of internationalizing a compute oversight scheme, and outline
some of the key challenges that need further attention and analysis
before this proposal is ready for adoption.

Unlike proposed US foreign customer identification rules for
IaaS providers \citep{federalregister2024}, we focus on oversight of
only frontier AI model development and deployment, rather than all
compute use. While we explore these issues in the US context, similar
analyses could also be done for other jurisdictions, like the EU, and in
the international context. We encourage further policy analysis in this
space.

\begin{figure}
    \centering
    \centerline{\includegraphics[width=1.3\linewidth]{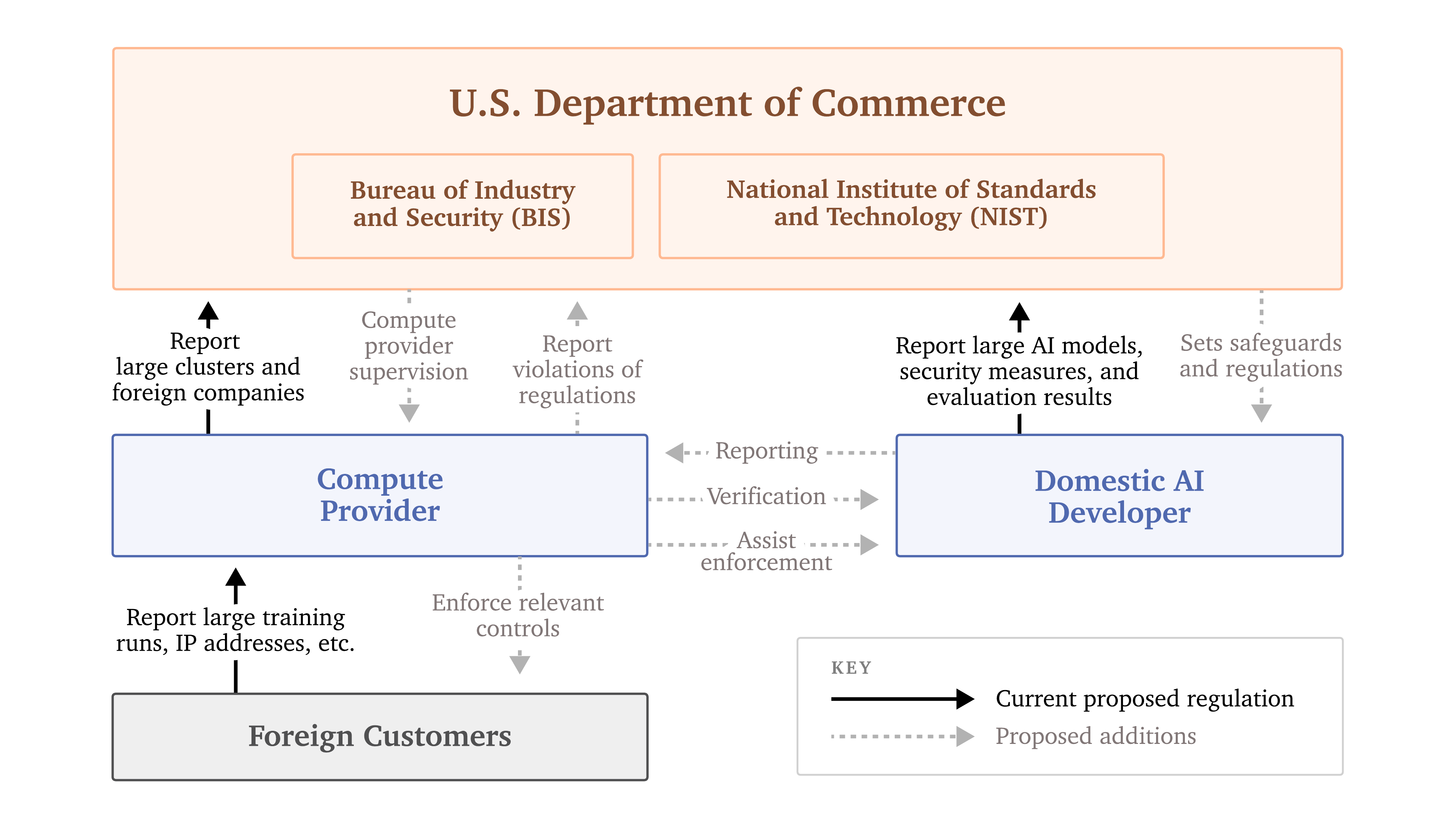}}
    \caption*{\textbf{Figure 2:} Additional measures, implemented by the Department of
Commerce, would strengthen the intermediary role of compute providers
and enable a compute oversight scheme.}
    \label{fig:Additional-Measures-Commerce}
\end{figure}

\subsection{Case Study: Compute Providers' Intermediary Role in the
US}\label{case-study-compute-providers-intermediary-role-in-the-us}

\subsubsection{Record Keeping and Reporting in the AI Executive
Order}\label{record-keeping-and-reporting-in-the-ai-executive-order}

The AI Executive Order acknowledges the importance of compute providers
in AI regulation, and begins to impose record keeping and reporting
requirements. First, under the authority of the Defense Production Act
(DPA), it requires firms owning or developing clusters capable of
collectively providing more than 10\textsuperscript{20} OP/s with more
than 100 Gbit/s networking bandwidth\footnote{Threshold as stipulated by
  the AI Executive Order. Note it is subject to updates.} to report its existence and location. Once
implemented, this measure would provide the US Government with
visibility of the most significant compute infrastructure in the US.
While the high reporting threshold will not capture all compute clusters
capable of training frontier AI, it will nevertheless be useful in
identifying relevant industry stakeholders that would play a role in
frontier AI governance.

Additionally, leveraging the powers of the International Emergency
Economic Powers Act (IEEPA), the AI Executive Order requires that
compute providers identify and report to the Department of Commerce when
a foreign person uses their services to train a frontier AI
model.\footnote{In this case, models trained on with more than
  10\textsuperscript{26} operations and occurring on frontier
  infrastructure.} The Department of Commerce has proposed rules to
implement this that would require US compute providers (and their
foreign resellers) to maintain Customer Identification
Programs.\footnote{The proposed rules also require US IaaS providers and
  their foreign resellers to verify the identities of \emph{all} foreign
  customers. This broader measure has come under criticism for being
  ineffective in addressing cyber threats, giving rise to privacy
  issues, and impacting the competitiveness of US compute provision
  \citep{nationalsecuritytelecommunicationsadvisorycommittee}. This
  paper does not engage in analysis of this broader measure, and instead
  focuses on the subsection of compute for frontier AI. As outlined in
  the introduction, this narrower focus only captures a small number of
  AI firms and compute providers, which mitigates much of the concern
  around regulatory burden (see discussion in \cref{limitations-and-future-research-directions}.}

\subsubsection{Potential for Verification
Capacity}\label{potential-for-verification-capacity}

The AI Executive Order offers the potential for compute providers to
exercise verification capacities. In a direct sense, the proposed rules
would require US compute providers to take a risk-based approach to
verifying the identities of foreign customers and their beneficial
owners, whether the customer is running a training workload with
particular technical properties, and reporting large training runs by
foreign customers to the Department of Commerce
\citep{federalregister2024}. In an indirect sense, the AI Executive
Order introduces a range of requirements for AI developers, which will
eventually require verification capacities to be effective. For example,
AI developers are required to report on the training of large models,
the cybersecurity protections taken to secure model weights, and the
results from red-team testing according to guidelines developed by the
National Institute of Standards and Technology (NIST). While the AI
Executive Order does not address how these measures will be verified,
compute providers could become governments' natural partners for
checking compliance.

The AI Executive Order and the proposed rules for IaaS providers give
compute providers a role in \emph{enforcing} regulations---enabling
the Government to require the compute provider to prohibit or limit
access to an account or an entire jurisdiction---but only when it
involves foreign customers with demonstrated patterns of using US IaaS
for malicious cyber-enabled activity. There is currently no explicit
authority that requires compute providers to stop a foreign customer
training a frontier model that might be used maliciously in the future. The existing authorities also would not apply in cases where risks arise from models being
developed domestically.

\subsubsection{Future Additions}\label{future-additions}

The AI Executive Order lays the groundwork for an institutional
structure that could support a compute oversight scheme. Importantly, it
begins to build greater AI understanding and regulatory capacity within
the US government, establishes clearer government accountabilities for
frontier AI governance, and establishes foundational elements of a
broader regulatory framework. Yet it falls short of a comprehensive
oversight scheme. Additional steps are required to effectively identify
and manage emerging risks and enable proportionate oversight of frontier
AI development. We outline potential measures to achieve this, including
ensuring proportionate information security practices, expanding and
streamlining record keeping requirements, enhancing verification roles,
and establishing enforcement capabilities.

\textbf{Ensuring proportionate information security practices}
--- Given the potential risks associated with frontier AI, further work may
be warranted to ensure cybersecurity standards are proportionate to
risk. Because compute providers are custodians of sensitive data and IP
for a broad range of clients, actions they take to improve security will
benefit all their AI developer customers and their respective
users.\footnote{Note, however, that strong cyber security at the
  infrastructure level alone is insufficient. AI firms will still need
  to implement and maintain strong cyber security practices on their own
  systems (\cref{security}).}
Further cooperation between industry and government (including bodies
like NIST and CISA) is required to consider what cybersecurity standards
are most appropriate for entities that hold sensitive frontier
AI-related IP. This could be informed by the risk profiles underpinning
the cybersecurity standards and practices that are currently being
developed by leading AI companies, as per their voluntary White House
commitments to manage AI risks
\citep{thewhitehouseFACTSHEETBidenHarris2023}.

\textbf{Expanding and streamlining record
keeping}\label{expanding-and-streamlining-record-keeping}
--- While the proposed rules require US compute providers to keep records on
foreign AI developers, significant AI-related risks may also arise from
domestic AI development. This may make it appropriate to expand the role
of compute providers to also undertake KYC on domestic developers of
frontier AI to enable a more comprehensive oversight scheme
\citep{smith2023}. Under the AI Executive Order, domestic AI
developers are themselves accountable for reporting their own frontier
training runs to the Department of Commerce. Using the compute provider
to collect information that could then be used to validate these
processes would help improve their effectiveness and ensure compliance.
In this way, KYC could be implemented as a cohesive scheme, drawing on
lessons from the financial sector
\citep{lennartOversightFrontierAI2023}.

In addition to using training compute thresholds, other measures should
be employed to create a more precise risk-management system. The current
10\textsuperscript{26} operations threshold minimizes regulations on
existing systems while capturing next-generation models that may pose
significant dual-use risks. However, below-threshold compute could also
be relevant in identifying problematic trends (e.g., from entities in
particular geographic regions) and entities trying to break up workloads
over multiple compute providers \citep{heimAccessingControlledAI2023}.
Furthermore, the threshold should be constantly re-evaluated as frontier
AI evolves and algorithmic innovations reduce the cost of training
powerful systems and if our understanding of how to predict risks from
compute in particular systems grows
\citep{pilzIncreasedComputeEfficiency2024}. On the training side,
indicators like training data (similar to the biological sequence data
criterion referenced in the AI Executive Order), the architecture of the
AI model, or the way training is conducted could all be useful proxies
for identifying risk levels from new AI systems. On the deployment side,
which does not currently fall under the purview of the AI Executive
Order, factors like the use of customer data (e.g., voice or images),
the scale of deployment, the level of access to the outside world (e.g.,
via the internet or physical effectors), and the ability to act with
limited direct supervision could be used to set a range of regulatory
thresholds \citep{shavitPracticesGoverningAgentic2023}. Developing
more nuanced thresholds, beyond blunt compute capacity and usage, will
require further research and collaboration between government, compute
providers, AI developers, and broader civil society.

\textbf{Enhancing verification
roles}\label{enhancing-verification-roles}
--- There is an opportunity to leverage compute providers' verification
capabilities to help ensure both foreign and domestic AI developers are
complying with AI safety standards and requirements. The AI Executive
Order requires AI developers to report the total amount of training
compute and whether biological sequence data was used to train their
frontier models, as well as the outcomes of safety testing. However,
there are presently no verification methods specified for these
requirements. The ability of compute providers to capture insights from
the metrics described in \cref{technical-feasibility-of-compute-providers-governance-role} could enable verification to be
performed while minimizing privacy tradeoffs. For example, rather than
reporting all metrics to the US government, compute providers could
instead only report when they have reasonable grounds to suspect a
violation of regulations or standards has occurred. Verification
mechanisms will need to be resilient against evasion and exploitation.
For example, privacy-preserving information sharing between compute
providers could be used to help identify and manage attempts to break
training runs down into smaller segments, or to obfuscate compute
utilization patterns. As regulatory frameworks develop, there may be an
opportunity for compute providers to play a role in advancing both
responsible training requirements for frontier model developers,
including notification and use of secure infrastructure, and
pre-deployment safeguards, for example by requiring frontier model
providers to demonstrate they have government approvals prior to placing
a model on the market. Such approvals could be made contingent on the
evaluation of highly capable models to a specified standard
\citep{obrienDeploymentCorrectionsIncident2023}.

Privacy protections could also be aided by the application of
confidential computing standards, which technically limit the ability of
compute providers to `look in' at sensitive data \citep{ibm,
HowAuditAI2023}. Further work is required to explore the possibilities
of data verification in a privacy-preserving way, particularly
mechanisms to identify when a developer is using biological sequence
data.

\textbf{Establishing enforcement
capabilities}\label{establishing-enforcement-capabilities}
--- To effectively leverage compute providers in an intermediary enforcement
role, the US government would need to implement additional authorities
and requirements for them to effectively halt training runs or
deployment in response to violations of set controls and safety
standards. Such authority could be linked to a comprehensive regulatory
scheme with components such as developer licenses, training risk
assessments, pre-deployment risk assessments, and the monitoring of
criminal activity through models after deployment
\citep{anderljungFrontierAIRegulation2023}, supplemented by a process
for appeal and corrective action. To draw parallels from other
industries, financial and aerospace regulators use a range of factors,
like customer risk profiles, track records, and company practices, to
evaluate the level of oversight necessary. AI developers with a pattern
of risky behavior may warrant stricter oversight in order to purchase
computing power. The ability to deploy an AI model at a large scale may
also be linked to model licensing and proper privacy and safety
precautions on the part of the compute customer in the future.

Once appropriate authorities have been established, there are several
ways compute providers could help enforce rules. It could be desirable
to be able to prevent or pause runs if the total amount of compute used
exceeds the limit permitted by present approvals, if deployed models are
actively causing harm, or if the developer is on the Entity List.
Compute providers already have the ability to ration or cut off compute
access to their customers. Therefore, key work that will need to be done
here includes:

\begin{enumerate}
\def\labelenumi{\arabic{enumi}.}
\item
  establishing formal rules for when AI developers can and cannot access
  compute at a particular scale,

\item
  creating formal channels of communication between compute providers
  and regulators, and

\item
  clearly establishing respective roles and authorities between
  government and industry.

\end{enumerate}

\subsection{Regulating the Compute Providers
Themselves}\label{regulating-the-compute-providers-themselves}

For compute providers to effectively contribute to AI regulation
downstream, they themselves also need to be subject to appropriate
oversight and compliance measures. They are both intermediaries and
agents: they act as pass-throughs that offer computing power (in the
form of hardware, often acquired from other firms) to AI developers, and
they make autonomous choices on pricing, select customers, and provide
differentiated services.

The AI Executive Order requires that compute providers declare the
existence and location of large-scale compute clusters, and the total
amount of computing power available in each cluster. Ad hoc
investigations and spot checks would help ensure compute providers are
incentivized to prioritize this obligation. A chip registry, where
policymakers mandate the reporting of the sale and transfer of
cutting-edge AI chips, could also help ensure compliance
\citep{fistChineseFirmsAre2024, fist2023,
sastryComputingPowerGovernance2024}. But the need to verify compute
provider compliance would be significantly heightened should they take
up the key role of screening AI developers. Ongoing training and
government and industry engagement could help ensure expectations are
clear and requirements are met in a way that minimizes regulatory
burden. Additional government capacity would likely be required to
ensure appropriate spot checks and investigations can be carried out. To
help ensure compute providers play an appropriate role in providing
secure infrastructure and advancing verification and enforcement, there
may also be a role for the licensing of compute providers themselves,
similar to the way in which operators of critical infrastructure are
licensed in other domains
\citep{transportdepartmentofinfrastructure2024,
californiaenergycommission, thewaterservicesregulationauthority}.
Ensuring that government-established regulations are upheld and not
tampered with will be key.

In cases where the compute provider and AI developer are the same firm
or linked by a close partnership, additional steps should be considered.
These arrangements are common for many leading AI companies, which
either build AI data centers for their own products, for example,
Google, or maintain close financial ties with their compute providers,
like OpenAI with Microsoft and Anthropic with Amazon. The strength of
these ties has prompted recent antitrust concerns: in January 2024, the
US Federal Trade Commission began an inquiry into these partnerships
\citep{wardUSAntitrustInquiry2024}. In such cases, there are unusually
strong incentives to collude on false reporting, as rapid progress in AI
development is in both entities' financial interests. This warrants
additional external scrutiny and reporting. A whistleblower scheme could
assist in identifying and addressing noncompliance, but in the longer
term, prudential regulations may be needed to create an accountable
compute ecosystem and ensure that providers are fit for acting as
regulatory pass-throughs. Compliance with these regulations could be
linked to maintaining operating licenses or accessing cutting-edge AI
chips. These rules should ensure sufficient independence between compute
providers and their customers, and that advanced AI and compute
capabilities are safeguarded and accessed in an equitable manner. Steps
to limit the partnerships that create the strongest incentives for
collusion, including via antitrust law, could also help ensure proper
oversight. In many industries, there are restrictions on how platform
providers may operate on their own platforms, as such operations can
involve anti-competitive market practices; a similar approach may
mitigate unfair practices in the cloud computing market.

\subsection{Domestic Government
Capacity}\label{domestic-government-capacity}

Given the strong public interest in managing the risks of frontier AI, a
robust government involvement will be key to the success of an oversight
scheme \citep{anderljungFrontierAIRegulation2023}. Building on
existing efforts, the US could establish a centralized authority within
the Department of Commerce responsible for AI risk management and
engagement with compute providers and AI developers. Locating both
industry opportunity and risk management within the same department,
alongside other AI-relevant agencies like NIST, will help enable a
holistic, proportionate approach to frontier AI controls. It could also
support alignment with the Department's Bureau of Industry and
Security's (BIS) work on compute hardware controls, which leverages
channels of engagement with similar stakeholders. This authority should
work closely with industry, researchers, and other government
stakeholders to ensure a holistic approach to AI governance.

Additional regulatory and legislative authority will likely be required.
The IEEPA provides the US government with broad abilities to control
transactions with foreign persons in circumstances where the President
declares a National Emergency
\citep{InternationalEmergencyEconomic2024}. This could provide the
power for scrutiny and enforcement against \emph{foreign} persons
accessing US compute, but the requirement to maintain a constant state
of `national emergency' for these rules to apply could be criticized as
government overreach \citep{boylePresidentExtraordinarySanctions2021}.
The government is also limited in its ability to effectively utilize
compute providers as a node for domestic oversight. While the DPA allows
the government to require information from companies to undergo
industrial base assessments, this falls short of establishing ongoing
channels of information sharing and verification, and there are
currently no clear mechanisms for checking compliance. The development
of new legislation should involve thorough consultation with industry
and affected stakeholders, as well as international counterparts.

\subsection{International
Coordination}\label{international-coordination}

A harmonized international approach will be essential for the success of
a comprehensive compute oversight scheme: both to manage complex
cross-jurisdictional oversight and privacy issues, and to minimize the
risks of customers and businesses relocating to other jurisdictions to
avoid regulations. International standards that preserve privacy and
confidentiality should be explored as key components of the solution.

\textbf{Cross-jurisdictional oversight issues}
--- Existing cross-jurisdictional data, privacy, and oversight issues
associated with compute providers with a global presence will be
amplified in the context of AI compute oversight. Many large compute
providers employ globally distributed data centers to enable low-latency
international service provision and resilience to local disruptions. For
example, while headquartered in the US, major compute providers like
Microsoft and AWS have data centers spread across many regions and
jurisdictions \citep{awsAWSGlobalInfrastructure}. However, this
distributed architecture can result in contradictory local and domiciled
regulatory requirements. This has been evident in the law enforcement
context, with the US enacting the Clarifying Lawful Overseas Use of Data
(CLOUD) Act in 2018 in response to impositions on its ability to access
data from US compute providers when that data was stored in a foreign
data center \citep{rep.collinsText4943115th2018}.

In the context of AI compute providers, cross-jurisdictional oversight
is of particular sensitivity, due to both the commercial value of
frontier AI model weights, as well as the strategic significance of AI.
With both the US and China actively competing to be the world leader in
AI \citep{thewhitehouseExecutiveOrderSafe2023}, there would be strong
incentives for states to misuse any access they may attain through a
compute provider oversight regime. For example, China's 2017 National
Intelligence Law stipulates that PRC citizens and corporations must
assist with China's national intelligence efforts, if directed
\citep{translateZhongHuaRenMinGongHeGuoGuoJiaQingBaoFa2018XiuZheng2017},
and in 2022, the US government used its powers under the Foreign
Intelligence Surveillance Act to order Microsoft to give it access to
between 42,000 and 43,996 accounts
\citep{microsoftUSNationalSecurity}. This creates a low-trust
environment that could diminish the competitiveness of compute providers
in countries that have oversight regimes and diminish international
support for such a scheme.

\textbf{Privacy-preserving standards}
--- To support international cooperation on mutually beneficial AI safety
oversight, compute providers could play a leading role in working with
governments to establish privacy-protecting compute provision norms and
standards. The need to balance private interests with state interests,
and security with privacy, is not new. In 2017, Microsoft proposed
making major technology companies akin to a `Digital Switzerland' --
refusing to side with governments in any cyberattack and instead
upholding a neutral and trusted environment
\citep{smithNeedDigitalGeneva2017}. However, in 2022, major technology
companies took an active stance in response to Russia's invasion of
Ukraine: Microsoft played an active role in defending Ukrainian
government systems and networks \citep{farrellUndergroundEmpireHow2023} and SpaceX
immediately provided Ukraine with free access to its Starlink satellite
networks \citep{jonesAdvancingAdversityUkraine2023}. This has
demonstrated a growing willingness of private technology companies to
play an active role in geopolitics and this shift showcases the
limitations of voluntary rules and norms \citep{farrellUndergroundEmpireHow2023}. In
this environment, technical solutions could be key to establishing the
trust and reliability necessary for customers to trust
internationally-owned compute providers in the context of an oversight
scheme. In particular, confidential computing techniques
\citep{confidentialcomputingconsortium2022} could ensure compute
providers cannot be compelled by governments to hand over sensitive data
(\cref{detailed-workload-verification}).

Notwithstanding privacy-preserving standards to protect sensitive
customer data, requirements for compute providers to report on customers
in foreign jurisdictions will raise significant interjurisdictional
privacy challenges. Even between close partners like the US and EU,
privacy protections have already come into conflict with free data flows
across jurisdictions. For example, the EU-US Privacy Shield
\citep{europeancommissionEUUPrivacyShield2016} and its predecessor,
the International Safe Harbor Privacy Principles
\citep{europeanunionagencyforcybersecuritySafeHarborPrivacy}, designed
to allow data sharing between the EU and US, were both overturned by the
European Court of Justice because of their failure to appropriately
uphold EU citizens' privacy rights \citep{EUUSPrivacyShield2020}.
Their replacement, the 2022 EU-US Data Privacy Agreement, is also
under scrutiny, with privacy groups announcing their intention to
contest it in court \citep{kimPrivacyActivistsSlam2023}. Unilateral US
action to require reporting on foreign compute customers will likely
evoke similar privacy challenges that could inhibit US compute
providers' abilities to serve global customers. To manage privacy issues
with close allies, the US could develop tailored agreements that enable
its oversight requirements to be met through compute providers reporting
to the government of the allied country within which their customer is
domiciled. This could be made contingent on some level of information
exchange between the US and partner government. It could also serve to
incentivize uptake of a consistent scheme amongst like-minded countries.
We encourage further research to explore these issues in more depth.
Ultimately, to be successful, a comprehensive compute oversight scheme
will need to incorporate and gain a multilateral agreement for privacy
protections that balance privacy and national security.

\textbf{Consistent international approaches}
--- Unilateral action by any one country---even those with a significant IaaS market share like the US---also risks incentivizing compute
providers and customers to shift to lower regulatory environments. This
could erode the market share of jurisdictions with greater oversight,
diminishing the ability of governments to oversee and address emerging
risks \citep{lennartOversightFrontierAI2023}. The already broad
geographic spread of major compute providers' data centers, and the
ability to access them remotely, could potentially enable the global
compute market to be more easily restructured. Unilateral action would
also push malicious actors and high-risk projects into jurisdictions
with lower scrutiny, thereby undermining efforts to increase AI safety.
Achieving the greatest consistency possible across international
jurisdictions will be key to ensuring that efforts to increase oversight
are effective and robust to evasion.

Ultimately, given the global nature of the compute industry, some form
of international agreement will be needed for monitoring and
verification frameworks to be durable. International compute oversight
coordination could be developed in a similar form to the Financial
Action Task Force, which currently works to achieve international
consistency in the application of anti-money laundering and
counter-terrorism financing (AML/CTF) safeguards and regulations
\citep{financialactiontaskforceHome}. A similar body for compute
oversight could facilitate information sharing and help align regulatory
approaches and enforcement between international jurisdictions
\citep{lennartOversightFrontierAI2023}. Yet, while a majority of
jurisdictions have strong shared incentives to combat organized crime
and terrorism financing, there is a risk of differing incentives in the
compute oversight context. Some states, particularly those skeptical of
AI-related risks, may actively promote lower regulatory environments in
an attempt to attract economic activity and investment -- similar to how
Ireland has used low corporate tax rates to successfully attract
multinational corporations to be domiciled there
\citep{aldermanIrelandDaysTax2021}. Initiatives seeking to build a
shared understanding of AI-related risks, like the 2023 UK Safety Summit
\citep{gov.uk2024}, will continue to be important for strengthening
alignment on compute issues. Many countries, for example, India, UK, and
Japan, are also exploring options to build greater sovereign compute
capacity \citep{GovtConsideringProposal2023, nagao2023,
ukresearchandinnovation2023}. This increasing self-reliance could
lower interest in common international approaches that might impact
national goals. Ensuring international buy-in to a compute oversight
scheme would therefore require substantial diplomatic engagement
alongside broader economic incentives.

International compute oversight could be advanced through a ``club
approach'' \citep{tragerInternationalGovernanceCivilian2023} that
predicates access to cutting-edge chips on adherence to the scheme. As
the US, in partnership with its allies, currently has a chokehold on the
manufacturing of the advanced chips needed for frontier AI compute
\citep{allenJapanNetherlandsAnnounce2023}, it could restrict exports
to jurisdictions that are unwilling to implement appropriate oversight
measures. Just as unimpeded access to the US banking system is
conditioned on a jurisdiction's compliance with FATF
standards and regulations \citep{farrellUndergroundEmpireHow2023,
financialactiontaskforceHighriskOtherMonitored}, so too could access
to advanced AI chips require adherence with established compute
oversight standards \citep{sastryComputingPowerGovernance2024}. There
would be significant risks with this approach that require further
analysis. For example, the club approach could significantly impact the
US semiconductor industry and diminish US technology leadership. Its
success is also predicated on the US and aligned partners maintaining
the lead in advanced chip hardware. But if carefully implemented with
enough international consultation and engagement, it could provide
incentives for a broader range of countries to opt into the scheme.

Even in the context of narrower oversight clubs, there is a benefit in
supporting global standards on how compute provider governance in
general can support AI oversight and regulation. A key challenge will be
supporting cooperation between geostrategic rivals like the US and
China, as the strategic significance and dual-use capabilities of AI
incentivize strong competition. However, the foundation for increased
cooperation is already being established: both the US and China attended
the 2023 AI Safety Summit and signed the Bletchley Declaration
foreshadowing greater cooperation on safety issues
\citep{primeministersoffice10downingstreetBletchleyDeclarationCountries2023},
and the US and China have also bilaterally agreed to work together on
mitigating AI safety risks \citep{murgiaWhiteHouseScience2024}.

International cooperation could be further strengthened by centering
compute governance issues in G7 and G20 discussions -- bringing together
major economic players to agree upon shared priorities in this space and
building on existing agreements to cooperate on AI governance
\citep{thewhitehouseG7HiroshimaLeaders2023}. Engagement with large
regional bodies like the Indo-Pacific Economic Forum, the Association of
Southeast Asian Nations (ASEAN), and the Africa Union (AU), could
encourage buy-in for compute oversight harmonization. Working with
global and regional banks may also help unlock options to link
involvement with funding and investment.

The details of an international harmonization scheme require further
research, consultation, and analysis. Further work is needed on issues
of verification and enforcement between countries. Nevertheless, we
encourage policymakers to give greater consideration to the need for
cooperative compute provider governance in international discussions, as
a first step in working toward a mutually beneficial scheme.

\section{Key Challenges}\label{key-challenges}

While certain technical and institutional capabilities for conducting infrastructure governance already exist and could be employed, many areas still require further research to ensure robustness and scalability.
Furthermore, as compute becomes an increasingly important resource in the global economy, regulators will need to make sure that access to compute is equitable, competitive, and privacy- and confidentiality-protected. Regulation also needs to strike the right balance with public safety as AI capabilities develop, ideally with differentiated regulatory levers that enable specific and nuanced policies. 
Below, we present some key technical and governance challenges that should be explored in the near term. The general challenges and considerations of using compute as a policy lever are discussed in \citet{sastryComputingPowerGovernance2024} and are applicable here.

\subsection{Technical Challenges}\label{technical-challenges}

\textbf{Variation in workload signatures due to changes in algorithms
and hardware at the frontier.}
\\
Certain signatures (e.g., network bandwidth utilization, AI accelerator
core utilization) could be used to classify customer workloads. However,
these signatures depend on the model architecture and training algorithm
used to train frontier models, which will likely change over time
\citep{rabinovitsjOpeningAIInfrastructure}.\footnote{For example,
  techniques to reduce the memory footprint of training are an active
  area of research (e.g., \citep{dettmersLLMInt88bit2022}).}
Additionally, algorithmic progress within one type of model (e.g.,
utilizing AI accelerators more efficiently) would also change aspects of
this signature. In order to ensure that infrastructure governance
measures are effective, regulators and compute providers need to be kept
aware of how the state-of-the-art algorithms and hardware are changing;
classification techniques need to be updated over time to track these
advancements.

\textbf{Methods for workload classification that are robust to
adversarial gaming.}
\\
In addition to this natural variation in workload signatures,
adversarial actors may attempt to circumvent detection by intentionally
changing the computational pattern\footnote{A customer might attempt to
  hide the fact that they are engaging in AI training by using
  non-standard number representations, affecting traffic to/from outside
  networks, or deliberately using less efficient algorithms with
  different workload characteristics, such as under-utilizing memory or
  computation. However, the more a customer attempts to disguise
  workloads in this fashion, the greater the cost in terms of lost
  efficiency. In the context of frontier model training, where a single
  workload can cost tens of millions of dollars, perhaps these losses
  could be significant enough to make many forms of obfuscation too
  costly.} of their workload.\footnote{In the event of adversarial
  actors attempting to obscure their activities, it should be noted that
  such gaming of the system typically comes at a cost, potentially
  affecting the efficiency or performance of the AI system being
  trained. Compute providers and regulators will need to consider
  whether the penalties for non-compliance are substantial enough to
  deter such behavior, weighing if the cost of circumventing outweighs
  the cost of compliance.} Compute providers will have to develop
classification methods that are robust to deliberate obfuscation and
adopt red-teaming techniques to check against these attacks. These
changes could be much smaller and more unpredictable than the natural
variation of algorithms, so these gaming techniques need to be
anticipated prior to deploying the classification scheme.

\textbf{Preventing evasion by structuring training runs across multiple
data centers or compute providers.}
\\
As with money laundering techniques observed in the financial industry,
malicious customers may split their workloads (e.g., training runs)
across multiple data centers or compute providers to evade regulations
(a practice described as ``structuring'' in the finance industry
\citep{sanctionscannerWhatDifferenceSmurfing}). Regulators and compute
providers will have to develop and constantly update technical and
governance methods to prevent evasion. Over time, techniques for
avoiding detection will evolve, as will more sophisticated approaches to
detection. This should become an active and resourced area of technical
governance research.

Detecting training runs distributed across the data centers of a single
provider is relatively straightforward due to the
provider's consolidated insight into infrastructure and
relevant customer metrics. However, when workloads are distributed
across multiple providers, compute providers face significant challenges
in consolidating data for analysis. Yet, in practice, the number of
providers capable of delivering the requisite level of compute for
high-risk models is limited \citep{InfographicAmazonMaintains2024}.
This issue can be addressed by a technical framework that facilitates
the secure and privacy-preserving information exchange between compute
providers---being mindful of the commercial sensitivities and potential
antitrust implications associated with sharing details about training
runs and other client data.

Moreover, the strategic implementation of a compute threshold (Pistillo
et al., forthcoming) could serve as a deterrent against ``structured
training'' practices. Next-generation AI models generally necessitate an
order of magnitude more training compute than their predecessors to
achieve meaningful advancements in performance.\footnote{Scaling laws
  suggest that achieving substantial enhancements in performance on
  downstream tasks necessitates exponential increases in training
  compute. This principle underscores that compute investments grow
  exponentially for comparatively linear improvements in task
  performance \citep{owenHowPredictableLanguage2024}.} A compute
threshold set slightly above the current models, for example, the one
used in the AI Executive Order, reflects the reality that future, more
advanced models will not just require marginal incremental compute
increases but rather an exponential scale-up (usually an order of
magnitude). Practically, this leads to the necessity of involving more
than ten compute providers to distribute the workload sufficiently to
circumvent the threshold. This requirement imposes substantial
logistical and performance challenges. Simply identifying more than ten
compute providers with sufficient compute capacity over which to
structure a workload is a formidable challenge, and doing so without
detection is even more difficult. Furthermore, next-generation
models---given the exponential increase in training compute---increase
the challenges even further. Should the exponential scaling of training
compute persist, it would necessitate a proportionally exponential
increase in the number of compute providers involved, exacerbating the
challenges even further.

\textbf{Using privacy-preserving technologies for detailed workload
verification.}
\\
Emerging privacy and verification mechanisms such as trusted execution
environments and proof-of-training may allow multiple parties to
cooperate without the risk of exposing confidential or sensitive
information to each other. For example, a compute provider, data
provider, and software provider could cooperate in a mutually beneficial
training run without needing to share trade secrets or personally
identifiable information, producing a tamper-evident and
privacy-preserving training record that can be later scrutinized by a
verifier. More work is needed in this area to develop robust hardware
mechanisms and algorithmic techniques that can provably account for and
verify aspects of compute usage. Any methods of de-identifying personal
or confidential data for such exercises must be robust enough to meet
the standards imposed by key privacy laws, such as the EU GDPR, to avoid
personal data breaches or privacy conflicts.

\textbf{Creating a robust customer identification scheme.}
\\
Work must be done to robustly link cloud accounts to particular
individuals and entities. This is needed for key functions like
liability tracing, developer licensing, export control enforcement, and
defending against ``structured'' training runs where an adversarial
actor distributes its workload across multiple compute providers.
Furthermore, advances in AI may continue to increase the challenge of
verifying customers' identities. For example, websites are already
claiming to use neural networks to generate photos of highly plausible
fake IDs, \citep{cox2024} which could potentially be applied to
generate fake business documentation.\footnote{The complexity of
  verification processes is further increased by the diversity of
  national ID and business registration regulations when considered in
  an international context (e.g., India has more advanced ID
  verification systems than other countries
  \citep{indiasministryofelectronicsandinformationtechnology2023}).}
This issue is not unique to compute provider oversight and would need to
be addressed on a broader scale to investigate how AI could help
facilitate fraud, money laundering, and other criminal activity.

However, as discussed in \cref{limitations-and-future-research-directions},
customer identification within the context of frontier AI predominantly
involves a limited number of entities, primarily large-scale
corporations. Rather than necessitating widespread identification
efforts, our focus is on conducting thorough verifications for this
select group, ensuring that these in-depth analyses are both effective
and targeted.

\textbf{Challenges from technological developments.}
\\
The utility of compute providers as intermediary regulators is supported
by the need for significant computational resources for the training and
deployment of frontier AI models
\citep{sastryComputingPowerGovernance2024}. While current compute and
AI trends appear to reinforce the primacy of large-scale compute for
such models, a variety of technical developments may challenge this norm
\citep{heimCrucialConsiderationsCompute2024,
pilzIncreasedComputeEfficiency2024,
sastryComputingPowerGovernance2024}. It will be important to ensure
that compute governance policy is complemented by research into other
effective mechanisms for AI governance.

\subsection{Governance Challenges}\label{governance-challenges}

\textbf{Maintaining equitable access.}
\\
Increasing regulatory requirements on both compute providers and
customers can increase the barriers to entry, preventing smaller, less
well-resourced actors from entering the market. This risks intensifying
power concentration among hyperscalers. Key challenges include making
legislation legible, ensuring that secure implementation methods are
accessible, and striking a balance between safety and regulatory burdens
at various threat levels. For example, the technical tools needed for
compute verification should be accessible to potential providers.
Regulatory agencies, such as NIST, can play a role in determining
standards and facilitating open-source solutions.

\textbf{Strengthening regulatory structures for in-house compute.}
\\
Many AI developers currently operate their own data centers for training
and deploying AI systems. This provides additional challenges for
oversight. Since the AI developer and the compute provider are
essentially the same entity, they may be incentivized to collude,
weakening the \emph{verification} and \emph{enforcement} capabilities
provided by an independent compute provider. In such cases, prudential
regulations, whistleblower schemes, or even direct inspections by the
regulator may be necessary to ensure adequate supervision. Experience
may be drawn from the financial world, where regulators face a similar
challenge with corporations running their own banks
\citep{consumerfinancialprotectionbureauCFPBProposesNew2023}. More
work should be done on developing both regulatory and technical tools that enable robust and low-cost oversight in these situations.\footnote{E.g., zero-knowledge proofs that regulators can trust to verify compute
usage without having direct access to data centers or sensitive data.}

\textbf{Ensuring compliance with privacy commitments and related laws on
privacy and data transfer.}
\\
Although we discussed privacy-preserving methods to perform compute
accounting and workload classification, it is critical that any
implementation of an infrastructure governance scheme is in compliance
with existing privacy laws and commitments (e.g., that personal data is
not processed for purposes beyond that for which there exists a solid
legal basis). There may also be issues surrounding international
application for foreign privacy law. For example, if US providers are
required to collect information on customers in the EU, the US
government is likely to face similar challenges from EU institutions,
member states, civil society, and US companies who have long dealt with
EU-US data transfer legal uncertainties caused by US national security
data collection and Schrems legal challenges regarding such collection
for the past decade. In the long-term, there also needs to be a
conversation about how to balance privacy commitments with public
safety, differentiating between AI systems with varying risk levels.

\textbf{Preventing regulatory flight across borders.}
\\
We have outlined international harmonization as a key step to minimize
the risk of compute providers and AI developers and deployers relocating
their businesses or simply moving their workloads to jurisdictions with
lower regulatory standards. However, further work is needed to explore
the most effective mechanisms for supporting broad buy-in to a
coordinated regulatory scheme, as well as effective approaches for
incentivizing businesses to remain in higher-regulatory environments.
This work would be aided by analysis on the elasticity of the global
compute and AI markets, and the extent to which businesses will attempt
to avoid oversight measures.

\section{Conclusion}\label{conclusion}

As governments move to take action on addressing AI risks, compute
providers can play a key role in ensuring regulations are upheld and
enforced. Their concentrated role in the AI supply chain allows them to
use record keeping, verification and enforcement capabilities to help
secure sensitive AI-related IP and data, engage in greater oversight of
emerging AI risks, provide post-incident attribution and forensics,
prevent bad actors from training frontier AI models, and ensure AI
developers adhere to set standards. In this way, they have the ability
to act as \emph{securers}, \emph{record keepers}, \emph{verifiers}, and
\emph{enforcers}.

Implementing this regulatory model in a way that preserves privacy and
enables innovation will be key. Our analysis indicates that selected
mechanisms of these governance capabilities are technically feasible and
possible to implement in a privacy-preserving way. However, further work
is required to ensure such a governance model remains resilient to
evasion efforts, and to further refine particular technical mechanisms
and standards that preserve privacy and innovation while allowing
sufficient oversight. Greater oversight would also be required of the
compute providers themselves. This will require close collaboration
between compute providers, governments, and the research community.

When scaled internationally, this governance model has the potential to
support global AI governance architecture. International coordination
will be essential for addressing cross-border oversight and data issues,
as well as to reduce the risk of compute providers and AI developers
relocating to lower scrutiny jurisdictions.

Using compute providers as intermediary regulators will be most
effective at addressing risks arising from large-scale AI training and
deployment, rather than all AI-related risks. Moreover, the viability of
compute as a governance node may be challenged by developments in
hardware and software, which could reduce the need for large-scale data
center usage \citep{sastryComputingPowerGovernance2024,pilzIncreasedComputeEfficiency2024}. These
proposals should therefore be complemented by other regulatory measures
as required.

Compute providers can play a key role in a governance regime that
protects privacy and innovation, while ensuring sufficient oversight to
mitigate critical AI-related risks. We urge policymakers, regulators,
compute providers, and the research community to work together on the next steps.

\clearpage

\section*{Acknowledgments}\label{acknowledgments}

We would like to express our thanks to the people who have offered
feedback and input on the ideas in this paper, including Aris
Richardson, Boxi Wu, Christian Troncoso, Christoph Winter, Christopher
Hoff, Chris Phenicie, Hadrien Pouget, Haydn Belfield, Keegan McBride,
Konstantin Pilz, Maia Hamin, Markus Anderljung, Mauricio Baker, Owen
Larter, Pam Dixon, Peter Cihon, Trey Herr, and Zaheed Kara.

We thank Beth Eakman for copy editing and José Medina for formatting assistance.

OpenAI's ChatGPT and Google's Gemini Advanced were used for editing
assistance throughout this paper.

\appendix

\section{Overview of Compute Provider
Technologies}\label{a.-overview-of-compute-provider-technologies}

\begin{figure}[ht]
    \centering
    \centerline{\includegraphics[width=1.3\linewidth]{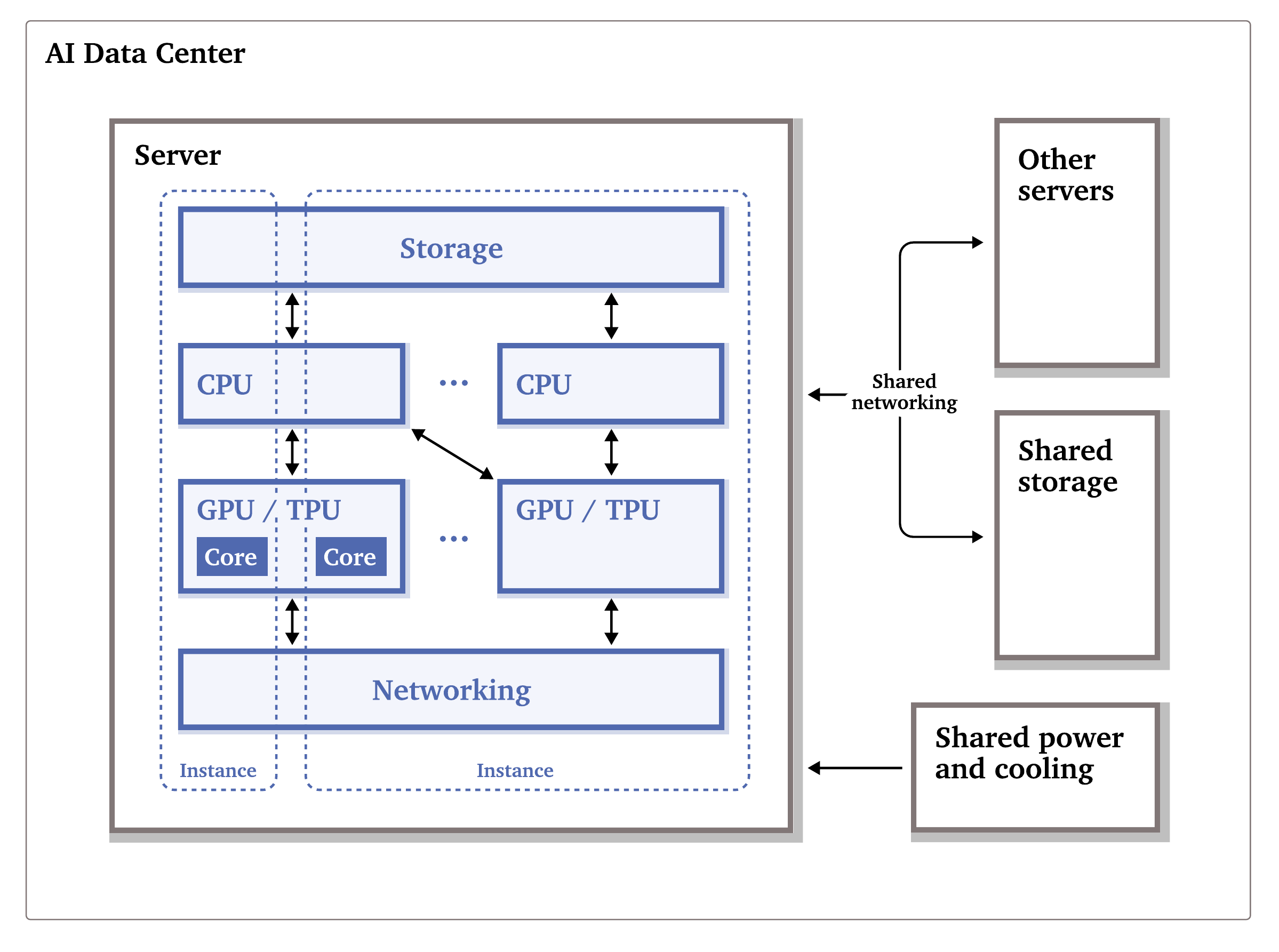}}
    \caption{A logical diagram of the software and hardware
components and interactions in an AI data center. A user provides their
AI code and data, interacting with a software stack that differs
depending on provider infrastructure but inherently includes the
hardware interactions depicted above. A server can be partitioned into
virtual instances, where each instance has a fraction of the physical
resources: CPU, GPU/TPU, networking, and storage.}
    \label{fig:logical-diagram-data-center}
\end{figure}

AI computing infrastructure is typically provided through ``data
centers'', buildings designed to power, house, and operate large amounts
of computing hardware. AI data centers contain many ``servers,''
computers optimized for AI computational workloads (we abstractly refer
to servers as ``nodes'' in \cref{technical-feasibility-of-compute-providers-governance-role}).\footnote{Typically a
  node and a server are equivalent, but ``server'' refers to the
  physical hardware, while a ``node'' is a unit of computational
  resources within the cluster's infrastructure.} Each server contains
a variable number of CPUs (general-purpose processors), AI accelerators
(specialized AI processors such as GPUs and TPUs), networking to allow
these devices to communicate, and shared data storage (\cref{fig:logical-diagram-data-center}). A relatively
large number of AI accelerators, and the capacity for those devices to
communicate at high speed, are the primary attributes differentiating AI
data centers from other kinds of data centers.

AI accelerators contain many ``cores,'' sub-processors that are
optimized for simple, parallel computation. AI accelerators reach high
throughput as measured by operations per unit time due to both the
number of cores and the capability of these cores to perform ``vector
processing'': simple math operations on an entire array of numbers in a
single step. Contemporary deep learning workloads are efficiently implemented in software using matrix
multiplication, and the vector processing hardware is an efficient
mechanism for computing these matrices. Different types of cores may
exist within a single AI accelerator, where each type is optimized for
handling different types of instructions or input data, such as graphics
and video games, or mixed-precision operations beyond simple parallel
operations such as matrix multiplication.

The primary goal of a compute provider is to map customer requests for
compute (in the form of workloads or ``machine instances'') to physical
hardware resources efficiently: achieving high utilization of the
(expensive) hardware while maintaining security and high performance. To
achieve this, providers typically provide ``virtualization.'' This means
dynamically creating virtual (rather than physical) versions of the
hardware within a server, each with their own operating system and
software, and isolated code and data. These virtual servers are often
referred to as ``instances,'' and are used to partition physical
hardware resources (e.g., cores, storage) to be efficiently shared
across separate computational workloads and different customers. For the purposes of the kinds of workloads considered in this paper (large-scale AI workloads run on dedicated single-tenant infrastructure), sharing resources across workloads rather than customers is more relevant.
Virtualization exists in part to subdivide a provider's physical
hardware resources, but also as a mechanism to isolate the provider from
the customer. In cases where the tenant is the sole tenant of a server,
two options exist: the provider can manage the server in the form of a
hypervisor layer to provide services such as additional security,
networking configuration, and software management through an API, or the
provider can provide a ``bare metal'' instance to the customer, in which
case the customer is responsible for all software. Even in the bare
metal instance, the provider maintains control over their network.

On the software side, customer code is often run in combination
with or on top of a standardized set of software packages such as
frameworks and APIs provided by the infrastructure provider, set up by
default on each instance. Code for a particular workload (e.g., training
a model) is orchestrated by the CPU, which schedules tasks to run on AI
accelerators. The main code executed on an AI accelerator is ``compute
kernels,'' small programs written to execute a specific computational
task on a core, typically involving operations on matrices. Compute
kernels are often provided by AI accelerator firms (e.g., NVIDIA's CUDA
library) or generated by a compiler (e.g., Triton), with support for
customers writing their own custom kernel. AI accelerator firms also
generally provide performance monitoring and debugging tools, which
measure various aspects of an AI accelerator's performance, such as
which cores are active, whether the cores are running at full capacity,
and the amount of data being transmitted to/from the AI accelerator's
memory. Typically these tools rely on hardware-based ``performance
counters,'' which track different metrics on the AI accelerator.

The infrastructure provider runs management and scheduling software,
which accepts customer requests and issues new instances to the
customer, and allocates available hardware resources to workloads based
on their computational requirements. In tandem, another software system
monitors and maintains the cluster, tracking machine health (power
consumption, temperature, network status, etc), and yet other software.
This management software allows infrastructure providers to track which
resources are being used by which customers, and bill customers for the
resources they are using.

\section{Observable Data
Attributes}\label{b.-observable-data-attributes}

\centering
\small
\setlength{\tabcolsep}{4.5pt}
\renewcommand{\arraystretch}{2}
\rowcolors{1}{white}{lightgray}
\begin{longtable}{C{0.14\linewidth}D{0.27\linewidth}D{0.17\linewidth}D{0.19\linewidth}D{0.115\linewidth}}
\hiderowcolors
\caption{An overview of observable data attributes.} \label{tab:long} \\
\showrowcolors
\toprule
\begin{tabular}[c]{@{}C{\linewidth}@{}}\bfseries \footnotesize Visible attribute\end{tabular} & %
\begin{tabular}[c]{@{}C{\linewidth}@{}}\bfseries\footnotesize Uses \end{tabular} & %
\begin{tabular}[c]{@{}C{\linewidth}@{}}\bfseries \footnotesize Involves collection of data not already widely collected? \end{tabular} & 
\begin{tabular}[c]{@{}C{\linewidth}@{}}\bfseries \footnotesize Ease of implementation \end{tabular}& 
\begin{tabular}[c]{@{}C{\linewidth}@{}}\bfseries \footnotesize Ease of circumvention\end{tabular}\\
\hline
\endfirsthead
\hiderowcolors
\multicolumn{5}{c}%
{{\textbf{\tablename\ \thetable{}} -- Continued from previous page.}} \\
\toprule
\begin{tabular}[c]{@{}C{\linewidth}@{}}\bfseries \footnotesize Visible attribute\end{tabular} & %
\begin{tabular}[c]{@{}C{\linewidth}@{}}\bfseries\footnotesize Uses \end{tabular} & %
\begin{tabular}[c]{@{}C{\linewidth}@{}}\bfseries \footnotesize Involves collection of data not already widely collected?\end{tabular} & 
\begin{tabular}[c]{@{}C{\linewidth}@{}}\bfseries \footnotesize Ease of implementation \end{tabular}& 
\begin{tabular}[c]{@{}C{\linewidth}@{}}\bfseries \footnotesize Ease of circumvention\end{tabular}\\
\hline
\endhead
\showrowcolors

\hline \multicolumn{5}{r}{{Continued on next page.}} \\ 
\endfoot

\hline \hline
\endlastfoot

Hardware configuration requested by the customer & \begin{tabular}[c]{@{}C{\linewidth}@{}}\emph{Workload classification.}\\[-2pt] The quantity of AI accelerators requested and networking setup are strongly suggestive of the workloads a customer intends to run.\end{tabular} & No, already collected. &  Already collected by compute providers to set up and provision infrastructure. &  Highly difficult or impossible. \\
\begin{tabular}[c]{@{}C{\linewidth}@{}} Number of hours that resources (e.g., AI accelerators) are in use \end{tabular} & \begin{tabular}[c]{@{}C{\linewidth}@{}}\emph{Workload classification, compute accounting.}\\[-2pt] Allows high-level boundary setting on workload type/size.\end{tabular} & No, already collected. & Already collected by compute providers for billing purposes. & Highly difficult or impossible. \\
Power draw & \begin{tabular}[c]{@{}C{\linewidth}@{}}\emph{Workload classification, compute accounting.} \\[-2pt] Allows high-level boundary setting on workload type/size, as increased power draw corresponds to increased throughput for a particular device. Power consumption over time may allow differentiation of inference from training \citep{patelPOLCAPowerOversubscription2023}.\end{tabular} & No, already collected. & Possible to collect using existing tooling. Already collected by some compute providers. & Possible, but would involve substantial cost efficiency penalties. \\
Network bandwidth between AI accelerator servers & \begin{tabular}[c]{@{}C{\linewidth}@{}}\emph{Workload classification, compute accounting.}\\[-2pt] Large AI training workloads require high bandwidth between servers. Different communication patterns correspond to different kinds of workloads, and bandwidth utilization is related to the quantity of computation performed on each server.\end{tabular} & No, already collected. & Possible to collect using existing tooling. Already collected by some compute providers. & Possible, but could involve substantial cost efficiency penalties. \\
Network bandwidth within AI accelerator servers & \begin{tabular}[c]{@{}C{\linewidth}@{}}\emph{Workload classification, compute accounting.} \\[-2pt] Different bandwidth patterns correspond to different kinds of workloads, and bandwidth utilization is related to the quantity of computation performed within each server.\end{tabular} & No, already collected. & Possible to collect using existing tooling. Already collected by some compute providers. & Possible, but could involve substantial cost efficiency penalties. \\
AI accelerator core \& memory bandwidth utilization & \begin{tabular}[c]{@{}C{\linewidth}@{}}\emph{Workload classification, compute accounting.} \\[-2pt] Large AI workloads typically have high memory bandwidth utilization, and core utilization will tend to be constant for training, while inference is typically variable.\end{tabular} & No, already collected. & Possible to collect using existing tooling. Difficult to collect for bare-metal services. & Possible, but could involve substantial cost efficiency penalties. \\
Performance counters by numerical precision & \begin{tabular}[c]{@{}C{\linewidth}@{}}\emph{Workload classification, compute accounting.}\\[-2pt] Lower precision is common in AI workloads and allows differentiation from most scientific computing workloads and possibly gaming. Counters also provide a direct measurement of operations consumed by a workload.\end{tabular} & Potentially. This degree of telemetry on an individual customer is unusual. Policies for collection and analysis would need to be clearly outlined in provider’s terms of service. & Possible to collect using existing tooling. Difficult to collect for bare metal services. & Possible, but could involve moderate cost efficiency penalties. \\
Modification of weights in memory & \begin{tabular}[c]{@{}C{\linewidth}@{}}\emph{Workload classification, compute accounting.}\\[-2pt] Model training requires changing the weights in memory using a backward pass. Typically the only large data structures in memory are the weights and activations, so it should be possible to observe whether stores are made to that region of memory. The magnitude and frequency of memory updates are related to the quantity of compute consumed.\end{tabular} & Potentially & Not currently possible. & Difficult: training requires modifying weights in memory to be highly performant. \\
Workload hyperparameters & \emph{Workload classification, compute accounting, detailed workload verification.} & Yes. Can potentially be made privacy-preserving using confidential computing techniques. & Possible to collect with customer consent. & Unclear (highly dependent on implementation). \\
Training dataset & \emph{Workload classification, compute accounting, detailed workload verification.} & Yes. Can potentially be made privacy-preserving using confidential computing techniques. & Possible to collect with customer consent. & Unclear (highly dependent on implementation). \\
\end{longtable}

\clearpage

\addcontentsline{toc}{section}{References}
\begin{CJK*}{UTF8}{gbsn}
\bibliography{references}

\begin{thebibliography}{172}
\providecommand{\natexlab}[1]{#1}
\providecommand{\url}[1]{\texttt{#1}}
\expandafter\ifx\csname urlstyle\endcsname\relax
  \providecommand{\doi}[1]{doi: #1}\else
  \providecommand{\doi}{doi: \begingroup \urlstyle{rm}\Url}\fi

\bibitem[Aarne et~al.(2024)Aarne, Fist, and Withers]{aarneSecureGovernableChips2024}
Aarne, O., Fist, T., and Withers, C.
\newblock Secure, {Governable} {Chips}.
\newblock Technical report, Center for a New American Security, January 2024.
\newblock URL \url{https://www.cnas.org/publications/reports/secure-governable-chips}.

\bibitem[Abbott et~al.(2017{\natexlab{a}})Abbott, Levi-Faur, and Snidal]{abbottRegulatoryIntermediariesAge2017}
Abbott, K.~W., Levi-Faur, D., and Snidal, D. (eds.).
\newblock \emph{Regulatory intermediaries in the age of governance}, volume 670 of \emph{The {Annals} of the {American} {Academy} of {Political} and {Social} {Science}}.
\newblock SAGE, March 2017{\natexlab{a}}.
\newblock ISBN 978-1-5063-9011-6.
\newblock URL \url{https://www.jstor.org/stable/i26361533}.

\bibitem[Abbott et~al.(2017{\natexlab{b}})Abbott, Levi-faur, and Snidal]{abbottTheorizingRegulatoryIntermediaries2017}
Abbott, K.~W., Levi-faur, D., and Snidal, D.
\newblock Theorizing {Regulatory} {Intermediaries}: {The} {RIT} {Model}.
\newblock \emph{The ANNALS of the American Academy of Political and Social Science}, 670\penalty0 (1):\penalty0 14--35, March 2017{\natexlab{b}}.
\newblock ISSN 0002-7162, 1552-3349.
\newblock \doi{10.1177/0002716216688272}.
\newblock URL \url{http://journals.sagepub.com/doi/10.1177/0002716216688272}.

\bibitem[Advani et~al.(2023)Advani, Elming, and Shaw]{advani2023}
Advani, A., Elming, W., and Shaw, J.
\newblock The {Dynamic} {Effects} of {Tax} {Audits}.
\newblock \emph{The Review of Economics and Statistics}, 105\penalty0 (3):\penalty0 545--561, May 2023.
\newblock ISSN 0034-6535.
\newblock \doi{10.1162/rest_a_01101}.
\newblock URL \url{https://doi.org/10.1162/rest_a_01101}.

\bibitem[Ahmed \& Wahed(2020)Ahmed and Wahed]{ahmedDedemocratizationAIDeep2020}
Ahmed, N. and Wahed, M.
\newblock The {De}-democratization of {AI}: {Deep} {Learning} and the {Compute} {Divide} in {Artificial} {Intelligence} {Research}, October 2020.
\newblock URL \url{http://arxiv.org/abs/2010.15581}.
\newblock arXiv:2010.15581 [cs].

\bibitem[Alderman(2021)]{aldermanIrelandDaysTax2021}
Alderman, L.
\newblock Ireland’s {Days} as a {Tax} {Haven} {May} {Be} {Ending}, but {Not} {Without} a {Fight}.
\newblock \emph{The New York Times}, July 2021.
\newblock ISSN 0362-4331.
\newblock URL \url{https://www.nytimes.com/2021/07/08/business/ireland-minimum-corporate-tax.html}.

\bibitem[{Alibaba}(2024)]{AlibabaClusterTrace2024}
{Alibaba}.
\newblock Alibaba {Cluster} {Trace} {Program}, March 2024.
\newblock URL \url{https://github.com/alibaba/clusterdata}.
\newblock original-date: 2017-09-05T03:16:34Z.

\bibitem[Allen(2022)]{allen2022}
Allen, G.~C.
\newblock Choking off {China}’s {Access} to the {Future} of {AI}.
\newblock Technical report, Center for Strategic and International Studies, October 2022.
\newblock URL \url{https://www.csis.org/analysis/choking-chinas-access-future-ai}.

\bibitem[Allen et~al.(2023)Allen, Benson, and Putnam]{allenJapanNetherlandsAnnounce2023}
Allen, G.~C., Benson, E., and Putnam, M.
\newblock Japan and the {Netherlands} {Announce} {Plans} for {New} {Export} {Controls} on {Semiconductor} {Equipment}.
\newblock Commentary, Center for Strategic and International Studies, April 2023.
\newblock URL \url{https://www.csis.org/analysis/japan-and-netherlands-announce-plans-new-export-controls-semiconductor-equipment}.

\bibitem[{Amazon.com, Inc.}(2023{\natexlab{a}})]{amazon.cominc.2023}
{Amazon.com, Inc.}
\newblock An update on {Amazon}’s efforts to combat child sexual abuse material, April 2023{\natexlab{a}}.
\newblock URL \url{https://www.aboutamazon.com/news/policy-news-views/amazon-csam-transparency-report-2022}.
\newblock Section: Policy news \& views.

\bibitem[{Amazon.com, Inc.}(2023{\natexlab{b}})]{amazon.cominc.AmazonAnthropicAnnounce2023}
{Amazon.com, Inc.}
\newblock Amazon and {Anthropic} {Announce} {Strategic} {Collaboration} to {Advance} {Generative} {AI}, September 2023{\natexlab{b}}.
\newblock URL \url{https://press.aboutamazon.com/2023/9/amazon-and-anthropic-announce-strategic-collaboration-to-advance-generative-ai}.

\bibitem[{Amazon.com, Inc.}(2024)]{amazon.cominc.Cloud}
{Amazon.com, Inc.}
\newblock Sustainability {Amazon}, 2024.
\newblock URL \url{https://sustainability.aboutamazon.com/products-services/the-cloud}.

\bibitem[{AMD}(2024)]{amd}
{AMD}.
\newblock {AMD} {Regulatory} {Trade} {Compliance}, 2024.
\newblock URL \url{https://www.amd.com/en/legal/compliance/trade-compliance.html}.

\bibitem[Anderljung et~al.(2023{\natexlab{a}})Anderljung, Barnhart, Korinek, Leung, O'Keefe, Whittlestone, Avin, Brundage, Bullock, Cass-Beggs, Chang, Collins, Fist, Hadfield, Hayes, Ho, Hooker, Horvitz, Kolt, Schuett, Shavit, Siddarth, Trager, and Wolf]{anderljungFrontierAIRegulation2023}
Anderljung, M., Barnhart, J., Korinek, A., Leung, J., O'Keefe, C., Whittlestone, J., Avin, S., Brundage, M., Bullock, J., Cass-Beggs, D., Chang, B., Collins, T., Fist, T., Hadfield, G., Hayes, A., Ho, L., Hooker, S., Horvitz, E., Kolt, N., Schuett, J., Shavit, Y., Siddarth, D., Trager, R., and Wolf, K.
\newblock Frontier {AI} {Regulation}: {Managing} {Emerging} {Risks} to {Public} {Safety}, November 2023{\natexlab{a}}.
\newblock URL \url{http://arxiv.org/abs/2307.03718}.
\newblock arXiv:2307.03718 [cs].

\bibitem[Anderljung et~al.(2023{\natexlab{b}})Anderljung, Smith, O'Brien, Soder, Bucknall, Bluemke, Schuett, Trager, Strahm, and Chowdhury]{anderljung2023}
Anderljung, M., Smith, E.~T., O'Brien, J., Soder, L., Bucknall, B., Bluemke, E., Schuett, J., Trager, R., Strahm, L., and Chowdhury, R.
\newblock Towards {Publicly} {Accountable} {Frontier} {LLMs}: {Building} an {External} {Scrutiny} {Ecosystem} under the {ASPIRE} {Framework}.
\newblock Technical report, Centre for the Governance of AI, November 2023{\natexlab{b}}.
\newblock URL \url{https://www.governance.ai/research-paper/towards-publicly-accountable-frontier-llms}.
\newblock arXiv:2311.14711 [cs].

\bibitem[{Anthropic}(2023)]{anthropicAnthropicPartnersGoogle2023}
{Anthropic}.
\newblock Anthropic {Partners} with {Google} {Cloud}, February 2023.
\newblock URL \url{https://www.anthropic.com/news/anthropic-partners-with-google-cloud}.

\bibitem[{Australian Attorney-General’s Department}(2015)]{australianattorney-generalsdepartmentDataRetentionGuideline2015}
{Australian Attorney-General’s Department}.
\newblock Data {Retention}: {Guideline} for {Service} {Providers}.
\newblock Technical report, Australian Government, July 2015.
\newblock URL \url{https://www.homeaffairs.gov.au/nat-security/files/data-retention-guidelines-service-providers.pdf}.

\bibitem[{Australian DITRDCA}(2024)]{transportdepartmentofinfrastructure2024}
{Australian DITRDCA}.
\newblock International {Air} {Services} {Information} {Memorandum}, January 2024.
\newblock URL \url{https://www.infrastructure.gov.au/infrastructure-transport-vehicles/aviation/international-aviation/air-services-agreements-arrangements/international-air-services-information-memorandum}.
\newblock Last Modified: 2024-01-18 Publisher: Department of Infrastructure, Transport, Regional Development, Communications and the Arts.

\bibitem[{Aviation Transport Security Act}(2005)]{AviationTransportSecurity2005}
{Aviation Transport Security Act}.
\newblock Aviation {Transport} {Security} {Regulations}, 2005.
\newblock URL \url{https://www8.austlii.edu.au/cgi-bin/viewdb/au/legis/cth/consol_reg/atsr2005457/}.

\bibitem[{AWS}(2024{\natexlab{a}})]{aws}
{AWS}.
\newblock General {Data} {Protection} {Regulation} ({GDPR}) {Center}, 2024{\natexlab{a}}.
\newblock URL \url{https://aws.amazon.com/compliance/gdpr-center/}.

\bibitem[{AWS}(2024{\natexlab{b}})]{awsAWSCustomerAgreement2024}
{AWS}.
\newblock {AWS} {Customer} {Agreement}, 2024{\natexlab{b}}.
\newblock URL \url{https://aws.amazon.com/agreement/#:~:text=You%20consent%20to%20the%20storage,order%20of%20a%20governmental%20body}.

\bibitem[{AWS}(2024{\natexlab{c}})]{awsAWSGlobalInfrastructure}
{AWS}.
\newblock {AWS} {Global} {Infrastructure}, 2024{\natexlab{c}}.
\newblock URL \url{https://aws.amazon.com/about-aws/global-infrastructure/}.

\bibitem[{AWS}(2024{\natexlab{d}})]{awsAWSNitroEnclaves}
{AWS}.
\newblock {AWS} {Nitro} {Enclaves}, 2024{\natexlab{d}}.
\newblock URL \url{https://aws.amazon.com/ec2/nitro/nitro-enclaves/}.

\bibitem[{AWS}(2024{\natexlab{e}})]{awsAmazonEC2OnDemand}
{AWS}.
\newblock Amazon {EC2} {On}-{Demand} {Pricing}, 2024{\natexlab{e}}.
\newblock URL \url{https://aws.amazon.com/ec2/pricing/on-demand/}.

\bibitem[{AWS}(2024{\natexlab{f}})]{awsAmazonEC2Reserved}
{AWS}.
\newblock Amazon {EC2} {Reserved} {Instances}, 2024{\natexlab{f}}.
\newblock URL \url{https://aws.amazon.com/ec2/pricing/reserved-instances/}.

\bibitem[{AWS}(2024{\natexlab{g}})]{awsPrivacyNotice2024}
{AWS}.
\newblock {AWS} {Privacy} {Notice}, 2024{\natexlab{g}}.
\newblock URL \url{https://aws.amazon.com/privacy/}.

\bibitem[{AWS}(2024{\natexlab{h}})]{awsSecurityAWSInfrastructure}
{AWS}.
\newblock Security of the {AWS} {Infrastructure}, 2024{\natexlab{h}}.
\newblock URL \url{https://docs.aws.amazon.com/whitepapers/latest/introduction-aws-security/security-of-the-aws-infrastructure.html}.

\bibitem[Banjongkan et~al.(2018)Banjongkan, Pongsena, Chanklan, Kerdprasop, and Kerdprasop]{banjongkanMultilabelClassificationHigh2018}
Banjongkan, A., Pongsena, W., Chanklan, R., Kerdprasop, N., and Kerdprasop, K.
\newblock Multi-label classification of high performance computing workload with variable transformation.
\newblock \emph{International Journal of Machine Learning and Computing}, 8:\penalty0 536--541, December 2018.
\newblock \doi{10.18178/ijmlc.2018.8.6.742}.
\newblock URL \url{https://www.ijmlc.org/vol8/742-ML0023.pdf}.

\bibitem[{BBC}(2020)]{EUUSPrivacyShield2020}
{BBC}.
\newblock {EU}-{US} {Privacy} {Shield} for data struck down by court.
\newblock \emph{BBC News}, July 2020.
\newblock URL \url{https://www.bbc.com/news/technology-53418898}.

\bibitem[Belfield \& Hua(2022)Belfield and Hua]{belfield2022}
Belfield, H. and Hua, S.-S.
\newblock Compute and {Antitrust}: {Regulatory} implications of the {AI} hardware supply chain, from chip design to cloud {APIs}.
\newblock \emph{Verfassungsblog}, August 2022.
\newblock URL \url{https://verfassungsblog.de/compute-and-antitrust/}.

\bibitem[Besiroglu et~al.(2024)Besiroglu, Bergerson, Michael, Heim, Luo, and Thompson]{besiroglu2024}
Besiroglu, T., Bergerson, S.~A., Michael, A., Heim, L., Luo, X., and Thompson, N.
\newblock The {Compute} {Divide} in {Machine} {Learning}: {A} {Threat} to {Academic} {Contribution} and {Scrutiny}?, January 2024.
\newblock URL \url{http://arxiv.org/abs/2401.02452}.
\newblock arXiv:2401.02452 [cs].

\bibitem[Biden(2008)]{biden1738110thCongress2008}
Biden, J.~R.
\newblock S.1738 - 110th {Congress} (2007-2008): {PROTECT} {Our} {Children} {Act} of 2008, October 2008.
\newblock URL \url{https://www.congress.gov/bill/110th-congress/senate-bill/1738}.
\newblock Archive Location: 2007-06-28.

\bibitem[Boyle \& Lau(2021)Boyle and Lau]{boylePresidentExtraordinarySanctions2021}
Boyle, A. and Lau, T.
\newblock The {President}’s {Extraordinary} {Sanctions} {Powers}, July 2021.
\newblock URL \url{https://www.brennancenter.org/our-work/research-reports/presidents-extraordinary-sanctions-powers}.

\bibitem[Buttarelli(2018)]{buttarelli2018}
Buttarelli, G.
\newblock The {EU}-{U}.{S}. {Privacy} {Shield} two years on, March 2018.
\newblock URL \url{https://www.edps.europa.eu/press-publications/press-news/blog/eu-us-privacy-shield-two-years}.

\bibitem[{California Energy Commission}(2024)]{californiaenergycommission}
{California Energy Commission}.
\newblock Power {Plant} {Licensing}, 2024.
\newblock URL \url{https://www.energy.ca.gov/programs-and-topics/topics/power-plants/power-plant-licensing}.
\newblock Publisher: California Energy Commission.

\bibitem[{Center for Open Science}(2019)]{centerforopenscience2019}
{Center for Open Science}.
\newblock {BrainsCAN} {Computational} {Core} {Neuroimaging} {Wiki}, March 2019.
\newblock URL \url{https://osf.io/k89fh/}.
\newblock Publisher: OSF.

\bibitem[{China Law Translate}(2017)]{translateZhongHuaRenMinGongHeGuoGuoJiaQingBaoFa2018XiuZheng2017}
{China Law Translate}.
\newblock 中华人民共和国国家情报法 (2018修正), June 2017.
\newblock URL \url{https://www.chinalawtranslate.com/national-intelligence-law-of-the-p-r-c-2017/}.

\bibitem[Choi et~al.(2023)Choi, Shavit, and Duvenaud]{choiToolsVerifyingNeural2023}
Choi, D., Shavit, Y., and Duvenaud, D.
\newblock Tools for {Verifying} {Neural} {Models}' {Training} {Data}, July 2023.
\newblock URL \url{http://arxiv.org/abs/2307.00682}.
\newblock arXiv:2307.00682 [cs].

\bibitem[{Code of Federal Regulations}(2024)]{codeoffederalregulations22CFRPart}
{Code of Federal Regulations}.
\newblock 22 {CFR} {Part} 120 - {Purpose} and {Definitions}, 2024.
\newblock URL \url{https://www.ecfr.gov/current/title-22/part-120}.

\bibitem[{Confidential Computing Consortium}(2022)]{confidentialcomputingconsortium2022}
{Confidential Computing Consortium}.
\newblock Confidential {Computing}: {Hardware}-{Based} {Trusted} {Execution} for {Applications} and {Data}.
\newblock Technical report, Confidential Computing Consortium, November 2022.
\newblock URL \url{https://confidentialcomputing.io/wp-content/uploads/sites/10/2023/03/CCC_outreach_whitepaper_updated_November_2022.pdf}.

\bibitem[{Congressional Research Service}(2023)]{congressionalresearchservice2023}
{Congressional Research Service}.
\newblock Who {Regulates} {Whom}? {An} {Overview} of the {U}.{S}. {Financial} {Regulatory} {Framework}.
\newblock {CRS} {Report} R44918, Congressional Research Service, October 2023.
\newblock URL \url{https://sgp.fas.org/crs/misc/R44918.pdf}.

\bibitem[{Congressional Research Service}(2024)]{InternationalEmergencyEconomic2024}
{Congressional Research Service}.
\newblock The {International} {Emergency} {Economic} {Powers} {Act}: {Origins}, {Evolution}, and {Use}.
\newblock {CRS} {Report} R45618, Congressional Research Service, January 2024.
\newblock URL \url{https://sgp.fas.org/crs/natsec/R45618.pdf}.

\bibitem[{Consumer Financial Protection Bureau}(2023)]{consumerfinancialprotectionbureauCFPBProposesNew2023}
{Consumer Financial Protection Bureau}.
\newblock {CFPB} {Proposes} {New} {Federal} {Oversight} of {Big} {Tech} {Companies} and {Other} {Providers} of {Digital} {Wallets} and {Payment} {Apps}, November 2023.
\newblock URL \url{https://www.consumerfinance.gov/about-us/newsroom/cfpb-proposes-new-federal-oversight-of-big-tech-companies-and-other-providers-of-digital-wallets-and-payment-apps/}.

\bibitem[Copos \& Peisert(2020)Copos and Peisert]{coposCatchMeIf2020}
Copos, B. and Peisert, S.
\newblock Catch {Me} {If} {You} {Can}: {Using} {Power} {Analysis} to {Identify} {HPC} {Activity}, May 2020.
\newblock URL \url{http://arxiv.org/abs/2005.03135}.
\newblock arXiv:2005.03135 [cs].

\bibitem[{CoreWeave}(2022)]{coreweavePrivacyPolicy2022}
{CoreWeave}.
\newblock Privacy {Policy}, October 2022.
\newblock URL \url{https://docs.coreweave.com/policies/terms-of-service/privacy-policy}.

\bibitem[{CoreWeave}(2023)]{coreweaveSecurityCompliance}
{CoreWeave}.
\newblock Security \& {Compliance}, 2023.
\newblock URL \url{https://docs.coreweave.com/policies/terms-of-service/security-and-compliance}.

\bibitem[Cottier(2023)]{epoch2023trendsinthedollartrainingcostofmachinelearningsystems}
Cottier, B.
\newblock Trends in the {Dollar} {Training} {Cost} of {Machine} {Learning} {Systems}, January 2023.
\newblock URL \url{https://epochai.org/blog/trends-in-the-dollar-training-cost-of-machine-learning-systems}.

\bibitem[{Council of the European Union}(2024)]{council_of_the_european_union_proposal_2024}
{Council of the European Union}.
\newblock Proposal for a {Regulation} of the {European} {Parliament} and of the {Council} laying down harmonised rules on artificial intelligence ({Artificial} {Intelligence} {Act}) and amending certain {Union} legislative acts, January 2024.
\newblock URL \url{https://data.consilium.europa.eu/doc/document/ST-5662-2024-INIT/en/pdf}.

\bibitem[{Country Legal Frameworks Resource}(2023)]{countrylegalframeworksresourceFrance2023}
{Country Legal Frameworks Resource}.
\newblock Provision of {Real}-time {Lawful} {Interception} {Assistance}, 2023.
\newblock URL \url{https://clfr.globalnetworkinitiative.org/country/france/}.

\bibitem[Cox(2024)]{cox2024}
Cox, J.
\newblock Inside the {Underground} {Site} {Where} ‘{Neural} {Networks}’ {Churn} {Out} {Fake} {IDs}, February 2024.
\newblock URL \url{https://www.404media.co/inside-the-underground-site-where-ai-neural-networks-churns-out-fake-ids-onlyfake/}.

\bibitem[Dettmers et~al.(2022)Dettmers, Lewis, Belkada, and Zettlemoyer]{dettmersLLMInt88bit2022}
Dettmers, T., Lewis, M., Belkada, Y., and Zettlemoyer, L.
\newblock {LLM}.int8(): 8-bit {Matrix} {Multiplication} for {Transformers} at {Scale}, November 2022.
\newblock URL \url{http://arxiv.org/abs/2208.07339}.
\newblock arXiv:2208.07339 [cs].

\bibitem[{ECFR}(2024)]{AcceptanceScreeningIndividuals2024}
{ECFR}.
\newblock Acceptance and screening of individuals and accessible property., 2024.
\newblock URL \url{https://www.ecfr.gov/current/title-49/part-1544/section-1544.201}.

\bibitem[Egan \& Heim(2023)Egan and Heim]{lennartOversightFrontierAI2023}
Egan, J. and Heim, L.
\newblock Oversight for {Frontier} {AI} through a {Know}-{Your}-{Customer} {Scheme} for {Compute} {Providers}.
\newblock Technical report, Centre for the Governance of AI, October 2023.
\newblock URL \url{https://www.governance.ai/research-paper/oversight-for-frontier-ai-through-kyc-scheme-for-compute-providers}.

\bibitem[{EU Agency for Cybersecurity}(2000)]{europeanunionagencyforcybersecuritySafeHarborPrivacy}
{EU Agency for Cybersecurity}.
\newblock Safe {Harbor} {Privacy} {Principles}, July 2000.
\newblock URL \url{https://www.enisa.europa.eu/topics/risk-management/current-risk/laws-regulation/data-protection-privacy/safe-harbor-privacy-principles}.

\bibitem[{European Commission}(2016)]{europeancommissionEUUPrivacyShield2016}
{European Commission}.
\newblock {EU}-{U}.{S}. {Privacy} {Shield}: {Frequently} {Asked} {Questions}, July 2016.
\newblock URL \url{https://ec.europa.eu/commission/presscorner/detail/hr/MEMO_16_2462}.

\bibitem[{European Commission}(2023{\natexlab{a}})]{europeancomission2023}
{European Commission}.
\newblock Commission welcomes {G7} leaders' agreement on {Guiding} {Principles} and a {Code} of {Conduct} on {Artificial} {Intelligence}, October 2023{\natexlab{a}}.
\newblock URL \url{https://ec.europa.eu/commission/presscorner/detail/en/ip_23_5379}.

\bibitem[{European Commission}(2023{\natexlab{b}})]{europeancomission2023a}
{European Commission}.
\newblock Shaping {Europe}’s digital future: {Hiroshima} {Process} {International} {Code} of {Conduct} for {Advanced} {AI} {Systems}, October 2023{\natexlab{b}}.
\newblock URL \url{https://digital-strategy.ec.europa.eu/en/library/hiroshima-process-international-code-conduct-advanced-ai-systems}.

\bibitem[{European Union}(2011)]{europeanunion2011}
{European Union}.
\newblock Directive 2011/93/{EU} of the {European} {Parliament} and of the {Council} of 13 {December} 2011 on combating the sexual abuse and sexual exploitation of children and child pornography, and replacing {Council} {Framework} {Decision} 2004/68/{JHA}, July 2011.
\newblock URL \url{https://eur-lex.europa.eu/eli/dir/2011/93/oj}.

\bibitem[{European Union}(2016)]{europeanunion2016}
{European Union}.
\newblock Regulation ({EU}) 2016/679 of the {European} {Parliament} and of the {Council} of 27 {April} 2016 on the protection of natural persons with regard to the processing of personal data and on the free movement of such data, and repealing {Directive} 95/46/{EC} ({General} {Data} {Protection} {Regulation}), May 2016.
\newblock URL \url{http://data.europa.eu/eli/reg/2016/679/oj}.
\newblock 119.

\bibitem[{European Union}(2021)]{europeanunion2021}
{European Union}.
\newblock Regulation ({EU}) 2021/784 of the {European} {Parliament} and of the {Council} of 29 {April} 2021 on addressing the dissemination of terrorist content online.
\newblock \emph{Official Journal of the European Union L 172/79}, pp.\  79--109, May 2021.
\newblock URL \url{https://eur-lex.europa.eu/eli/reg/2021/784/oj}.

\bibitem[Farrell \& Newman(2023)Farrell and Newman]{farrellUndergroundEmpireHow2023}
Farrell, H. and Newman, A.
\newblock \emph{Underground {Empire}: {How} {America} {Weaponized} the {World} {Economy}}.
\newblock Henry Holt and Co., September 2023.
\newblock ISBN 978-1-250-84055-4.
\newblock URL \url{https://us.macmillan.com/books/9781250840554/undergroundempire}.

\bibitem[{Federal Register}(2024)]{federalregister2024}
{Federal Register}.
\newblock Taking {Additional} {Steps} {To} {Address} the {National} {Emergency} {With} {Respect} to {Significant} {Malicious} {Cyber}-{Enabled} {Activities}: {A} {Proposed} {Rule} by the {Commerce} {Department}, January 2024.
\newblock URL \url{https://www.federalregister.gov/documents/2024/01/29/2024-01580/taking-additional-steps-to-address-the-national-emergency-with-respect-to-significant-malicious}.

\bibitem[{Federal Trade Commission}(2024)]{federaltradecommissionFTCLaunchesInquiry2024}
{Federal Trade Commission}.
\newblock {FTC} {Launches} {Inquiry} into {Generative} {AI} {Investments} and {Partnerships}, January 2024.
\newblock URL \url{https://www.ftc.gov/news-events/news/press-releases/2024/01/ftc-launches-inquiry-generative-ai-investments-partnerships}.

\bibitem[{FedRAMP}(2024)]{Baselines}
{FedRAMP}.
\newblock Understanding {Baselines} and {Impact} {Levels} for {FedRAMP} {Authorization}, 2024.
\newblock URL \url{https://www.fedramp.gov/baselines/}.

\bibitem[{Financial Action Task Force}(2003)]{financialactiontaskforceFATF40Recommendations2003}
{Financial Action Task Force}.
\newblock {FATF} 40 {Recommendations} {October} 2003 (incorporating all subsequent amendments until {October} 2004).
\newblock Technical report, FATF, October 2003.
\newblock URL \url{https://www.fatf-gafi.org/content/dam/fatf-gafi/recommendations/FATF%20Standards%20-%2040%20Recommendations%20rc.pdf}.

\bibitem[{Financial Action Task Force}(2023)]{financialactiontaskforceInternationalStandardsCombating2023}
{Financial Action Task Force}.
\newblock International {Standards} on {Combating} {Money} {Laundering} and the {Financing} of {Terrorism} and {Proliferation}.
\newblock Technical report, FATF, Paris, France, November 2023.
\newblock URL \url{https://www.fatf-gafi.org/content/dam/fatf-gafi/recommendations/FATF%20Recommendations%202012.pdf.coredownload.inline.pdf}.

\bibitem[{Financial Action Task Force}(2024{\natexlab{a}})]{financialactiontaskforceHighriskOtherMonitored}
{Financial Action Task Force}.
\newblock High-risk and other monitored jurisdictions, 2024{\natexlab{a}}.
\newblock URL \url{https://www.fatf-gafi.org/en/topics/high-risk-and-other-monitored-jurisdictions.html}.

\bibitem[{Financial Action Task Force}(2024{\natexlab{b}})]{financialactiontaskforceHome}
{Financial Action Task Force}.
\newblock Financial {Action} {Task} {Force} {Home}, 2024{\natexlab{b}}.
\newblock URL \url{https://www.fatf-gafi.org/en/home.html}.

\bibitem[{Financial Crimes Enforcement Network}(2005)]{financialcrimesenforcementnetwork}
{Financial Crimes Enforcement Network}.
\newblock {FACT} {SHEET} for {Section} 312 of the {USA} {PATRIOT} {Act} {Final} {Regulation} and {Notice} of {Proposed} {Rulemaking}.
\newblock Technical report, U.S. Treasury, December 2005.
\newblock URL \url{https://www.fincen.gov/fact-sheet-section-312-usa-patriot-act-final-regulation-and-notice-proposed-rulemaking}.

\bibitem[{Financial Crimes Enforcement Network}(2024)]{financialcrimesenforcementnetworka}
{Financial Crimes Enforcement Network}.
\newblock Financial {Crimes} {Enforcement} {Network} - {Mission}, 2024.
\newblock URL \url{https://www.fincen.gov/about/mission}.

\bibitem[Fist \& Grunewald(2023)Fist and Grunewald]{fist2023}
Fist, T. and Grunewald, E.
\newblock Preventing {AI} {Chip} {Smuggling} to {China}.
\newblock Working {Paper}, Center for a New American Security, October 2023.
\newblock URL \url{https://www.cnas.org/publications/reports/preventing-ai-chip-smuggling-to-china}.

\bibitem[Fist et~al.(2023)Fist, Heim, and Schneider]{fistChineseFirmsAre2024}
Fist, T., Heim, L., and Schneider, J.
\newblock Chinese {Firms} {Are} {Evading} {Chip} {Controls}, June 2023.
\newblock URL \url{https://foreignpolicy.com/2023/06/21/china-united-states-semiconductor-chips-sanctions-evasion/}.

\bibitem[{FluidStack}(2022)]{fluidstackFluidStackPrivacyNotice2022}
{FluidStack}.
\newblock {FluidStack} {Privacy} {Notice}, February 2022.
\newblock URL \url{https://uploads-ssl.webflow.com/64e49a6c77bc12449a05e6a2/6502008478387e54d7816888_Privacy%20Policy.pdf}.

\bibitem[{Future of Life Institute}(2024)]{futureoflifeinstituteEUArtificialIntelligence}
{Future of Life Institute}.
\newblock {EU} {Artificial} {Intelligence} {Act}: {The} {Act} {Texts}, 2024.
\newblock URL \url{https://artificialintelligenceact.eu/the-act/}.

\bibitem[Gatlan(2023)]{gatlan2023}
Gatlan, S.
\newblock Microsoft breach led to theft of 60,000 {US} {State} {Dept} emails, September 2023.
\newblock URL \url{https://www.bleepingcomputer.com/news/security/microsoft-breach-led-to-theft-of-60-000-us-state-dept-emails/}.

\bibitem[{Google}(2024)]{GoogleClusterdata2024}
{Google}.
\newblock google/cluster-data, March 2024.
\newblock URL \url{https://github.com/google/cluster-data}.
\newblock original-date: 2015-07-29T17:52:23Z.

\bibitem[{Google Cloud}(2021)]{googlecloud2021}
{Google Cloud}.
\newblock Google {Cloud} \& the {General} {Data} {Protection} {Regulation} ({GDPR}), 2021.
\newblock URL \url{https://cloud.google.com/privacy/gdpr}.

\bibitem[{Google Cloud}(2023)]{googlecloudCloudDataProcessing}
{Google Cloud}.
\newblock Cloud {Data} {Processing} {Addendum}, November 2023.
\newblock URL \url{https://cloud.google.com/terms/data-processing-addendum}.

\bibitem[{Google Cloud}(2024{\natexlab{a}})]{googlecloud}
{Google Cloud}.
\newblock All networking pricing, 2024{\natexlab{a}}.
\newblock URL \url{https://cloud.google.com/vpc/network-pricing}.

\bibitem[{Google Cloud}(2024{\natexlab{b}})]{googlecloudClimateSustainability}
{Google Cloud}.
\newblock Climate {Sustainability}, 2024{\natexlab{b}}.
\newblock URL \url{https://cloud.google.com/gov/sustainability}.

\bibitem[{Google Cloud}(2024{\natexlab{c}})]{googlecloudGoogleCloudPrivacy2024}
{Google Cloud}.
\newblock Google {Cloud} {Privacy} {Notice}, 2024{\natexlab{c}}.
\newblock URL \url{https://cloud.google.com/terms/cloud-privacy-notice}.

\bibitem[{Government of India}(2023)]{GovtConsideringProposal2023}
{Government of India}.
\newblock Govt considering proposal to set up {25K} {GPUs}, October 2023.
\newblock URL \url{https://indbiz.gov.in/govt-considering-proposal-to-set-up-25k-gpus/}.

\bibitem[Hacker(2023)]{hacker2023}
Hacker, P.
\newblock The {European} {AI} liability directives – {Critique} of a half-hearted approach and lessons for the future.
\newblock \emph{Computer Law \& Security Review}, 51:\penalty0 105871, November 2023.
\newblock ISSN 0267-3649.
\newblock \doi{10.1016/j.clsr.2023.105871}.
\newblock URL \url{https://www.sciencedirect.com/science/article/pii/S026736492300081X}.

\bibitem[Hay \& Shleifer(1998)Hay and Shleifer]{hayPrivateEnforcementPublic1998}
Hay, J.~R. and Shleifer, A.
\newblock Private enforcement of public laws: a theory of legal reform.
\newblock \emph{American Economic Review Papers and Proceedings}, 88\penalty0 (2):\penalty0 398--403, 1998.
\newblock URL \url{https://www.jstor.org/stable/116955}.

\bibitem[Heim(2024)]{heimCrucialConsiderationsCompute2024}
Heim, L.
\newblock Crucial {Considerations} for {Compute} {Governance}, February 2024.
\newblock URL \url{https://blog.heim.xyz/crucial-considerations-for-compute-governance/}.

\bibitem[Heim \& Egan(2023)Heim and Egan]{heimAccessingControlledAI2023}
Heim, L. and Egan, J.
\newblock Accessing {Controlled} {AI} {Chips} via {Infrastructure}-as-a-{Service} ({IaaS}): {Implications} for {Export} {Controls}: {Comment} on {BIS}–2022–0025 ({RIN} 0694–{AI94}) — {Question} 1.
\newblock Technical report, Centre for the Governance of AI, December 2023.
\newblock URL \url{https://cdn.governance.ai/Accessing_Controlled_AI_Chips_via_Infrastructure-as-a-Service.pdf}.

\bibitem[{IBM}(2024)]{ibm}
{IBM}.
\newblock What is {Confidential} {Computing}?, 2024.
\newblock URL \url{https://www.ibm.com/topics/confidential-computing}.

\bibitem[{India's Ministry of Electronics and Information Technology}(2023)]{indiasministryofelectronicsandinformationtechnology2023}
{India's Ministry of Electronics and Information Technology}.
\newblock Proposed {Digital} {India} {Act}, 2023, March 2023.
\newblock URL \url{https://www.meity.gov.in/writereaddata/files/DIA_Presentation%2009.03.2023%20Final.pdf}.

\bibitem[Janardhan(2023)]{janardhanReimaginingOurInfrastructure2023}
Janardhan, S.
\newblock Reimagining {Our} {Infrastructure} for the {AI} {Age}, May 2023.
\newblock URL \url{https://about.fb.com/news/2023/05/metas-infrastructure-for-ai/}.

\bibitem[Jeon et~al.(2019)Jeon, Venkataraman, Phanishayee, Qian, Xiao, and Yang]{jeonAnalysisLargeScaleMultiTenant2019}
Jeon, M., Venkataraman, S., Phanishayee, A., Qian, J., Xiao, W., and Yang, F.
\newblock Analysis of {Large}-{Scale} {Multi}-{Tenant} {GPU} {Clusters} for {DNN} {Training} {Workloads}, August 2019.
\newblock URL \url{http://arxiv.org/abs/1901.05758}.
\newblock arXiv:1901.05758 [cs].

\bibitem[Jiang \& Cao(2023)Jiang and Cao]{jiangChinaCreateImplement2023}
Jiang, B. and Cao, A.
\newblock China to create and implement national standard for large language models in move to regulate {AI}, while using its power to transform industries.
\newblock \emph{South China Morning Post}, July 2023.
\newblock URL \url{https://www.scmp.com/tech/policy/article/3226942/china-create-and-implement-national-standard-large-language-models-move-regulate-ai-while-using-its}.

\bibitem[Jones et~al.(2023)Jones, Egan, and Rosenbach]{jonesAdvancingAdversityUkraine2023}
Jones, G., Egan, J., and Rosenbach, E.
\newblock Advancing in {Adversity}: {Ukraine}’s {Battlefield} {Technologies} and {Lessons} for the {U}.{S}.
\newblock Policy {Brief}, Belfer Center for Science and International Affairs, July 2023.
\newblock URL \url{https://www.belfercenter.org/publication/advancing-adversity-ukraines-battlefield-technologies-and-lessons-us}.

\bibitem[Kim(2023)]{kimPrivacyActivistsSlam2023}
Kim, C.
\newblock Privacy activists slam {EU}-{US} pact on data sharing.
\newblock \emph{BBC News}, July 2023.
\newblock URL \url{https://www.bbc.com/news/world-us-canada-66161135}.

\bibitem[Korolov(2023)]{korolov2023}
Korolov, M.
\newblock Data centers unprepared for new {European} energy efficiency regulations, December 2023.
\newblock URL \url{https://www.networkworld.com/article/1251883/data-centers-unprepared-for-new-european-energy-efficiency-regulations.html}.

\bibitem[Kulp et~al.(2024)Kulp, Gonzales, Smith, Heim, Puri, Vermeer, and Winkelman]{kulpHardwareEnabledGovernanceMechanisms2024}
Kulp, G., Gonzales, D., Smith, E., Heim, L., Puri, P., Vermeer, M. J.~D., and Winkelman, Z.
\newblock Hardware-{Enabled} {Governance} {Mechanisms}: {Developing} {Technical} {Solutions} to {Exempt} {Items} {Otherwise} {Classified} {Under} {Export} {Control} {Classification} {Numbers} {3A090} and {4A090}.
\newblock Technical report, RAND Corporation, January 2024.
\newblock URL \url{https://www.rand.org/pubs/working_papers/WRA3056-1.html}.

\bibitem[Köhler et~al.(2021)Köhler, Wenzel, Plauth, Böning, Gampe, Geier, and Polze]{kohlerRecognizingHPCWorkloads2021}
Köhler, S., Wenzel, L., Plauth, M., Böning, P., Gampe, P., Geier, L., and Polze, A.
\newblock Recognizing {HPC} {Workloads} {Based} on {Power} {Draw} {Signatures}.
\newblock In \emph{2021 {Ninth} {International} {Symposium} on {Computing} and {Networking} {Workshops} ({CANDARW})}, pp.\  278--284, December 2021.
\newblock \doi{10.1109/CANDARW53999.2021.00053}.
\newblock URL \url{https://ieeexplore.ieee.org/document/9644213}.

\bibitem[{Lambda Labs}(2022)]{lambdaLambdaPrivacyPolicy2022}
{Lambda Labs}.
\newblock Lambda {Privacy} {Policy}, August 2022.
\newblock URL \url{https://lambdalabs.com/legal/privacy-policy}.

\bibitem[{legislation.gov.uk}(2014)]{legislation.gov.ukDataRetentionInvestigatory}
{legislation.gov.uk}.
\newblock Data {Retention} and {Investigatory} {Powers} {Act} 2014, 2014.
\newblock URL \url{https://www.legislation.gov.uk/ukpga/2014/27/crossheading/retention-of-relevant-communications-data/enacted}.
\newblock Publisher: King's Printer of Acts of Parliament.

\bibitem[Linn(2010)]{linnRedefiningBankSecrecy2010}
Linn, C.~J.
\newblock Redefining the {Bank} {Secrecy} {Act}: {Currency} {Reporting} and the {Crime} of {Structuring}.
\newblock \emph{Santa Clara Law Review}, 50\penalty0 (2):\penalty0 407--513, January 2010.
\newblock URL \url{https://digitalcommons.law.scu.edu/lawreview/vol50/iss2/4/}.

\bibitem[Lohr(2023)]{lohr2023}
Lohr, A.
\newblock Intelligence community and {Defense} {Department} to share classified cloud services, July 2023.
\newblock URL \url{https://federalnewsnetwork.com/defense-main/2023/07/intelligence-community-and-defense-department-to-share-classified-cloud-services/}.

\bibitem[Merritt(2023)]{merritt2023}
Merritt, R.
\newblock What {Is} {NVLink}?, March 2023.
\newblock URL \url{https://blogs.nvidia.com/blog/what-is-nvidia-nvlink/}.

\bibitem[{Microsoft}(2022)]{microsoftUSNationalSecurity}
{Microsoft}.
\newblock {US} {National} {Security} {Orders} {Reports} {\textbar} {Microsoft} {CSR}, 2022.
\newblock URL \url{https://www.microsoft.com/en-us/corporate-responsibility/fisa}.

\bibitem[{Microsoft}(2023)]{microsoftlearnWhatConfidentialComputing2023}
{Microsoft}.
\newblock What is confidential computing?, October 2023.
\newblock URL \url{https://learn.microsoft.com/en-us/azure/confidential-computing/overview}.

\bibitem[{Microsoft}(2024{\natexlab{a}})]{microsoftMicrosoftPrivacyStatement2024}
{Microsoft}.
\newblock Microsoft {Privacy} {Statement} – {Microsoft} privacy, February 2024{\natexlab{a}}.
\newblock URL \url{https://privacy.microsoft.com/en-us/privacystatement}.

\bibitem[{Microsoft}(2024{\natexlab{b}})]{microsoftlearnConfidentialComputingAzure2024}
{Microsoft}.
\newblock Confidential {Computing} on {Azure}, February 2024{\natexlab{b}}.
\newblock URL \url{https://learn.microsoft.com/en-us/azure/confidential-computing/overview-azure-products}.

\bibitem[{Microsoft Azure}(2024{\natexlab{a}})]{azureAzureSustainability}
{Microsoft Azure}.
\newblock Azure {Sustainability}, 2024{\natexlab{a}}.
\newblock URL \url{https://azure.microsoft.com/en-us/explore/global-infrastructure/sustainability}.

\bibitem[{Microsoft Azure}(2024{\natexlab{b}})]{microsoftazure}
{Microsoft Azure}.
\newblock Bandwidth pricing, 2024{\natexlab{b}}.
\newblock URL \url{https://azure.microsoft.com/en-us/pricing/details/bandwidth/}.

\bibitem[{Microsoft Corporate Blogs}(2023)]{microsoftcorporateblogsMicrosoftOpenAIExtend2023}
{Microsoft Corporate Blogs}.
\newblock Microsoft and {OpenAI} extend partnership, January 2023.
\newblock URL \url{https://blogs.microsoft.com/blog/2023/01/23/microsoftandopenaiextendpartnership/}.

\bibitem[Milmo(2023)]{milmoCMAInvestigateUK2023}
Milmo, D.
\newblock {CMA} to investigate {UK} cloud computing market amid {Microsoft} and {Amazon} concerns.
\newblock \emph{The Guardian}, October 2023.
\newblock ISSN 0261-3077.
\newblock URL \url{https://www.theguardian.com/business/2023/oct/05/amazon-and-microsofts-uk-cloud-computing-dominance-faces-investigation}.

\bibitem[{Mithril Security}(2024{\natexlab{a}})]{MithrilsecurityAicert2024}
{Mithril Security}.
\newblock mithril-security/aicert, January 2024{\natexlab{a}}.
\newblock URL \url{https://github.com/mithril-security/aicert}.
\newblock original-date: 2023-07-04T07:24:29Z.

\bibitem[{Mithril Security}(2024{\natexlab{b}})]{MithrilsecurityBlindai2024}
{Mithril Security}.
\newblock mithril-security/blindai, February 2024{\natexlab{b}}.
\newblock URL \url{https://github.com/mithril-security/blindai}.
\newblock original-date: 2022-02-06T14:07:35Z.

\bibitem[Mulani \& Whittlestone(2023)Mulani and Whittlestone]{mulani2023}
Mulani, N. and Whittlestone, J.
\newblock Proposing a {Foundation} {Model} {Information}-{Sharing} {Regime} for the {UK}, June 2023.
\newblock URL \url{https://www.governance.ai/post/proposing-a-foundation-model-information-sharing-regime-for-the-uk}.

\bibitem[Murgia(2024)]{murgiaWhiteHouseScience2024}
Murgia, M.
\newblock White {House} science chief signals {US}-{China} co-operation on {AI} safety.
\newblock \emph{Financial Times}, January 2024.
\newblock URL \url{https://www.ft.com/content/94b9878b-9412-4dbc-83ba-aac2baadafd9}.

\bibitem[Nagao(2023)]{nagao2023}
Nagao, R.
\newblock Japan to pay for half of \$100m generative {AI} supercomputer - {Nikkei} {Asia}.
\newblock \emph{Nikkei Asia}, June 2023.
\newblock URL \url{https://asia.nikkei.com/Business/Technology/Japan-to-pay-for-half-of-100m-generative-AI-supercomputer}.

\bibitem[{National Energy Research Scientific Computing}(2024)]{nationalenergyresearchscientificcomputing}
{National Energy Research Scientific Computing}.
\newblock {NERSC} {Documentation}: {Roofline} {Performance} {Model}, 2024.
\newblock URL \url{https://docs.nersc.gov/tools/performance/roofline/}.

\bibitem[{National Security Telecommunications Advisory Committee}(2023)]{nationalsecuritytelecommunicationsadvisorycommittee}
{National Security Telecommunications Advisory Committee}.
\newblock {NSTAC} {Report} to the {President}: {Addressing} the {Abuse} of {Domestic} {Infrastructure} by {Foreign} {Malicious} {Actors}.
\newblock Technical report, National Security Telecommunications Advisory Committee, 2023.
\newblock URL \url{https://www.cisa.gov/sites/default/files/2024-01/NSTAC_Report_to_the_President_on_Addressing_the_Abuse_of_Domestic_Infrastructure_by_Foreign_Malicious_Actors_508c.pdf}.

\bibitem[Nevo et~al.(2023)Nevo, Lahav, Karpur, Alstott, and Matheny]{nevo2023}
Nevo, S., Lahav, D., Karpur, A., Alstott, J., and Matheny, J.
\newblock Securing {Artificial} {Intelligence} {Model} {Weights}: {Interim} {Report}.
\newblock Technical report, RAND Corporation, October 2023.
\newblock URL \url{https://www.rand.org/pubs/working_papers/WRA2849-1.html}.

\bibitem[{NVIDIA}(2023)]{nvidiaConfidentialComputeNVIDIA2023}
{NVIDIA}.
\newblock Confidential {Compute} on {NVIDIA} {Hopper} {H100}, July 2023.
\newblock URL \url{https://images.nvidia.com/aem-dam/en-zz/Solutions/data-center/HCC-Whitepaper-v1.0.pdf}.

\bibitem[{NVIDIA}(2024{\natexlab{a}})]{nvidia}
{NVIDIA}.
\newblock {NVIDIA} {BlueField} {Networking} {Platform}, 2024{\natexlab{a}}.
\newblock URL \url{https://resources.nvidia.com/en-us-accelerated-networking-resource-library-ms/}.

\bibitem[{NVIDIA}(2024{\natexlab{b}})]{nvidiaNVIDIANsightPerf}
{NVIDIA}.
\newblock {NVIDIA} {Nsight} {Perf} {SDK}, 2024{\natexlab{b}}.
\newblock URL \url{https://developer.nvidia.com/nsight-perf-sdk}.

\bibitem[{NVIDIA}(2024{\natexlab{c}})]{zotero-3386}
{NVIDIA}.
\newblock Redfish {APIs} {Support}, 2024{\natexlab{c}}.
\newblock URL \url{https://docs.nvidia.com/dgx/dgxh100-user-guide/redfish-api-supp.html}.

\bibitem[O'Brien et~al.(2023)O'Brien, Ee, and Williams]{obrienDeploymentCorrectionsIncident2023}
O'Brien, J., Ee, S., and Williams, Z.
\newblock Deployment corrections: {An} incident response framework for frontier {AI} models.
\newblock Technical report, Institute for AI Policy and Strategy, September 2023.
\newblock URL \url{https://arxiv.org/abs/2310.00328}.

\bibitem[{OECD}(2022)]{oecdMeasuringEnvironmentalImpacts2022}
{OECD}.
\newblock Measuring the environmental impacts of artificial intelligence compute and applications: {The} {AI} footprint.
\newblock {OECD} {Digital} {Economy} {Papers} 341, OECD, November 2022.
\newblock URL \url{https://www.oecd-ilibrary.org/science-and-technology/measuring-the-environmental-impacts-of-artificial-intelligence-compute-and-applications_7babf571-en}.
\newblock Series: OECD Digital Economy Papers Volume: 341.

\bibitem[{Office for Civil Rights}(2016)]{officeforcivilrights2016}
{Office for Civil Rights}.
\newblock Guidance on {HIPAA} \& {Cloud} {Computing}, October 2016.
\newblock URL \url{https://www.hhs.gov/hipaa/for-professionals/special-topics/health-information-technology/cloud-computing/index.html}.
\newblock Last Modified: 2023-02-02T10:24:50-0500.

\bibitem[{OFWAT}(2024)]{thewaterservicesregulationauthority}
{OFWAT}.
\newblock The {Water} {Services} {Regulation} {Authority}, 2024.
\newblock URL \url{https://www.ofwat.gov.uk/regulated-companies/ofwat-industry-overview/licences/}.

\bibitem[{OpenMined}(2023)]{HowAuditAI2023}
{OpenMined}.
\newblock How to {Audit} an {AI} {Model} {Owned} by {Someone} {Else} ({Part} 1): {An} introduction to state-of-the-art {AI} auditing infrastructure, June 2023.
\newblock URL \url{https://blog.openmined.org/ai-audit-part-1/}.

\bibitem[{OpenMined}(2024)]{OpenMinedPySyft2024}
{OpenMined}.
\newblock {OpenMined}/{PySyft}, March 2024.
\newblock URL \url{https://github.com/OpenMined/PySyft}.
\newblock original-date: 2017-07-18T20:41:16Z.

\bibitem[Owen(2024)]{owenHowPredictableLanguage2024}
Owen, D.
\newblock How predictable is language model benchmark performance?, January 2024.
\newblock URL \url{http://arxiv.org/abs/2401.04757}.
\newblock arXiv:2401.04757 [cs].

\bibitem[Pal et~al.(2021)Pal, Gorczynski, and Schmidt]{pal2021}
Pal, B., Gorczynski, S., and Schmidt, D.
\newblock Overview of {Data} {Transfer} {Costs} for {Common} {Architectures} {\textbar} {AWS} {Architecture} {Blog}, June 2021.
\newblock URL \url{https://aws.amazon.com/blogs/architecture/overview-of-data-transfer-costs-for-common-architectures/}.
\newblock Section: Amazon EC2.

\bibitem[Patel et~al.(2023)Patel, Choukse, Zhang, Goiri, Warrier, Mahalingam, and Bianchini]{patelPOLCAPowerOversubscription2023}
Patel, P., Choukse, E., Zhang, C., Goiri, I., Warrier, B., Mahalingam, N., and Bianchini, R.
\newblock {POLCA}: {Power} {Oversubscription} in {LLM} {Cloud} {Providers}, August 2023.
\newblock URL \url{http://arxiv.org/abs/2308.12908}.
\newblock arXiv:2308.12908 [cs].

\bibitem[Pilz \& Heim(2023)Pilz and Heim]{pilz2023}
Pilz, K. and Heim, L.
\newblock Compute at {Scale}: {A} {Broad} {Investigation} into the {Data} {Center} {Industry}, November 2023.
\newblock URL \url{https://arxiv.org/abs/2311.02651v4}.

\bibitem[Pilz et~al.(2024)Pilz, Heim, and Brown]{pilzIncreasedComputeEfficiency2024}
Pilz, K., Heim, L., and Brown, N.
\newblock Increased {Compute} {Efficiency} and the {Diffusion} of {AI} {Capabilities}, February 2024.
\newblock URL \url{http://arxiv.org/abs/2311.15377}.
\newblock arXiv:2311.15377 [cs].

\bibitem[{Prime Minister's Office} et~al.(2023){Prime Minister's Office}, {FCDO UK}, and {DSIT UK}]{primeministersoffice10downingstreetBletchleyDeclarationCountries2023}
{Prime Minister's Office}, {FCDO UK}, and {DSIT UK}.
\newblock The {Bletchley} {Declaration} by {Countries} {Attending} the {AI} {Safety} {Summit}, 1-2 {November} 2023, November 2023.
\newblock URL \url{https://www.gov.uk/government/publications/ai-safety-summit-2023-the-bletchley-declaration/the-bletchley-declaration-by-countries-attending-the-ai-safety-summit-1-2-november-2023}.

\bibitem[Rabinovitsj(2023)]{rabinovitsjOpeningAIInfrastructure}
Rabinovitsj, D.
\newblock Opening {AI} {Infrastructure}: {Ushering} {In} {The} {Age} {Of} {GenAI}, December 2023.
\newblock URL \url{https://drive.google.com/file/d/1ud1JZqco2868AvmkNkrA-Axp-74PvwWx/view?usp=embed_facebook}.

\bibitem[Rep.~Collins(2018)]{rep.collinsText4943115th2018}
Rep.~Collins, D. R.-G.-.
\newblock Text - {H}.{R}.4943 - 115th {Congress} (2017-2018): {CLOUD} {Act}, February 2018.
\newblock URL \url{https://www.congress.gov/bill/115th-congress/house-bill/4943/text}.
\newblock Archive Location: 2018-02-06.

\bibitem[Richter(2024)]{InfographicAmazonMaintains2024}
Richter, F.
\newblock Amazon {Maintains} {Cloud} {Lead} as {Microsoft} {Edges} {Closer}, February 2024.
\newblock URL \url{https://www.statista.com/chart/18819/worldwide-market-share-of-leading-cloud-infrastructure-service-providers}.

\bibitem[{Sanction Scanner}(2024)]{sanctionscannerWhatDifferenceSmurfing}
{Sanction Scanner}.
\newblock What is the {Difference} {Between} {Smurfing} and {Structuring}?, 2024.
\newblock URL \url{https://sanctionscanner.com/blog/what-is-the-difference-between-smurfing-and-structuring-594}.

\bibitem[Sastry et~al.(2024)Sastry, Heim, Belfield, Anderljung, Brundage, Hazell, O'Keefe, Hadfield, Ngo, Pilz, Gor, Bluemke, Shoker, Egan, Trager, Avin, Weller, Bengio, and Coyle]{sastryComputingPowerGovernance2024}
Sastry, G., Heim, L., Belfield, H., Anderljung, M., Brundage, M., Hazell, J., O'Keefe, C., Hadfield, G.~K., Ngo, R., Pilz, K., Gor, G., Bluemke, E., Shoker, S., Egan, J., Trager, R.~F., Avin, S., Weller, A., Bengio, Y., and Coyle, D.
\newblock Computing {Power} and the {Governance} of {Artificial} {Intelligence}, February 2024.
\newblock URL \url{http://arxiv.org/abs/2402.08797}.
\newblock arXiv:2402.08797 [cs].

\bibitem[{Secretary of State for Science, Innovation and Technology}(2023)]{secretaryofstateforscienceinnovationandtechnologyProinnovationApproachAI2023}
{Secretary of State for Science, Innovation and Technology}.
\newblock A pro-innovation approach to {AI} regulation.
\newblock Policy paper Command Paper Number: 815, GOV.UK., August 2023.
\newblock URL \url{https://www.gov.uk/government/publications/ai-regulation-a-pro-innovation-approach/white-paper}.
\newblock ISBN: 978-1-5286-4009-1.

\bibitem[{Senator Scott Wiener}(2024)]{SenatorWienerIntroduces2024}
{Senator Scott Wiener}.
\newblock Senator {Wiener} {Introduces} {Legislation} to {Ensure} {Safe} {Development} of {Large}-{Scale} {Artificial} {Intelligence} {Systems} and {Support} {AI} {Innovation} in {California}, February 2024.
\newblock URL \url{https://sd11.senate.ca.gov/news/20240208-senator-wiener-introduces-legislation-ensure-safe-development-large-scale-artificial}.

\bibitem[Sevilla et~al.(2022)Sevilla, Heim, Ho, Besiroglu, Hobbhahn, and Villalobos]{sevillaComputeTrendsThree2022}
Sevilla, J., Heim, L., Ho, A., Besiroglu, T., Hobbhahn, M., and Villalobos, P.
\newblock Compute {Trends} {Across} {Three} {Eras} of {Machine} {Learning}.
\newblock In \emph{2022 {International} {Joint} {Conference} on {Neural} {Networks} ({IJCNN})}, pp.\  1--8, July 2022.
\newblock \doi{10.1109/IJCNN55064.2022.9891914}.
\newblock URL \url{http://arxiv.org/abs/2202.05924}.
\newblock arXiv:2202.05924 [cs].

\bibitem[Shavit et~al.(2023)Shavit, Agarwal, Brundage, and Adler]{shavitPracticesGoverningAgentic2023}
Shavit, Y., Agarwal, S., Brundage, M., and Adler, S.
\newblock Practices for {Governing} {Agentic} {AI} {Systems}.
\newblock Research {Paper}, OpenAI, December 2023.
\newblock URL \url{https://cdn.openai.com/papers/practices-for-governing-agentic-ai-systems.pdf}.

\bibitem[Sheehan(2024)]{sheehanTracingRootsChina2024}
Sheehan, M.
\newblock Tracing the {Roots} of {China}’s {AI} {Regulations}.
\newblock Technical report, Carnegie Endowment for International Peace, February 2024.
\newblock URL \url{https://carnegieendowment.org/2024/02/27/tracing-roots-of-china-s-ai-regulations-pub-91815}.

\bibitem[Shoeybi et~al.(2020)Shoeybi, Patwary, Puri, LeGresley, Casper, and Catanzaro]{shoeybi2020}
Shoeybi, M., Patwary, M., Puri, R., LeGresley, P., Casper, J., and Catanzaro, B.
\newblock Megatron-{LM}: {Training} {Multi}-{Billion} {Parameter} {Language} {Models} {Using} {Model} {Parallelism}, March 2020.
\newblock URL \url{http://arxiv.org/abs/1909.08053}.
\newblock arXiv:1909.08053 [cs].

\bibitem[Sliwko \& Getov(2016)Sliwko and Getov]{sliwkoAGOCSAccurateGoogle2016}
Sliwko, L. and Getov, V.
\newblock {AGOCS} — {Accurate} {Google} {Cloud} {Simulator} {Framework}.
\newblock In \emph{2016 {Intl} {IEEE} {Conferences} on {Ubiquitous} {Intelligence} \& {Computing}, {Advanced} and {Trusted} {Computing}, {Scalable} {Computing} and {Communications}, {Cloud} and {Big} {Data} {Computing}, {Internet} of {People}, and {Smart} {World} {Congress} ({UIC}/{ATC}/{ScalCom}/{CBDCom}/{IoP}/{SmartWorld})}, pp.\  550--558, January 2016.
\newblock \doi{10.1109/UIC-ATC-ScalCom-CBDCom-IoP-SmartWorld.2016.0095}.
\newblock URL \url{https://ieeexplore.ieee.org/document/7816891}.

\bibitem[Smith(2017)]{smithNeedDigitalGeneva2017}
Smith, B.
\newblock The need for a {Digital} {Geneva} {Convention}, February 2017.
\newblock URL \url{https://blogs.microsoft.com/on-the-issues/2017/02/14/need-digital-geneva-convention/}.

\bibitem[Smith(2023)]{smith2023}
Smith, B.
\newblock How do we best govern {AI}?, May 2023.
\newblock URL \url{https://blogs.microsoft.com/on-the-issues/2023/05/25/how-do-we-best-govern-ai/}.

\bibitem[Tang et~al.(2022)Tang, Chen, Weiss, Frey, McDonald, Bestor, Yee, Arcand, Byun, Edelman, Hubbell, Jones, Kepner, Klein, Michaleas, Michaleas, Milechin, Mullen, Prout, Reuther, Rosa, Bowne, McEvoy, Li, Tiwari, Gadepally, and Samsi]{tangMITSupercloudWorkload2022}
Tang, B.~J., Chen, Q., Weiss, M.~L., Frey, N., McDonald, J., Bestor, D., Yee, C., Arcand, W., Byun, C., Edelman, D., Hubbell, M., Jones, M., Kepner, J., Klein, A., Michaleas, A., Michaleas, P., Milechin, L., Mullen, J., Prout, A., Reuther, A., Rosa, A., Bowne, A., McEvoy, L., Li, B., Tiwari, D., Gadepally, V., and Samsi, S.
\newblock The {MIT} {Supercloud} {Workload} {Classification} {Challenge}.
\newblock In \emph{2022 {IEEE} {International} {Parallel} and {Distributed} {Processing} {Symposium} {Workshops} ({IPDPSW})}, pp.\  708--714, May 2022.
\newblock \doi{10.1109/IPDPSW55747.2022.00122}.
\newblock URL \url{http://arxiv.org/abs/2204.05839}.
\newblock arXiv:2204.05839 [cs].

\bibitem[Terai et~al.(2017)Terai, Kashiwaki, and Shoji]{teraiWorkloadClassificationPerformance2017}
Terai, M., Kashiwaki, R., and Shoji, F.
\newblock Workload {Classification} and {Performance} {Analysis} using {Job} {Metrics} in the {K} computer.
\newblock {IPSJ} {SIG} {Technical} {Report}~13, Information Processing Society of Japan, December 2017.
\newblock URL \url{https://ipsj.ixsq.nii.ac.jp/ej/?action=repository_uri&item_id=184893&file_id=1&file_no=1}.
\newblock Vol. 2017-HPC-162.

\bibitem[{The European High Performance Computing Joint Undertaking}(2023)]{theeuropeanhighperformancecomputingjointundertaking2023}
{The European High Performance Computing Joint Undertaking}.
\newblock Open {Call} to {Support} {HPC}-powered {Artificial} {Intelligence} ({AI}) {Applications} - {European} {Commission}, November 2023.
\newblock URL \url{https://eurohpc-ju.europa.eu/open-call-support-hpc-powered-artificial-intelligence-ai-applications-2023-11-28_en}.

\bibitem[{The White House}(2023{\natexlab{a}})]{thewhitehouse2023}
{The White House}.
\newblock Voluntary {AI} {Commitments}, September 2023{\natexlab{a}}.
\newblock URL \url{https://www.whitehouse.gov/wp-content/uploads/2023/09/Voluntary-AI-Commitments-September-2023.pdf}.

\bibitem[{The White House}(2023{\natexlab{b}})]{thewhitehouseExecutiveOrderSafe2023}
{The White House}.
\newblock Executive {Order} on the {Safe}, {Secure}, and {Trustworthy} {Development} and {Use} of {Artificial} {Intelligence}.
\newblock Technical report, The White House, October 2023{\natexlab{b}}.
\newblock URL \url{https://www.whitehouse.gov/briefing-room/presidential-actions/2023/10/30/executive-order-on-the-safe-secure-and-trustworthy-development-and-use-of-artificial-intelligence/}.

\bibitem[{The White House}(2023{\natexlab{c}})]{thewhitehouseFACTSHEETBidenHarris2023}
{The White House}.
\newblock {FACT} {SHEET}: {Biden}-{Harris} {Administration} {Secures} {Voluntary} {Commitments} from {Leading} {Artificial} {Intelligence} {Companies} to {Manage} the {Risks} {Posed} by {AI}, July 2023{\natexlab{c}}.
\newblock URL \url{https://www.whitehouse.gov/briefing-room/statements-releases/2023/07/21/fact-sheet-biden-harris-administration-secures-voluntary-commitments-from-leading-artificial-intelligence-companies-to-manage-the-risks-posed-by-ai/}.

\bibitem[{The White House}(2023{\natexlab{d}})]{thewhitehouseG7HiroshimaLeaders2023}
{The White House}.
\newblock G7 {Hiroshima} {Leaders}’ {Communiqué}, May 2023{\natexlab{d}}.
\newblock URL \url{https://www.whitehouse.gov/briefing-room/statements-releases/2023/05/20/g7-hiroshima-leaders-communique/}.

\bibitem[Tirmazi et~al.(2020)Tirmazi, Barker, Deng, Haque, Qin, Hand, Harchol-Balter, and Wilkes]{tirmaziBorgNextGeneration2020}
Tirmazi, M., Barker, A., Deng, N., Haque, M.~E., Qin, Z.~G., Hand, S., Harchol-Balter, M., and Wilkes, J.
\newblock Borg: the next generation.
\newblock In \emph{Proceedings of the {Fifteenth} {European} {Conference} on {Computer} {Systems}}, pp.\  1--14, Heraklion Greece, April 2020. ACM.
\newblock ISBN 978-1-4503-6882-7.
\newblock \doi{10.1145/3342195.3387517}.
\newblock URL \url{https://dl.acm.org/doi/10.1145/3342195.3387517}.

\bibitem[Trager et~al.(2023)Trager, Harack, Reuel, Carnegie, Heim, Ho, Kreps, Lall, Larter, Ó~hÉigeartaigh, Staffell, and Villalobos]{tragerInternationalGovernanceCivilian2023}
Trager, R., Harack, B., Reuel, A., Carnegie, A., Heim, L., Ho, L., Kreps, S., Lall, R., Larter, O., Ó~hÉigeartaigh, S., Staffell, S., and Villalobos, J.~J.
\newblock International {Governance} of {Civilian} {AI}: {A} {Jurisdictional} {Certification} {Approach}.
\newblock Whitepaper, Oxford Martin AI Governance initiative and Centre for the Governance of AI, August 2023.
\newblock URL \url{https://cdn.governance.ai/International_Governance_of_Civilian_AI_OMS.pdf}.

\bibitem[{UK DSIT}(2023)]{departmentforscienceinnovationandtechnologyCapabilitiesRisksFrontier2023}
{UK DSIT}.
\newblock Capabilities and risks from frontier {AI}: {A} discussion paper on the need for further research into {AI} risk.
\newblock Technical report, DSIT UK, October 2023.
\newblock URL \url{https://assets.publishing.service.gov.uk/media/65395abae6c968000daa9b25/frontier-ai-capabilities-risks-report.pdf}.

\bibitem[{UK DSIT} \& Donelan(2023){UK DSIT} and Donelan]{departmentforscienceinnovationandtechnology2023}
{UK DSIT} and Donelan, M.
\newblock Technology {Secretary} announces investment boost making {British} {AI} supercomputing 30 times more powerful, November 2023.
\newblock URL \url{https://www.gov.uk/government/news/technology-secretary-announces-investment-boost-making-british-ai-supercomputing-30-times-more-powerful}.

\bibitem[{UK Government}(2024)]{gov.uk2024}
{UK Government}.
\newblock {AI} {Safety} {Summit} 2023 - {GOV}.{UK}, February 2024.
\newblock URL \url{https://www.gov.uk/government/topical-events/ai-safety-summit-2023}.

\bibitem[{UK Research and Innovation}(2023)]{ukresearchandinnovation2023}
{UK Research and Innovation}.
\newblock £300 million to launch first phase of new {AI} {Research} {Resource}, November 2023.
\newblock URL \url{https://www.ukri.org/news/300-million-to-launch-first-phase-of-new-ai-research-resource/}.

\bibitem[{US BIS} \& {US DOC}(2023){US BIS} and {US DOC}]{bureauofindustryandsecurity2023}
{US BIS} and {US DOC}.
\newblock Implementation of {Additional} {Export} {Controls}: {Certain} {Advanced} {Computing} {Items}; {Supercomputer} and {Semiconductor} {End} {Use}; {Updates} and {Corrections}, October 2023.
\newblock URL \url{https://www.federalregister.gov/documents/2023/10/25/2023-23055/implementation-of-additional-export-controls-certain-advanced-computing-items-supercomputer-and}.

\bibitem[{U.S. National Science Foundation}(2024)]{u.s.nationalsciencefoundation2024}
{U.S. National Science Foundation}.
\newblock Democratizing the future of {AI} {R}\&{D}: {NSF} to launch {National} {AI} {Research} {Resource} pilot {\textbar} {NSF} - {National} {Science} {Foundation}, January 2024.
\newblock URL \url{https://new.nsf.gov/news/democratizing-future-ai-rd-nsf-launch-national-ai}.

\bibitem[{U.S. Securities and Exchange Commission}(2003)]{u.s.securitiesandexchangecommissionRetentionRecordsRelevant2003}
{U.S. Securities and Exchange Commission}.
\newblock Retention of {Records} {Relevant} to {Audits} and {Reviews}, March 2003.
\newblock URL \url{https://www.sec.gov/rules/2003/01/retention-records-relevant-audits-and-reviews}.

\bibitem[Vailshery(2024)]{vailsheryWorldwideInfrastructureService2024}
Vailshery, L.~S.
\newblock Worldwide infrastructure as a service ({IaaS}) and platform as a service ({PaaS}) hyperscaler market share from 2020 to 2023, by vendor, February 2024.
\newblock URL \url{https://www.statista.com/statistics/1202770/hyperscaler-iaas-paas-market-share/}.

\bibitem[Vanian(2024)]{vanian2024}
Vanian, J.
\newblock {HPE} hacked by same {Russian} intelligence group that hit {Microsoft}, January 2024.
\newblock URL \url{https://www.cnbc.com/2024/01/24/hpe-hit-by-russian-intelligence-group-that-hacked-microsoft.html}.

\bibitem[Ward \& Hu(2024)Ward and Hu]{wardUSAntitrustInquiry2024}
Ward, J. and Hu, K.
\newblock {US} antitrust inquiry targets {OpenAI} and {Anthropic}'s deals with {Big} {Tech}.
\newblock \emph{Reuters}, January 2024.
\newblock URL \url{https://www.reuters.com/technology/ftc-launches-inquiry-into-generative-ai-investments-partnerships-2024-01-25/}.

\bibitem[Weiss et~al.(2022)Weiss, McDonald, Bestor, Yee, Edelman, Jones, Prout, Bowne, McEvoy, Gadepally, and Samsi]{weissEvaluationLowOverhead2022}
Weiss, M.~L., McDonald, J., Bestor, D., Yee, C., Edelman, D., Jones, M., Prout, A., Bowne, A., McEvoy, L., Gadepally, V., and Samsi, S.
\newblock An {Evaluation} of {Low} {Overhead} {Time} {Series} {Preprocessing} {Techniques} for {Downstream} {Machine} {Learning}, September 2022.
\newblock URL \url{http://arxiv.org/abs/2209.05300}.
\newblock arXiv:2209.05300 [cs].

\bibitem[Weng et~al.(2022)Weng, Xiao, Yu, Wang, Wang, He, Li, Zhang, Lin, and Ding]{wengMLaaSWildWorkload2022}
Weng, Q., Xiao, W., Yu, Y., Wang, W., Wang, C., He, J., Li, Y., Zhang, L., Lin, W., and Ding, Y.
\newblock {MLaaS} in the wild: {Workload} analysis and scheduling in {Large}-{Scale} heterogeneous {GPU} clusters.
\newblock In \emph{19th {USENIX} symposium on networked systems design and implementation ({NSDI} 22)}, pp.\  945--960, Renton, WA, April 2022. USENIX Association.
\newblock ISBN 978-1-939133-27-4.
\newblock URL \url{https://www.usenix.org/conference/nsdi22/presentation/weng}.

\bibitem[Whittlestone \& Clark(2021)Whittlestone and Clark]{whittlestoneWhyHowGovernments2021}
Whittlestone, J. and Clark, J.
\newblock Why and {How} {Governments} {Should} {Monitor} {AI} {Development}, August 2021.
\newblock URL \url{http://arxiv.org/abs/2108.12427}.
\newblock arXiv:2108.12427 [cs].

\bibitem[Whittlestone et~al.(2023)Whittlestone, Avin, Heim, Anderljung, and Sastry]{whittlestone2023}
Whittlestone, J., Avin, S., Heim, L., Anderljung, M., and Sastry, G.
\newblock Response to the {UK}'s {Future} of {Compute} {Review}.
\newblock Technical report, Centre for the Governance of AI, March 2023.
\newblock URL \url{https://www.governance.ai/research-paper/response-to-the-uks-future-of-compute-review}.

\bibitem[{Wikipedia contributors}(2023)]{enwiki:1169157291}
{Wikipedia contributors}.
\newblock Hardware performance counter, August 2023.
\newblock URL \url{https://en.wikipedia.org/w/index.php?title=Hardware_performance_counter&oldid=1169157291}.

\bibitem[Williams et~al.(2008)Williams, Waterman, and Patterson]{williams2008}
Williams, S., Waterman, A., and Patterson, D.
\newblock Roofline: {An} {Insightful} {Visual} {Performance} {Model} for {Floating}-{Point} {Programs} and {Multicore} {Architectures}, October 2008.
\newblock URL \url{https://people.eecs.berkeley.edu/~kubitron/cs252/handouts/papers/RooflineVyNoYellow.pdf}.

\end{thebibliography}
\bibliographystyle{icml2024}
\end{CJK*}

\end{document}